\documentclass[aps,twocolumn,pra,superscriptaddress,amsmath,amssymb]{revtex4-2} 
\usepackage{dcolumn}
\usepackage{bm}
\usepackage{amsmath}
\usepackage{mathrsfs}
\usepackage{txfonts}
\usepackage[T1]{fontenc}
\usepackage{xspace}
\usepackage{ulem}
\usepackage{comment}
\usepackage{braket}
\usepackage{lipsum}
\usepackage{mathtools}
\newcommand{\chem}[1]{\mathrm{#1}}

\newcommand{\bn}[1]{\mathbf{#1}}
\ifx\pdfoutput\undefined
\usepackage[dvipdfmx]{graphicx}
\usepackage[dvipdfmx]{hyperref}
\usepackage[dvipdfmx]{color}
\usepackage[dvipdfmx]{xcolor}
\else
\usepackage{graphicx}
\usepackage{hyperref}
\usepackage{color}
\usepackage{xcolor}
\usepackage[version=3]{mhchem}
\fi
\hypersetup{
        colorlinks=true,
        citecolor=blue,
        urlcolor=blue,
        linkcolor=blue
}

\begin{document}
\let\emph\textit

\title{
Flavor-wave theory with quasiparticle damping at finite temperatures: Application to chiral edge modes in the Kitaev model
}
\author{Shinnosuke Koyama}
\author{Joji Nasu}
\affiliation{
  Department of Physics, Tohoku University, Sendai, Miyagi 980-8578, Japan
}

\date{\today}
\begin{abstract}

We propose a theoretical framework to investigate elementary excitations at finite temperatures within a localized electron model that describes the interactions between multiple degrees of freedom, such as quantum spin models and Kugel-Khomskii models. Thus far, their excitation structures have been mainly examined using the linear flavor-wave theory, an SU($N$) generalization of the linear spin-wave theory. These techniques introduce noninteracting bosonic quasiparticles as elementary excitations from the ground state, thereby elucidating numerous physical phenomena, including excitation spectra and transport properties characterized by topologically nontrivial band structures. Nevertheless, the interactions between quasiparticles cannot be ignored in systems exemplified by $S=1/2$ quantum spin models, where strong quantum fluctuations are present. Recent studies have investigated the effects of quasiparticle damping at zero temperature in such models. In our study, extending this approach to the flavor-wave theory for general localized electron models, we construct a comprehensive method to calculate excitation spectra with the quasiparticle damping at finite temperatures. We apply our method to the Kitaev model under magnetic fields, a typical example of models with topologically nontrivial magnon bands. Our calculations reveal that chiral edge modes undergo significant damping in weak magnetic fields, amplifying the damping rate by the temperature increase. This effect is caused by collisions with thermally excited quasiparticles. Since our approach starts from a general Hamiltonian, it will be widely applicable to other localized systems, such as spin-orbital coupled systems derived from multi-orbital Hubbard models in the strong correlation limit.
\end{abstract}
\maketitle


\section{Introduction}

In modern condensed matter physics, the topological nature of electronic systems is one of the crucial ingredients enriching physical phenomena.
For example, in quantum Hall systems, the Chern number, a topological invariant in electronic band dispersions, is closely related to a quantized value of the Hall conductance and the number of chiral edge modes~\cite{thouless1982,kohmoto1985}.
This concept has been applied to magnons, which are collective spin-wave excitations from a magnetically ordered state in insulating magnets~\cite{katsura2010,onose2010,mcclarty2022}; the Chern number of magnon bands possibly becomes nonzero, leading to the presence of chiral edge modes in the gap between the bands. Since magnons are charge-neutral bosonic excitations, quantum Hall effects do not occur.
Instead, one expects the emergence of thermal Hall effects~\cite{katsura2010,onose2010,mcclarty2022}, which is a phenomenon exhibiting a thermal current induced by a temperature gradient along the perpendicular direction when magnon bands possess nonzero Berry curvature~\cite{berry1984}.
Such magnons are called topological magnons, usually caused by anisotropic spin interactions beyond the Heisenberg coupling or non-collinear magnetic orders.
Thus far, the emergence of topological magnons has been theoretically proposed in spin systems with the Dzyaloshinskii-Moriya or Kitaev-type interactions~\cite{owerre2016,owerre2017,laurell2018,mcclarty2018,mcclarty2022,zhang2023_arxiv,chuan_zhang2021_prb}, and the thermal Hall effect has been observed the pyrochlore materials, $\mathrm{Lu_2V_2O_7}$, $\mathrm{Ho_2V_2O_7}$, $\mathrm{In_2Mn_2O_7}$ and $\mathrm{Tb_2Ti_2O_7}$~\cite{onose2010,ideue2012,hirschberger2015_science}, the layered honeycomb material $\mathrm{VI_3}$~\cite{czajka2023,zhang2021_prl},
and the layered Kagome materials, Cu(1,3-benzenedicarboxylate) and Cd-kapellasite~\cite{hirschberger2015_prl,Akazawa2020}.

Since magnons are collective excitations from a magnetically ordered state, they behave as bosons with zero chemical potential and possess positive energies.
These characteristics lead to crucially different behavior from fermionic systems.
In quantum Hall systems composed of electrons, the thermal Hall coefficient divided by temperature takes a nonzero quantized value determined by the Chern number of each band in the zero-temperature limit.
On the other hand, in its magnonic counterpart, the thermal Hall coefficient divided by temperature must vanish at zero temperature because chiral magnon edge modes appear not across zero energy but in between positive-energy bands~\cite{shindou2013,matsumoto2014}.
This consideration implies that thermally excited magnons play a crucial role in the transport phenomena originating from the chiral edge modes.

The topological nature of collective excitations has been discussed not only on magnons but also on other bosonic quasiparticles, such as triplons from a singlet-dimer covered ground state~\cite{romhanyi2015,mcclarty2017topological,zayed2014}.
A comprehensive approach to include these cases has been developed as the flavor-wave theory, an extension of the spin-wave theory.
This method was originally proposed for localized SU($N$) systems~\cite{joshi1999,Lauchli2006,Tsunetsugu2006,Kim_flavor-wave2017}, but it can be applied to other systems with interacting local degrees of freedom.
By using the generalized theory, it has been proposed that the Shastry–Sutherland model possesses the band structures of triplon excitations with nonzero Chern numbers due to Dzyaloshinskii–Moriya interactions~\cite{romhanyi2015}.

While the topological character of collective excitations has been examined under the assumption of noninteracting quasiparticles within a linear flavor-wave approximation, nonlinear terms beyond this approximation are inevitably present, which appear due to quantum fluctuations intrinsic to localized spins.
This effect gives rise to interactions between collective excitations and should be addressed appropriately, particularly for $S=1/2$ quantum spin systems.
In magnetically ordered systems, the interactions cause a finite lifetime to magnons, namely the damping of magnons, at zero temperature~\cite{zhitomirsky1999,zhitomirsky2013,mourigal2010,fuhrman2012} and finite temperatures~\cite{mook2021}.
The effects of the magnon damping have been introduced to understand the broad spectrum observed in inelastic neutron scattering measurements, and calculations incorporating the magnon damping have successfully explained experimental results.
Nevertheless, it is often challenging to address the impact of the magnon damping on the topological nature of magnons and related phenomena, such as the thermal Hall effect~\cite{chernyshev2016,mcclarty2022}.

Recently, it has been reported that the thermal Hall conductivity originating from collective excitations is much smaller than the theoretically predicted value obtained within the free-magnon picture in the quasi-two-dimensional quantum magnets SrCu$_2$(BO$_3$)$_2$ with triplon excitations~\cite{suetsugu2022} and Cr$_2$Ge$_2$Te$_6$ with magnon excitations~\cite{choi2023}.
These results suggest that the interactions between quasiparticles are crucial in topological transport phenomena.
Therefore, clarifying the damping effects of topological bosonic quasiparticles at zero and nonzero temperatures is highly desired.

Other well-known simple examples exhibiting topological excitations are the Kitaev-related systems under magnetic fields.
In the pure Kitaev model~\cite{kitaev}, magnetic fields along the [111] direction yield topologically nontrivial magnon bands~\cite{mcclarty2018,joshi2018}.
Moreover, recent theoretical studies based on the linear spin-wave theory have reported that the Chern number of each magnon band changes by introducing other interactions, such as the Heisenberg one and varying magnetic-field directions~\cite{mcclarty2018,joshi2018,li2022_prb,koyama2021,zhang2021_prb,chern2021_prl,chern2020_prr,chuan_zhang2021_prb,janssen2019_cm,cookmeyer2018,zhang2023}.
Correspondingly, the thermal Hall effect has been observed in the Kitaev candidate material $\alpha$-RuCl$_3$~\cite{czajka2023,young2023,hentrich2019,kasahara2018_prl,kasahara2018_nature,yokoi2021,yamashita2020,czajka2021_nature,bruin2022,kasahara2022,lefrancois2022,imamura2023_arxiv}.
Although the experimental results imply the presence of topological quasiparticles in this compound, there is a debate about which quasiparticles are responsible for the thermal Hall effect; several works have proposed that only magnons are responsible~\cite{czajka2021_nature,czajka2023}, while others have suggested Majorana fermions or phonons as possible origins~\cite{kasahara2018_nature,yokoi2021,bruin2022,lefrancois2022,imamura2023_arxiv}.

Meanwhile, it has been pointed out that strong magnon damping is crucial in understanding spin dynamics~\cite{winter2017_nc,maksimov2020}, suggesting that nonlinear terms beyond the linear spin-wave approximations are also crucial for the topological nature of the magnons in the Kitaev-related model under magnetic fields.
Topologically nontrivial magnon bands accompany chiral edge modes, and their stability has been examined at zero temperature~\cite{mcclarty2018}.
On the other hand, it has been known that possible magnon-damping processes at finite temperatures are entirely different from that at zero temperature~\cite{zhitomirsky2013}; the latter only originates from spontaneous decay by splitting into multiple magnons, but collisions with other thermally excited magnons also contribute to the finite-temperature processes~\cite{mook2021}.
Therefore, examining the finite-temperature effects of magnon damping on chiral edge modes is needed to clarify the role of topological magnons in thermal transport.

In this paper, we construct  a calculation framework capable of handling nonlinear terms beyond the linear flavor-wave theory for generalized many-body models with interacting local degrees of freedom at finite temperatures.
We apply this approach to the Kitaev model under magnetic fields. 
In the present scheme, we start from the Hamiltonian consisting of one-body and two-body terms for local degrees of freedom and apply the mean-field (MF) theory with the arbitrary number of sublattices.
Utilizing the generalized Holstein-Primakoff transformation, we rewrite the model Hamiltonian into a bosonic representation.
This bosonic Hamiltonian consists of bilinear terms of bosons and higher-order terms beyond these contributions.
The former describes noninteracting bosons.
We treat the latter contributions corresponding to interactions between bosons by the self-consistent imaginary Dyson equation (iDE) approach, which enables us to avoid the appearance of unphysical divergences in excitation spectra. 
Here, we apply the present scheme to the Kitaev model under magnetic fields on a honeycomb lattice, which possesses magnon bands with nonzero Chern numbers.
In this model, the MF solution is a forced ferromagnetic state regardless of the magnetic-field intensity.
To examine the effect of the magnon damping on edge states, we calculate magnon damping rates and magnon spectra on clusters with open boundaries.
The chiral edge modes are strongly damped at zero temperature in weak magnetic fields due to a magnon collapsing into two.
The damping rate decreases with increasing the intensity of an applied magnetic field and vanishes above a certain field intensity.
The disappearance of overlap between the magnon dispersions and the two-magnon continuum determines the critical value of the field intensity.
With increasing temperature, the damping rate of the magnon edge modes increases due to another magnon damping process, collisions with thermally excited magnons.
The damping rate is nonzero, even above the critical magnetic field.
We demonstrate that the nonlinear terms giving rise to the magnon damping strongly affect not only bulk spectra but also edge modes at finite temperatures.
Our results suggest that the effect of the magnon damping is relevant to finite-temperature topological transport phenomena such as the thermal Hall effect.

This paper is organized as follows. 
In the next section, we present the method used in the present study.
The MF theory and generalized Holstein-Primakoff transformation are described in Sections~\ref{method:MF} and \ref{method:GHP}, respectively.
We introduce the linear flavor-wave theory in Sec.~\ref{method:LSW}.
The extensions considering the nonlinear parts of flavor-waves are given in Sec.~\ref{method:NLSW}.
Section~\ref{method:iDE approach} provides the method we have developed to evaluate effects of the damping of collective modes in the nonlinear flavor-wave theory at finite temperatures.
In Sec.~\ref{method:kitaev model}, we introduce the $S=1/2$ Kitaev honeycomb model to which our method is applied in the present study.
The results are given in Sec.~\ref{result}.
First, we show the MF results for the two systems with different boundary conditions in Sec.~\ref{sec:mean-field}.
In Sec.~\ref{result:low magnetic field}, we show the magnon spectra at low magnetic fields at zero temperatures.
The results for higher magnetic fields at finite temperatures are given in Sec.~\ref{result:high magnetic field}.
In Sec.~\ref{discussion}, we discuss the magnetic-field dependence and temperature evolution of the magnon damping of chiral edge modes.
We also mention the relevance to real materials and observables such as the thermal Hall effect.
Finally, Sec.~\ref{summary} is devoted to the summary.

\section{Method}
\label{method}
\subsection{Mean-field theory}
\label{method:MF}

Before showing the details of the flavor-wave theory, we first explain the MF approximation.
We start from a general localized model, which is given by
\begin{align}
  \label{eq:H}
  \mathcal{H}= \frac{1}{2}\sum_{i,j} \sum_{\gamma \gamma^\prime}
  {J}^{\gamma\gamma^\prime}_{ij} \mathcal{O}_{i}^{\gamma} \mathcal{O}_{j}^{\gamma'} 
  - \sum_{i}\sum_{\gamma} h_{i}^{\gamma}\mathcal{O}^{\gamma}_{i},
\end{align}
where $\mathcal{O}_i^\gamma$ represents the $\gamma$ component of the local operator defined at site $i$ 
with the local dimension $\mathscr{N}$, $J_{ij}^{\gamma\gamma^\prime}$ stands for the interaction between the operators $\mathcal{O}_{i}^{\gamma}$ and $\mathcal{O}_{j}^{\gamma'}$.
The last term of Eq.~\eqref{eq:H} is the one-body term with the local field $h_{i}^\gamma$ for the operator $\mathcal{O}_{i}^{\gamma}$.
In the MF theory, each local operator is decomposed by the local average and the deviation from it as,
\begin{align}\label{eq:delO_1}
  \mathcal{O}^{\gamma}_{i}= \delta\mathcal{O}^{\gamma}_{i} + \braket{\mathcal{O}^\gamma}_{l}
\end{align}
where we assume that the local average $\braket{\mathcal{O}^\gamma}_{l}$ depends on sublattice $l$ to which the site $i$ belongs.
Here, we prepare $M$ sublattices to realize a stable MF solution.

The original Hamiltonian given in Eq.~\eqref{eq:H} is decomposed to the MF Hamiltonian $\mathcal{H}^{\text{MF}}$ and the deviation from it, $ \mathcal{H}^\prime$, as
\begin{align}
  \label{eq:H_decomp}
  \mathcal{H} = \mathcal{H}^{\text{MF}} + \mathcal{H}^\prime,
\end{align}
where $\mathcal{H}^{\text{MF}}$ is given by the sum of the local Hamiltonians as
\begin{align}
  \mathcal{H}^{\text{MF}} = \sum_{i} \mathcal{H}_{i}^{\text{MF}} + \text{const.}
\end{align}
The local MF Hamiltonian $\mathcal{H}^{\chem{MF}}$ at site $i$ is represented as
\begin{align}
  \mathcal{H}_{i}^{\text{MF}} = \sum_{\gamma} \left( \sum_{l'}\sum_{j\in l'} \sum_{\gamma^{\prime}} 
  J^{\gamma\gamma^{\prime}}_{ij} 
  \langle \mathcal{O}^{\gamma^{\prime}} \rangle_{l'} 
  - h_{i}^{\gamma} \right) \mathcal{O}^{\gamma}_{i}.
  \label{eq:lcoal-mf}
\end{align}
Here, $\langle \mathcal{O}^\gamma \rangle_{l} = \braket{0;i|\mathcal{O}^\gamma|0;i}$ 
denotes the expectation value for the ground state $\ket{0;i}$ 
of the local Hamiltonian $\mathcal{H}^{\chem{MF}}_{i}$ with site $i$ belonging to sublattice $l$.
To obtain the ground state, we diagonalize the $\mathscr{N}\times \mathscr{N}$-matrix representation of $\mathcal{H}^{\text{MF}}_{i}$ and obtain the ground state $\ket{0;i}$ with the eigenenergy $E_0^{l}$ and $m$-th excited states $\ket{m;i}$ with the eigenenergy $E_{m}^{l}$ for $m=1,2,\cdots , \mathscr{N} - 1$.
Note that the Hilbert space of the total Hamiltonian is spanned by the direct product of the local eigenstates $\ket{m;i}$ of $\mathcal{H}^{\text{MF}}_{i}$, but the eigenenergies and eigenstates at sites belonging to the same sublattice are equivalent, respectively.
Namely, the eigenenergy and eigenstate are labeled not by the site index but by the sublattice index $l$.

\subsection{Generalized Holstein-Primakoff transformation}
\label{method:GHP}

The excitation structure is described by the contribution beyond the MF Hamiltonian, $\mathcal{H}^{\prime}$, which is given by
\begin{align}
  \label{eq:delH}
  \mathcal{H}^{\prime} \ &= \ \frac{1}{2}\sum_{i,j}\sum_{\gamma\gamma^{\prime}}
  J_{ij}^{\gamma \gamma^{\prime}} \delta \mathcal{O}^{\gamma}_{i} \delta \mathcal{O}_{j}^{\gamma^{\prime}}.
\end{align}
We rewrite the above Hamiltonian using bosons to evaluate the elementary excitations from the MF ground state.
Here, we expand Eq.~\eqref{eq:delO_1} using the eigenstates of the local Hamiltonian at site $i$ in sublattice $l$ as
\begin{align}
  \label{eq:delO_2}
  \delta \mathcal{O}^\gamma_{i} \ = \ \sum_{m,m^\prime =0}^{\mathscr{N} -1}
  X^{mm^\prime}_{i} \delta\mathcal{O}_{mm';l}^{\gamma},
\end{align}
where $X^{mm^\prime}_{i}=\ket{m;i}\bra{m^\prime;i}$ is the local projection operator at site $i$ and $\delta\mathcal{O}_{mm';l}^{\gamma}=\braket{m;i|\delta\mathcal{O}^{\gamma}|m^\prime;i}$, which depends only on the sublattice index $l$ to which site $i$ belongs.
The projection operator is represented by bosons using the generalized Holstein-Primakoff transformation~\cite{joshi1999, kusunose2001, nasu2021, koyama2021}.
We introduce $\mathscr{N} - 1$ bosons described by the creation (annihilation) operators $a_{mi}^\dagger$ ($a_{mi}$) with $m=1,2,\cdots , \mathscr{N} - 1$.
For $m\geq 1$, $X^{0m}_{i}$ and $X^{m0}_{i}$ are written as
\begin{align}\label{eq:Xm0-operator}
  X_{i}^{m0} \ = \ a_{mi}^\dagger
  \left(
    \mathcal{S} - \sum_{n=1}^{\mathscr{N} - 1} a_{ni}^\dagger a_{ni}^{}
  \right)^{1/2} \quad
  X_{i}^{0m} \ = \ \left( X_i^{m0} \right)^\dagger,
\end{align}
and, for $1\leq m,m^\prime$, $X_i^{mm^\prime}$ is given by
\begin{align}
  \label{eq:excited-excited}
  X_i^{mm^\prime} \ = \ a_{mi}^\dagger a_{m^\prime i}^{}.
\end{align}
Note that ${\cal S}$ is introduced as 
\begin{align}\label{eq:sum-unity}
  \mathcal{S}= X_i^{00}+ \sum_{n=1}^{\mathscr{N} - 1} a_{ni}^\dagger a_{ni}^{},
\end{align}
and it should be unity because of the local constraint intrinsic to the projection operator, $\sum_{m=0}^{\mathscr{N} -1}X_i^{mm}=1$.
The bosonic expressions in Eqs.~\eqref{eq:Xm0-operator}--\eqref{eq:sum-unity} reproduce the commutation relations that the projection operators should satisfy.
The above bosonic representation for the local operators has been used perticularly for SU($N$) systems.
The approaches applied to these systems is known as the flavor-wave theory~\cite{joshi1999,Lauchli2006,Tsunetsugu2006,Kim_flavor-wave2017}.

Although the above bosonic representation using the generalized Holstein-Primakoff transformation is exact, the presence of the square root in Eq.~\eqref{eq:Xm0-operator} complicates further calculations.
When the number of excited bosons is small enough, the square root can be expanded with respect to $1/{\cal S}$ as~\cite{joshi1999,kusunose2001,nasu2021,koyama2021,nasu2022} 
\begin{align}
  X_{i}^{m0}  = \sqrt{\mathcal{S}}  a_{mi}^\dagger 
  \left( 
    1 - \frac{1}{2\mathcal{S}} \sum_{n=1}^{\mathscr{N} - 1} a_{ni}^\dagger a_{ni}^{}
  \right)
  + O(\mathcal{S}^{-3/2}).
\end{align}
Using this expression, ${\cal H}'$ in Eq.~\eqref{eq:delH} is represented by the bosons and expanded for $1/{\cal S}$ as
\begin{align}
    \label{eq:bosonic_Hp}
    \mathcal{H}'=
    \mathcal{S}\mathcal{H}'_2 + \sqrt{\mathcal{S}}\mathcal{H}'_3 + \mathcal{H}'_4 + O(\mathcal{S}^{-1/2}),
  \end{align}
where $\mathcal{H}'_2$, $\mathcal{H}'_3$, and $\mathcal{H}'_4$ are the terms composed of the products of two, three, and four boson creation/annihilation operators in $\mathcal{H}'$.

On the other hand, the local MF Hamiltonian is given by
\begin{align}
  \label{eq:bosonic_HMF}
  \mathcal{H}^{\text{MF}}_{i}=\sum_{m=0}^{\mathscr{N} - 1}E_m^{l} X^{mm}_{i}= \mathcal{S} E_{0}^{l}+ \sum_{m=1}^{\mathscr{N} - 1} \Delta E^{l}_m a_{mi}^\dagger a_{mi}^{}
\end{align}
where $\Delta E_{m}^{l}=E_{m}^{l}-E_{0}^{l}$ is the energy difference between the excited state
 and ground state of the local MF Hamiltonian at site $i$ belonging to sublattice $l$.
Note that the average $\langle \mathcal{O}^{\gamma} \rangle_{l}$ for the MF ground state is of $O(\mathcal{S})$, which is understood from Eq.~\eqref{eq:sum-unity}.
If one assumes that $h_i^\gamma$ is the order of $\mathcal{S}$, the local MF for $\mathcal{O}^{\gamma}_{i}$ in Eq.~\eqref{eq:lcoal-mf} should be the order of $O(\mathcal{S})$, and thereby, the energies $E_{m}^{l}$ ($m=0,1,\cdots, \mathscr{N} - 1$) are in this order.
Thus, the first and second terms of Eq.~\eqref{eq:bosonic_HMF} are in the orders of $O(\mathcal{S}^2)$ and $O(\mathcal{S})$, respectively.
Here, we introduce $\mathcal{H}_{\text{LFW}}$, which is given by the sum of the second term of Eq.~\eqref{eq:bosonic_HMF} and $\mathcal{S}\mathcal{H}'_2$ in Eq.~\eqref{eq:bosonic_Hp}.
Since these give contributions with $O(\mathcal{S})$, the total Hamiltonian is represented as
\begin{align}
  \label{eq:bosonic_H}
  \mathcal{H} =
    \mathcal{S}\left(\mathcal{H}_{\rm LFW}+  \frac{1}{\sqrt{\mathcal{S}}}\mathcal{H}'_3 + \frac{1}{\mathcal{S}}\mathcal{H}'_4 + O(\mathcal{S}^{-3/2})
  \right) + \text{const}.
\end{align}
Note that, while $a_{mi}^\dagger$ and $a_{mi}$ do not appear alone because of the stable condition of the MF solution, other odd-order terms are allowed to be present in the Hamiltonian~\cite{zhitomirsky2013}.

\subsection{Linear flavor-wave theory}
\label{method:LSW}
First, we consider $\mathcal{H}_{\text{LFW}}$, 
which is given by the bilinear from of $a_{mi}$ and $a_{mi}^\dagger$.
The approximation considering only $\mathcal{H}_{\text{LFW}}$ is called the linear flavor-wave theory.
By introducing the Fourier transformation of $a_{mi}$ with respect to $i$, the Hamiltonian $\mathcal{H}_{\text{LFW}}$ is formally written as
\begin{align}
  \label{eq:Hamil-A}
  \mathcal{H}_{\text{LFW}} = \frac{1}{2} \sum_{\bm{k}}^{\text{B.Z.}} \mathcal{A}_{\bm{k}}^\dagger {\cal M}_{\bm{k}}^{} \mathcal{A}_{\bm{k}}^{},
\end{align}
where ${\cal M}_{\bm{k}}$ is a $2N\times 2N$ Hermitian matrix and $N=(\mathscr{N} - 1)M$ is the number of collective mode branches (see Appendix~\ref{app:LSWT}).
The $2N$-dimensional vector $\mathcal{A}_{\bm{k}}^{\dagger}$ is given by
\begin{align}
  \mathcal{A}_{\bm{k}}^{\dagger} = \left(a_{1,\bm{k}}^{\dagger}\ a_{2,\bm{k}}^{\dagger}\ \cdots 
  \ a_{N,\bm{k}}^{\dagger} 
  \ a_{1,-\bm{k}}^{} \ a_{2,-\bm{k}}^{} \ \cdots \ a_{N,-\bm{k}}^{} \right),
\end{align}
where $a_{\iota,\bm{k}}$ with $\iota=(ml)$ being the composite index of local excited state $m$ and sublattice $l$ is the Fourier transformation of $a_{mi}^{}$, which is represented by
\begin{align}
  \label{eq:fourier}
  a_{\iota,\bm{k}}^{}= \sqrt{\dfrac{M}{N_t}}\sum_{i\in l} a_{mi}^{} e^{-i\bm{k}\cdot \bm{r}_i}.
\end{align}
Here, $\bm{r}_i$ is the position of site $i$, and $N_t$ is the number of sites.
We diagonalize ${\cal M}_{\bm{k}}^{}$ by applying the Bogoliubov transformation as $\mathcal{E}_{\bm{k}}^{}=T_{\bm{k}}^\dagger \mathcal{M}_{\bm{k}}^{} T_{\bm{k}}^{}$,
where $T_{\bm{k}}$ is a paraunitary matrix and $\mathcal{E}_{\bm{k}}$ is the diagonal matrix given by
$\mathcal{E}_{\bm{k}} = \mathrm{diag}\{\varepsilon_{1,\bm{k}}, \varepsilon_{2,\bm{k}}, \cdots, \varepsilon_{N,\bm{k}}, 
\varepsilon_{1,-\bm{k}}, \varepsilon_{2,-\bm{k}}, \cdots, \varepsilon_{N,-\bm{k}}\}$~\cite{colpa}.
Using this transformation, the Hamiltonian is rewritten as the following diagonalized form:
\begin{align}
  \label{eq:Hamil-B}
  \mathcal{H}_{\rm{LFW}}= \frac{1}{2}\sum_{\bm{k}}^{\text{B.Z.}} \mathcal{B}^{\dagger}_{\bm{k}} \mathcal{E}_{\bm{k}} \mathcal{B}_{\bm{k}}^{},
\end{align}
Here, we introduce the set of bosonic operators $\mathcal{B}_{\bm{k}}^{}  = T_{\bm{k}}^{-1} \mathcal{A}_{\bm{k}}^{}$, which is given by
\begin{align}
  \mathcal{B}_{\bm{k}}^{\dagger} =\left(b_{1,\bm{k}}^{\dagger}\ b_{2,\bm{k}}^{\dagger}\ \cdots 
  \ b_{N,\bm{k}}^{\dagger} 
  \ b_{1,-\bm{k}}^{} \ b_{2,-\bm{k}}^{} \ \cdots \ b_{N,-\bm{k}}^{} \right),
\end{align}
where $b_{\eta,\bm{k}}^{\dagger}$ is regarded as the creation operator of a quantized flavor-wave excitation with the energy $\varepsilon_{\eta,\bm{k}}$.
The sum of $\bm{k}$ is taken in the first Brillouin zone.

\subsection{Nonlinear flavor-wave theory}
\label{method:NLSW}

In the previous section, we consider only the bilinear terms of bosonic operators.
In this case, the Hamiltonian is written as a free boson system without interactions, as shown in Eq.~\eqref{eq:Hamil-B}.
Here, we address effects of the higher-order terms beyond the linear flavor-wave Hamiltonian $\mathcal{H}_{\rm{LFW}}$ in Eq.~\eqref{eq:bosonic_H}.
These contributions are treated as perturbation terms, and $\mathcal{H}_{\rm{LFW}}$ is regarded as an unperturbed term.
As discussed in Sec.~\ref{method:GHP}, ${\cal H}'$ is expanded with respect to $1/\mathcal{S}$, and we take account of $O(1/\mathcal{S})$ corrections from the bilinear term $\mathcal{H}_{\rm{LFW}}$~\cite{zhitomirsky1999,zhitomirsky2010,mourigal2010,chernyshev2009}.
In this sense, we need to deal with $\mathcal{H}'_3/\sqrt{\mathcal{S}}$ up to second-order perturbations and $\mathcal{H}'_4/\mathcal{S}$ up to first-order perturbations [see Eq.~\eqref{eq:Green perturbation}].

\begin{figure}[t]
  \begin{center}
    \includegraphics[width=\columnwidth,clip]{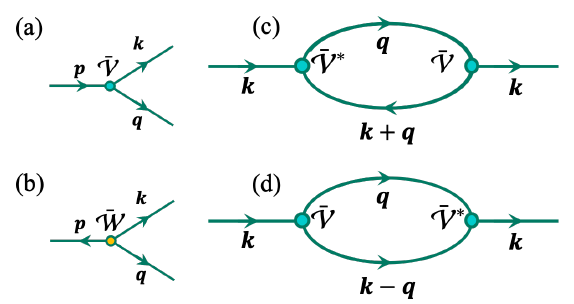}
    \caption{
      (a),(b) Schematic pictures of the three-quasiparticle interactions corresponding to Eqs.~\eqref{eq:decay} and \eqref{eq:source}, respectively.
      (c),(d) Lowest-order contributions of $\mathcal{H}_3$ to the self-energy $\tilde{\Sigma}$ in Eq.~\eqref{eq:self-energy-im}.
      At zero temperature, only the diagram shown in (d) contributes to the self-energy.
    }
    \label{fig:vertices_diagram}
  \end{center}
\end{figure}

In the present study, we focus on damping effects on the bosonic quasiparticles, and hence, we examine the imaginary part of the self-energy of the bosonic quasiparticles.
Note that the first-order perturbations contribute only to the real part of the self-energy~\cite{mahan,mourigal2010,chernyshev2009,fuhrman2012,maksimov2016_prl,pershoguba2018,mook2021}.
Thus, we concentrate second-order perturbations for the cubic term of bosons, $\mathcal{H}'_3/\sqrt{\mathcal{S}}$.
The cubic term can be decomposed into two terms, $\mathcal{H}'_3/\sqrt{\mathcal{S}} = \mathcal{H}_3^{(\text{d})}/\sqrt{\mathcal{S}} + \mathcal{H}_3^{(\text{s})}/\sqrt{\mathcal{S}}$.
The first term involves the process with a quasiparticle splitting into two particles and
second term stands for the process of creating (annihilating) three quasiparticles simultaneously.
The details are given in Appendix~\ref{app:NLSWT}, and only the results are presented here as
\begin{align}
  \label{eq:decay}
  \frac{1}{\sqrt{\mathcal{S}}}\mathcal{H}_{3}^{(\text{d})}&=
  \frac{1}{2!} \sqrt[]{\frac{M}{N_{t}\mathcal{S}}} \sum_{\eta\eta'\eta''}^{N}\sum_{\bm{k}\bm{q}\bm{p}}^{\bm{k}+\bm{q}=\bm{p}}
  \Biggl(
    \bar{\mathcal{V}}_{\bm{k},\bm{q}\leftarrow \bm{p}}^{\eta,\eta'\leftarrow\eta''} 
    b_{\eta,\bm{k}}^\dagger b_{\eta',\bm{q}}^\dagger b_{\eta'',\bm{p}}^{} + \text{H.c.}
  \Bigg),\\
  \label{eq:source}
  \frac{1}{\sqrt{\mathcal{S}}}\mathcal{H}_{3}^{(\text{s})}&=
  \frac{1}{3!} \sqrt[]{\frac{M}{N_{t}\mathcal{S}}} \sum_{\eta\eta'\eta''}^{N} \sum_{\bm{k}\bm{q}\bm{p}}^{\bm{k}+\bm{q}=-\bm{p}}
  \Biggl(
    \bar{\mathcal{W}}_{\bm{k},\bm{q},\bm{p}}^{\eta,\eta',\eta''} 
    b_{\eta,\bm{k}}^\dagger b_{\eta',\bm{q}}^\dagger b_{\eta'',\bm{p}}^{\dagger} + \text{H.c.} 
  \Bigg),
\end{align}
where $\bar{\mathcal{V}}_{\bm{k},\bm{q}\leftarrow \bm{p}}^{\eta,\eta'\leftarrow\eta''} $
and 
$\bar{\mathcal{W}}_{\bm{k},\bm{q},\bm{p}}^{\eta,\eta',\eta''} $
are the bare vertex functions, which is given explicitly in Eqs.~\eqref{appeq:Vbar} and \eqref{appeq:Wbar}, respectively.
Figures~\ref{fig:vertices_diagram}(a) and \ref{fig:vertices_diagram}(b) represent the schematic diagrams of the processes with $\bar{\mathcal{V}}_{\bm{k},\bm{q}\leftarrow \bm{p}}^{\eta,\eta'\leftarrow\eta''}$ and $\bar{\mathcal{W}}_{\bm{k},\bm{q},\bm{p}}^{\eta,\eta',\eta''} $, respectively.

Here, we treat Eqs.~\eqref{eq:decay} and~\eqref{eq:source} as perturbation terms, where Eq.~\eqref{eq:Hamil-B} is regarded as a unperturbed Hamiltonian.
We calculate the self-energy of bosonic quasiparticles, $\Sigma_{\bm{k}} (\omega,T)$, which is defined in Eq.~\eqref{appeq:self-energy} with $T$ being temperature.
In the present calculations, we focus on the quasiparticle damping caused within each collective mode branch.
To this end, we consider only the imaginary part of the diagonal components extracted from the self-energy $\Sigma_{\bm{k}} (\omega,T)$ up to $1/\mathcal{S}$ relative to $\mathcal{H}_{\rm{LFW}}$.
We denote the self-energy to which the above simplification is applied as $\tilde{\Sigma}_{\eta,\bm{k}} (\omega,T)$.
This self-energy is split into two contributions:
\begin{align}
  \tilde{\Sigma}_{\eta,\bm{k}} (\omega,T) = \Sigma_{\eta,\bm{k}}^{(\text{c})} (\omega,T) + \Sigma_{\eta,\bm{k}}^{(\text{d})} (\omega,T),
  \label{eq:self-energy-im}
\end{align}
where $\Sigma_{\eta,\bm{k}}^{(\text{c})} (\omega,T)$ and $\Sigma_{\eta,\bm{k}}^{(\text{d})} (\omega,T)$ are the contributions with $O(1/\mathcal{S})$ from the diagrams shown in Figs.~\ref{fig:vertices_diagram}(c) and~\ref{fig:vertices_diagram}(d), respectively.
The explicit representations are given in Eqs.~\eqref{appeq:self_collision} and \eqref{appeq:self_decay}.
As shown in Fig.~\ref{fig:vertices_diagram}(c), $\Sigma_{\eta,\bm{k}}^{(\text{c})} (\omega,T)$ comes from a collision with excited quasiparticles.
At zero temperature, there are no thermally excited quasiparticles, and hence, $\Sigma_{\eta,\bm{k}}^{(\text{c})} (\omega,T)$ is nonzero only at finite temperatures.
On the other hand,  $\Sigma_{\eta,\bm{k}}^{(\text{d})} (\omega,T)$ originates from a decay process to multiple magnons.
This process is not involved with excited quasiparticles and contributes to the self-energy even at zero temperature.
Note that there is another process involved with $\bar{\mathcal{W}}_{\bm{k},\bm{q},\bm{p}}^{\eta,\eta',\eta''} $ [Fig.~\ref{fig:vertices_diagram}(b)].
This contribution, defined as $\Sigma_{\eta,\bm{k}}^{\text{(s)}} (\omega,T)$, is represented by the diagram that the arrows of intermediate states in Fig.~\ref{fig:vertices_diagram}(d) are reversed~[see Appendix~\ref{app:NLSWT} and Eq.~\eqref{eq:self_source}].
Indeed, the imaginary part of $\Sigma_{\eta,\bm{k}}^{\text{(s)}} (\omega,T)$ vanishes due to the energy conservation law.

We comment on the first-order perturbation for the cubic term $\mathcal{H}'_3$, which does not contribute to the self-energy in the present scheme, as mentioned before.
This contribution can be considered by applying the Hartree-Fock (HF) decoupling to $\mathcal{H}'_3$.
It has been reported that the correction originating from this effect modifies the MF solution in the ground state~\cite{mourigal2010, mcclarty2018}.
Namely, an MF value $\braket{\delta\mathcal{O}^\gamma}_{l}$, which vanishes without the first-order perturbation of $\mathcal{H}'_3$, becomes nonzero due to contributions from HF decouplings like $\braket{a^\dagger a}a$.
Such terms destabilize the existing MF ground state of $\mathcal{H}^{\text{MF}}$ and compel us to find a new MF solution.
After solving the self-consistent equation determining MFs, we reconstruct a flavor-wave Hamiltonian with the newly found MFs.
Together with the HF decoupling for $\mathcal{H}_4$, we can take into account the HF corrections up to the order of $O(1/\mathcal{S})$.
If we consider the real part of the self-energy at zero temperature, we need to address them and $\Sigma_{\eta,\bm{k}}^{\text{(s)}} (\omega,T)$ properly~\cite{mourigal2010,chernyshev2009}.

\subsection{Imaginary Dyson Equation approach}
\label{method:iDE approach}

In this section, we show the details of the iDE approach used in the present calculations.
This method was developed in Refs.~\cite{chernyshev2009,maksimov2016_prb} and has been widely applied to magnon systems at zero temperature~\cite{winter2017_nc,smit2020,chernyshev2016,maksimov2020,maksimov2022,mook2021,rau2019}.
The details are shown in Appendix~\ref{app:NLSWT}.
We start from the Dyson equation shown in Eq.~\eqref{appeq:interaction temperature Green function}.
The pole of a bosonic Green's fuction is determined by
\begin{align}
  \label{eq:dyson eq}
  \text{det} \Big[ \omega\sigma_3 - \mathcal{E}_{\bm{k}} - \Sigma_{\bm{k}}(\omega,T) \Big] = 0,
\end{align}
where
$\sigma_3 = 
\begin{psmallmatrix} 
\bm{1}_{N\times N} & \\
 & -\bm{1}_{N\times N}
\end{psmallmatrix}$.
We introduce the damping rate as the imaginary part of the self-energy as follows:
\begin{align}
  \label{eq:damping rate}
  \Gamma_{\bm{k}} (\omega, T) \equiv - \mathrm{Im} \Sigma_{\bm{k}} (\omega, T).
\end{align}

As mentioned in Sec~\ref{method:NLSW}, we consider the diagonal part of the self-energy and neglect its real part, and such a self-energy has been introduced as $\tilde{\Sigma}_{\eta,\bm{k}}(\omega,T)$ in the previous section.
In the following, we restrict the range of $\eta$ to $1,2,\cdots,N$.
In this treatment, Eqs.~\eqref{eq:dyson eq} and~\eqref{eq:damping rate} are simplified as
\begin{align}
  \omega &=  \varepsilon_{\eta,\bm{k}} + \tilde{\Sigma}_{\eta,\bm{k}}(\omega,T),\\
  \Gamma_{\eta,\bm{k}} (\omega, T) \ &= \ - \mathrm{Im} \tilde{\Sigma}_{\eta,\bm{k}} (\omega, T).
\end{align}
The simplest approximation to eliminate the $\omega$ dependence is the on-shell approximation, where the argument $\omega$ of $\tilde{\Sigma}_{\eta,\bm{k}}$ is replaced to the one-particle energy $\varepsilon_{\eta,\bm{k}}$.
By applying the approximation, the damping rate up to $1/\mathcal{S}$ corrections is expressed as
$\Gamma_{\eta,\bm{k}} (T)\simeq -\mathrm{Im} \tilde{\Sigma}_{\eta,\bm{k}} (\varepsilon_{\eta,\bm{k}}, T)$~(see Appendix~\ref{app:on-shell approximation} for more details).
This approximation in the $1/\mathcal{S}$ correction corresponds to the Born approximation that considers only one loop in Figs.~\ref{fig:vertices_diagram}(c) and \ref{fig:vertices_diagram}(d).
Note that the imaginary part of $\tilde{\Sigma}_{\eta,\bm{k}} (\varepsilon_{\eta,\bm{k}},T)$ 
exhibits unphysical divergent behavior due to treating initial particle and two-particle states with different accuracy in the on-shell approximation~\cite{zhitomirsky2013,chernyshev2009, mook2021}.
In general, the issue may be alleviated by higher order $1/\mathcal{S}$ corrections, but such calculations are highly complicated, and performing them would involve significant computational costs.

Instead, to suppress singularities in the self-energy encountered within the $1/\mathcal{S}$ correction,
the iDE approach has been proposed~\cite{chernyshev2009,maksimov2016_prb}.
In this approach, a finite lifetime is introduced in the one-particle energy of the corresponding self-energy.
By considering the finite lifetime, one can relax the energy conservation law in the self-energy, which allows us to reduce the artificial singularity and enables regularizing the quasiparticle spectrum.
To determine the self-energy, we solve the following equation:
\begin{align}
  \omega = \varepsilon_{\eta,\bm{k}} + i\mathrm{Im} \tilde{\Sigma}_{\eta,\bm{k}} (\omega^{*}, T),
\end{align}
where the complex conjugate $\omega^{*}$ in $\Sigma_{\eta,\bm{k}}$ originates from causality~\cite{chernyshev2009}.
Note that, in this approach, the real part of $\omega$ remains unchanged, and the imaginary part is determined iteratively.
The obtained complex number value $\omega$ gives the pole of the bosonic Green's function into the upper half-plane, and the imaginary part corresponds to the damping rate $\Gamma_{\eta,\bm{k}} (T)$, which is expressed as
$\Gamma_{\eta,\bm{k}} (T) = -\mathrm{Im} \tilde{\Sigma}_{\eta,\bm{k}} (\omega^{*}, T)$.
This procedure is equivalent to changing the delta function in the on-shell approximation~[see Eq.~\eqref{appeq:damping rate_onshell}]
to the Lorentzian-type~[see Eqs.~\eqref{appeq:self_collision} and~\eqref{appeq:self_decay}].
It has been shown that the iDE approach can regularize the singularity of the damping rate appearing in the on-shell approximation~\cite{chernyshev2009,maksimov2016_prb,winter2017_nc}.
We numerically confirmed that the damping rate calculated using our scheme with the iDE approach is consistent with previous studies at zero temperature~\cite{chernyshev2016}.

In the present study, we extend the iDE method to finite-temperature calculations.
To demonstrate the validity of this method at finite temperatures, we calculate the dynamical spin structure factor in the Kitaev model, which is introduced in the next section.
The results are presented in  Appendix~\ref{app:Dsf}.
The present iDE results agree with those obtained by a continuous-time quantum Monte Carlo~(CTQMC) method~\cite{yoshitake2020} at the same temperature.
Therefore, we conclude that the iDE approach is valid, at least for the temperature region where the magnon picture is justified.

\begin{figure}[b]
  \begin{center}
    \includegraphics[width=\columnwidth,clip]{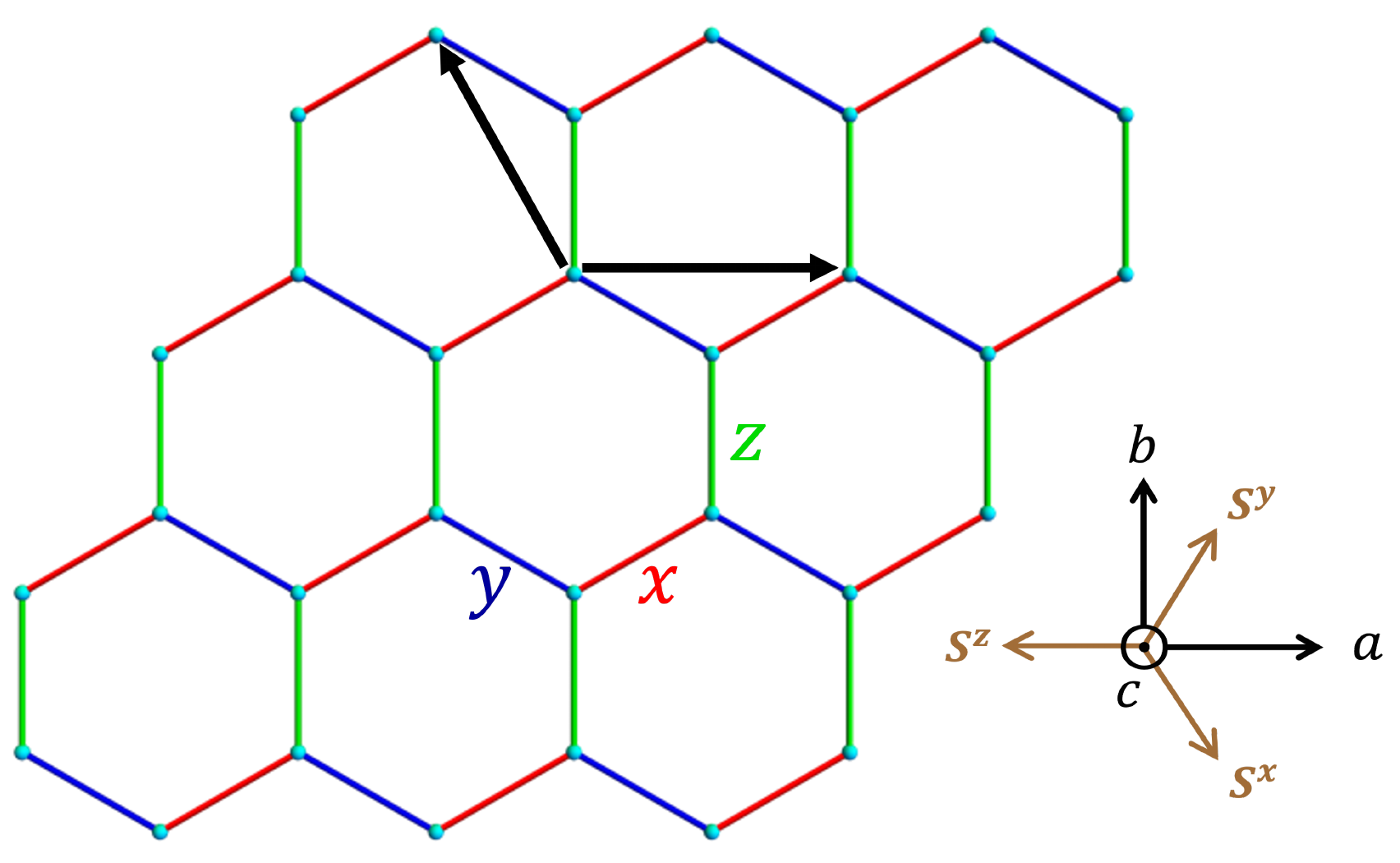}
    \caption{
    Schematic picture of the honeycomb lattice on which the $S=1/2$ Kitaev model is defined.
    The red, blue, and green lines stand for the $x$, $y$, and $z$ bonds, respectively.
    The black arrows represent the two primitive translation vectors. 
    The inset shows the relation between the coordinates of the spin space spanned by $(S^x,S^y,S^z)$ and the real space by $(a,b,c)$.
    }
    \label{fig:honeycomb}
  \end{center}
\end{figure}

\begin{figure}[b]
  \begin{center}
    \includegraphics[width=\columnwidth,clip]{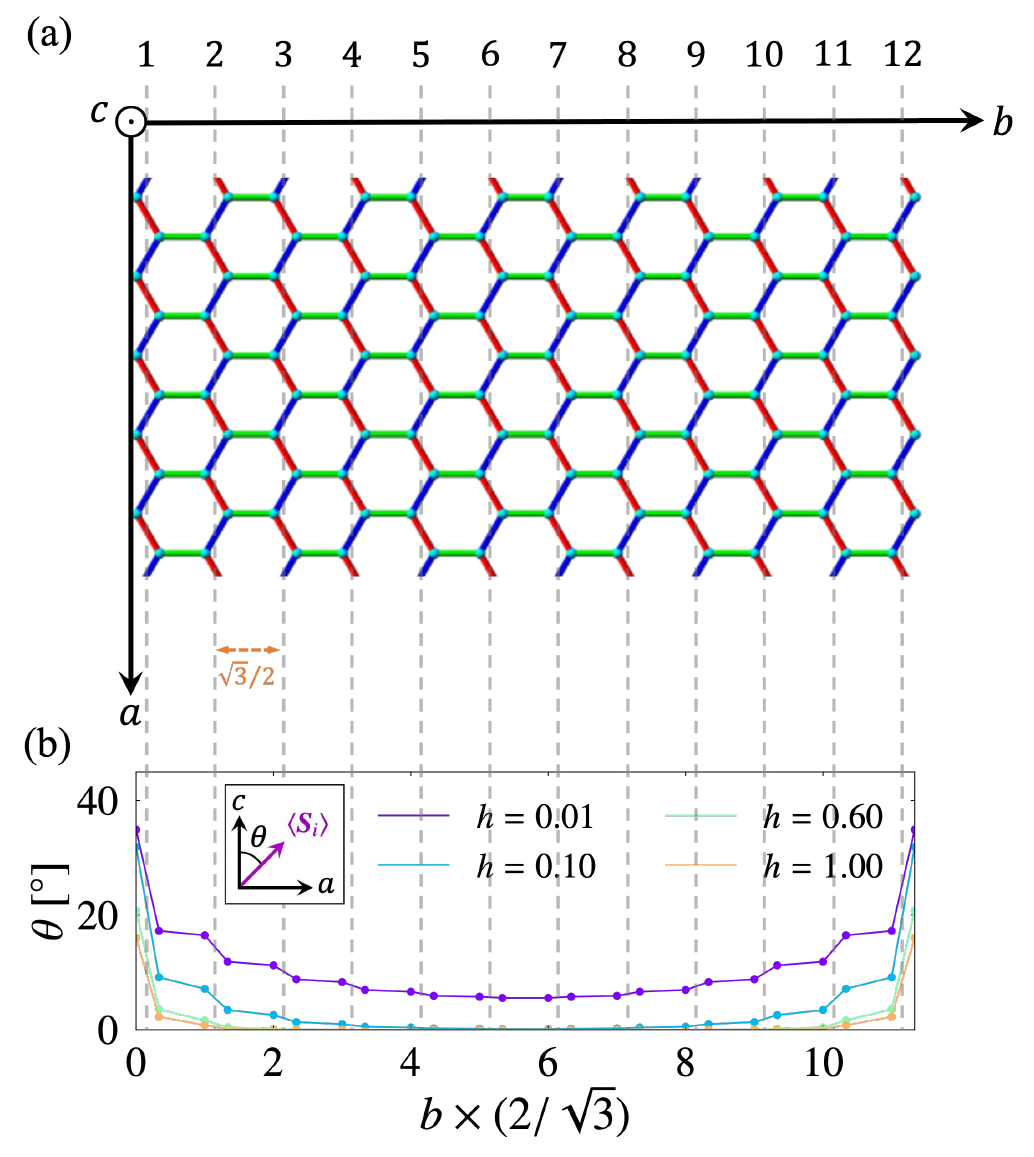}
    \caption{
    (a) Schematic picture of the 12 zigzag chains on the honeycomb cluster with zigzag edges where a periodic boundary condition is imposed along the $a$ direction.
    (b) Spatial distribution of the direction of spin moment at several magnetic fields.
    The inset shows the definition of $\theta$, which is the angle of the spin moment from the $c$ axis on the $a$-$c$ plane. 
    }
    \label{fig:openb}
  \end{center}
\end{figure}

\begin{figure*}[t]
  \begin{center}
    \includegraphics[width=120mm,clip]{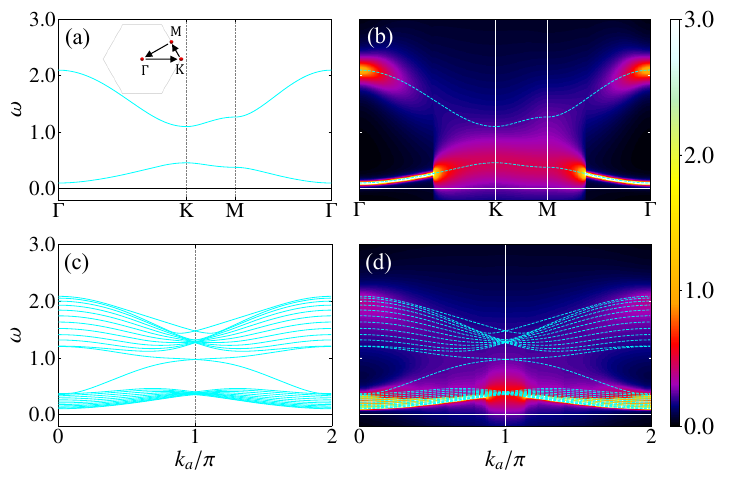}
    \caption{
    (a) Magnon dispersion presented along high-symmetry lines in the $\bm{k}$ space for $h=0.1$ in the system with the periodic boundary conditions imposed.
    (b) Magnon spectral functions at $h=0.1$ and $T=0$, where the magnon damping calculated by the iDE approach.
    In (b), the dashed lines represent the bare magnon dispersions shown in (a).
    (c),(d) Corresponding plots for the system with open boundaries where the lattice terminates with zigzag edges [see Fig.~\ref{fig:openb}(a)].
    }
    \label{fig:h=0.1}
  \end{center}
\end{figure*}

Here, we introduce the spectral function for bosonic excitations.
We approximate the retarded Green's function defined in Eq.~\eqref{appeq:interaction retarded Green function} as
\begin{align}
  \label{eq:retarded Green}
  G^{R}_{\eta,\bm{k}} (\omega,T) \simeq \frac{1}{\omega - \varepsilon_{\eta,\bm{k}} + i \Gamma_{\eta,\bm{k}}}.
\end{align}
By using Eq.~\eqref{eq:retarded Green}, the spectral function, $A_{\bm{k}}$, can be written as 
\begin{align}
  A_{\bm{k}} (\omega,T) &= -\frac{1}{\pi}\frac{1}{N}\sum_{\eta}^{N} \text{Im} \  G_{\eta,\bm{k}}^{\text{R}} (\omega,T)\nonumber\\
  &\simeq \frac{1}{N}\sum_{\eta}^{N}\frac{\Gamma_{\eta,\bm{k}}/\pi}{(\omega - \varepsilon_{\eta,\bm{k}})^2 + \Gamma_{\eta,\bm{k}}^2}.
  \label{eq:spectral function}
\end{align}

\section{Model}
\label{method:kitaev model}

We apply the iDE method explained in the previous section to the $S=1/2$ Kitaev model on a honeycomb lattice~\cite{kitaev,mcclarty2018,Motome2020rev,zhang2022,hickey2019}.
The Hamiltonian of this model is given by 
\begin{align}
  \label{eq:kitaev}
  \mathcal{H} = 2K\sum_{\gamma=x,y,z} \sum_{\langle i,j \rangle_{\gamma}}
  S_{i}^\gamma S_{j}^\gamma - \sum_{i} \bm{h} \cdot \bm{S}_{i},
\end{align}
where $S_{i}^\gamma (=x,y,z)$ represents the $S=1/2$ spin at site $i$,
and $K (<0)$ is the exchange constant of the ferromagnetic Kitaev interaction between spins on the 
nearest neighbor~(NN) sites.
The Kitaev interaction is bond-dependent, and $\langle i,j \rangle_{\gamma}$ denotes the NN $\gamma$ bond on the honeycomb lattice (see Fig.~\ref{fig:honeycomb}).
The last term of Eq.~\eqref{eq:kitaev} is the Zeeman term with the magnetic field $\bm{h}$.
The Kitaev model is believed to be realized in compounds with $4d$ or $5d$ transition metal ions~\cite{jackeli2009,motome2020_iop}.
Considering the connection to real materials, we introduce the spin coordinate such that the $[111]$ direction in the spin space is parallel to the
$c$ axis in the real space and the $S^z$ direction is on the $c$-$a$ plane~(see the inset of Fig.~\ref{fig:honeycomb}).
Hereafter, we set the Kitaev interaction to $K=-1$ and 
apply the magnetic field perpendicular to the honeycomb plane, $\bm{h}\parallel c$.

In our calculations, we introduce two clusters characterized by distinct boundary conditions:
One is a cluster with periodic boundary conditions along the two primitive translation vectors for the honeycomb lattice [see Fig.~\ref{fig:honeycomb}(a)], and the other possesses the open boundary with zigzag edges and the periodic condition imposed along the direction of the zigzag chain [see Fig.~\ref{fig:openb}(a)].
For the latter, we consider the cluster including 12 zigzag chains as shown in Fig.~\ref{fig:openb}(a). 

\section{Result}
\label{result}

\subsection{Mean-field solution}
\label{sec:mean-field}

First, we mention the MF ground states of the Kitaev model under magnetic fields on the two clusters.
In the cluster with the periodic boundary conditions, the MF ground state is the spin-polarized state parallel to the magnetic field direction regardless of the magnetic-field intensity.
On the other hand, in the case of the cluster with zigzag edges, the spin direction depends on the distance from the edge.
Spins near the center of the cluster are almost parallel to the magnetic field direction.
Meanwhile, in the vicinity of the edges, the spin direction is tilted from the magnetic field direction ($\bm{h}\parallel c$) to the $a$ axis, as shown in Fig.~\ref{fig:openb}(b).
This is due to the lack of $z$ bonds for edge sites; spins on edge sites tilt toward the $S^x$-$S^y$ plane to gain the exchange energies on $x$ and $y$ bonds.
The deviation from the applied field direction becomes small with increasing the field intensity.

\subsection{Magnon spectra for low-field regime at zero temperature}
\label{result:low magnetic field}

In this section, we show the results for the magnon spectra in the Kitaev model under the magnetic field with $h=0.1$ at zero temperature.
Figure~\ref{fig:h=0.1}(a) displays the magnon dispersions obtained by the linear flavor-wave theory on the cluster with the periodic boundary conditions.
The dispersions are plotted along the lines shown in Fig.~\ref{fig:h=0.1}(a) in the first Brillouin zone.
There are two branches because there are two spins in a unit cell of the honeycomb lattice.
The results are consistent with the magnon dispersions obtained by the previous studies~\cite{mcclarty2018, koyama2021}.
We also calculate the magnon dispersions on the cluster with zigzag edges.
As shown in Fig.~\ref{fig:h=0.1}(c), there are two modes connected between two bulk bands corresponding to the two magnon dispersions shown in Fig.~\ref{fig:h=0.1}(a).
These two ingap modes are chiral edge modes along the two zigzag edges, which result from topologically nontrivial magnon bands caused by an applied magnetic field.

Here, we introduce the damping effect of magnons.
Figures~\ref{fig:h=0.1}(b) and \ref{fig:h=0.1}(d) show the magnon spectral functions defined in Eq.~\eqref{eq:spectral function}, where the imaginary part of the self-energy is evaluated by the iDE method, in the cluster with the periodic boundary conditions and that with zigzag edges, respectively.
In the former case shown in Fig.~\ref{fig:h=0.1}(b), magnons around the $\Gamma$ point in the low-energy branch survive even in the presence of the magnon-magnon interactions.
Away from the $\Gamma$ point, the damping effect of magnons becomes more significant, which is reflected by the enhancement of the imaginary part of the self-energy.
For the higher-energy magnon branch, magnons appear to decay throughout the Brillouin zone.
In the case of the cluster with zigzag edges [Fig.~\ref{fig:h=0.1}(d)], low-energy magnons around $k_a=0$ are stable against introducing magnon-magnon interactions similar to the results in the cluster with the periodic boundaries.
On the other hand, low-energy magnons away from the $k_a=0$ point are strongly damped, as well as high-energy magnons.
It is worth noting that the chiral edge modes also exhibit strong damping at zero temperature.

\begin{figure}[t]
  \begin{center}
    \includegraphics[width=80mm,clip]{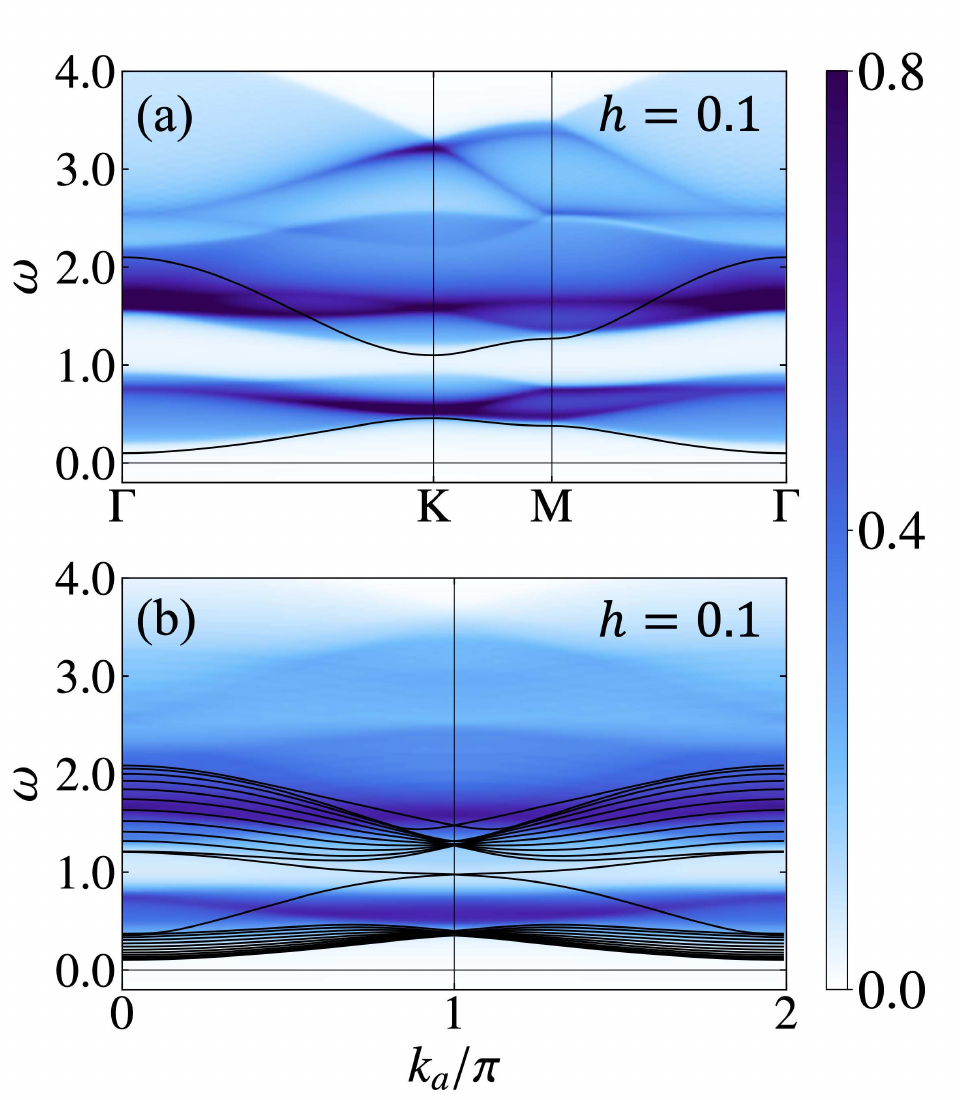}
    \caption{
    Two-magnon DOS at $h=0.1$ in (a) the system with the periodic boundary conditions imposed and (b) that with zigzag edges.
    The black lines represent the one-particle magnon dispersion.
    }
    \label{fig:2dos_h=0.1}
  \end{center}
\end{figure}

To clarify the $\bm{k}$ and energy dependence of the magnon damping, we focus on the self-energy of magnons.
There are two contributions in the self-energy as shown in Eq.~\eqref{eq:self-energy-im}, but the second term $\Sigma_{\eta,\bm{k}}^{(\text{d})} (\omega,T)$ corresponding to the decay process [Fig.~\ref{fig:vertices_diagram}(d)] only contributes at zero temperature.
From the explicit representation of $\Sigma_{\eta,\bm{k}}^{(\text{d})} (\omega,T)$ presented in Eq.~\eqref{appeq:self_decay}, it is naively expected that the imaginary part is zero when the condition $\omega - \varepsilon_{\eta',\bm{q}} - \varepsilon_{\eta'',\bm{k}-\bm{q}}=0$ is satisfied for $\omega=\varepsilon_{\eta,\bm{k}}$.
This consideration is correct for the on-shell Born approximation.
We apply it to the present results obtained by the iDE approach.
Here, we introduce two-magnon density of states~(DOS), which is defined as
\begin{align}
  \label{eq:split_dos}
  D^{\text{(d)}}_{\bm{k}} (\omega) \equiv \frac{1}{N^2} \sum_{\eta',\eta''}^{N}\frac{M}{N_t}\sum_{\bm{q}}^{\mathrm{B.Z.}} \delta(\omega -\varepsilon_{\eta',\bm{q}} - \varepsilon_{\eta'',\bm{k}-\bm{q}}).
\end{align}
When the dispersion relation $\omega=\varepsilon_{\eta,\bm{k}}$ overlaps with nonzero $D^{\text{(d)}}_{\bm{k}} (\omega)$, such magnons possibly collapse due to the decay process inherent in the self-energy $\Sigma_{\eta,\bm{k}}^{(\text{d})} (\omega,T)$.

The two-magnon DOS at $h=0.1$ is shown in Fig.~\ref{fig:2dos_h=0.1},
where Figs.~\ref{fig:2dos_h=0.1}(a) and \ref{fig:2dos_h=0.1}(b) correspond to the results in the system under the periodic boundary conditions and that with zigzag edges, respectively.
In both cases, the upper one-magnon branches overlap with the two-magnon continuum.
In the on-shell Born approximation, a necessary condition for the single-magnon with $\varepsilon_{\eta,\bm{k}}$ splitting into a two-magnon continuum is represented as
\begin{align}
  \label{eq:decay_constraint}
  \varepsilon_{\eta,\bm{k}} = \varepsilon_{\eta',\bm{q}} + \varepsilon_{\eta'',\bm{k}-\bm{q}}.
\end{align}
This condition indicates that the high-energy magnons decay as long as $\bar{\mathcal{V}}_{\bm{q},\bm{k}-\bm{q}\leftarrow \bm{k}}^{\eta,\eta'\leftarrow \eta''} \neq 0$, and the strong magnon decay obtained by the iDE approach [Figs.~\ref{fig:h=0.1}(b) and \ref{fig:h=0.1}(d)] is consistent with this consideration.
On the other hand, the chiral edge modes also decay strongly, as shown in Fig.~\ref{fig:h=0.1}(d), although the intensity of $D^{\text{(d)}}_{\bm{k}} (\omega)$ is weak compared to the high-energy region.
Indeed, we have numerically confirmed that the chiral edge modes rarely decay in the on-shell Born approximation.
Contrary to this approximation,  the criteria for magnon decay shown in Eq~\eqref{eq:decay_constraint} do not necessarily have to hold strictly in the iDE approach because this approach takes account of the energy fluctuations of magnons.

\subsection{Temperature dependence of magnon spectra with higher magnetic fields}
\label{result:high magnetic field}
\begin{figure*}[t]
  \begin{center}
    \includegraphics[width=135mm,clip]{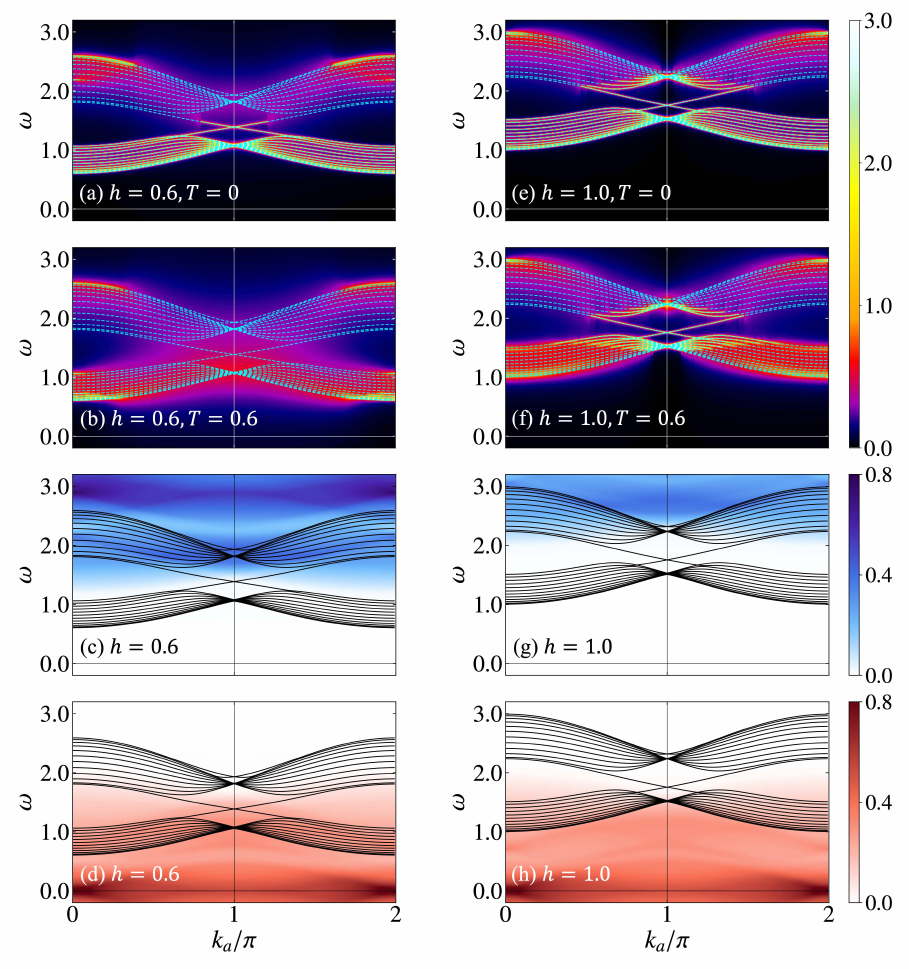}
    \caption{
    (a),(b) Magnon spectral function given in Eq.~\eqref{eq:spectral function} in the system with zigzag edges under the magnetic field with $h=0.6$ at (a) $T=0$ and (b) $T=0.6$.
    The dashed lines represent the one-particle magnon dispersions.
    (c) Two-magnon DOS and (d) two-magnon collision DOS at $h=0.6$. 
    The black lines represent the one-particle magnon dispersions.
    (e)--(h) Corresponding plots for $h=1.0$.
    }
    \label{fig:h=0.6and1.0}
  \end{center}
\end{figure*}

In this section, we focus on the results for higher magnetic fields in the cluster with zigzag edges.
Figure~\ref{fig:h=0.6and1.0}(a) shows the magnon spectrum for $h=0.6$ at zero temperature calculated by the iDE approach.
As shown in this figure, the chiral edge modes appear to survive particularly below $\omega\lesssim 1.5$ in contrast to the case with $h=0.1$.
Moreover, we also find that magnons in the low-energy band survive, although those in the high-energy band collapse due to the nonzero lifetime of the magnons. 
These results are understood from the overlap between the single-magnon dispersion relation $\varepsilon_{\eta,\bm{k}}$ and two-magnon DOS $D^{\text{(d)}}_{\bm{k}} (\omega)$;
the lower-energy magnons and chiral edge modes do not overlap with nonzero $D^{\text{(d)}}_{\bm{k}} (\omega)$ but this DOS takes larger values in the energy window of the higher-energy magnons, as shown in Fig.~\ref{fig:h=0.6and1.0}(c).
Further increase of the magnetic field leads to stable chiral edge modes in the wider energy region, as shown in Fig.~\ref{fig:h=0.6and1.0}(e).
This is understood from the high-energy shift of two-magnon DOS [Fig.~\ref{fig:h=0.6and1.0}(g)].

Here, we examine the effect of thermal fluctuations on the magnon damping.
Figure~\ref{fig:h=0.6and1.0}(b) shows the magnon spectrum for $h=0.6$ at $T=0.6$.
As shown in this figure, the chiral edge modes are smeared strongly by thermal fluctuations.
To clarify the origin of this effect, we discuss the contribution from $\Sigma_{\eta,\bm{k}}^{(\text{c})} (\omega,T)$ in Eq.~\eqref{eq:self-energy-im}, which is nonzero at finite temperatures.
This contribution originates from the collision with thermally excited magnons.
Here, we introduce the two-magnon collision DOS as
\begin{align}
  \label{eq:collision_dos}
  D^{\text{(c)}}_{\bm{k}} (\omega) \equiv  \frac{1}{N^2}\sum_{\eta',\eta''}^{N} \frac{M}{N_t}\sum_{\bm{q}}^{\mathrm{B.Z.}} 
  \delta(\omega + \varepsilon_{\eta',\bm{q}} - \varepsilon_{\eta'',\bm{k}+\bm{q}}).
\end{align}
Similar to the case of $D^{\text{(d)}}_{\bm{k}} (\omega)$, the overlap between the single-magnon dispersion and the two-magnon collision DOS gives the necessity condition of nonzero ${\rm Im}\Sigma_{\eta,\bm{k}}^{(\text{c})} (\omega,T)$ within the on-shell Born approximation [see Eq.~\eqref{appeq:self_collision}].
Figure~\ref{fig:h=0.6and1.0}(d) shows the two-magnon collision DOS $D^{\text{(c)}}_{\bm{k}} (\omega)$.
Unlike the case of $D^{\text{(d)}}_{\bm{k}} (\omega)$ shown in Fig.~\ref{fig:h=0.6and1.0}(c), the chiral edge modes overlap with the two-magnon collision DOS.
This result indicates that the decay of the chiral edge modes at $h=0.6$ and $T=0.6$ is due to the collision process with thermally excited magnons in the bulk bands.
The present consideration is also applied to the case with higher magnetic fields.
Figure~\ref{fig:h=0.6and1.0}(f) displays the magnon spectrum for $h=1.0$ at $T=0.6$.
In this case, the chiral edge modes appear to survive even at finite temperatures.
The existence of the stable edge modes against thermal fluctuations is understood from the two-magnon collision DOS as shown in Fig.~\ref{fig:h=0.6and1.0}(h); the edge modes do not overlap the two-magnon collision DOS.
Therefore, in the high-field case, the chiral edge magnon is robust in the presence of the magnon-magnon interactions at both zero and finite temperatures, and the topological magnon picture in terms of the noninteracting limit is well established.

\section{Discussion}
\label{discussion}

\begin{figure}[t]
  \begin{center}
    \includegraphics[width=85mm,clip]{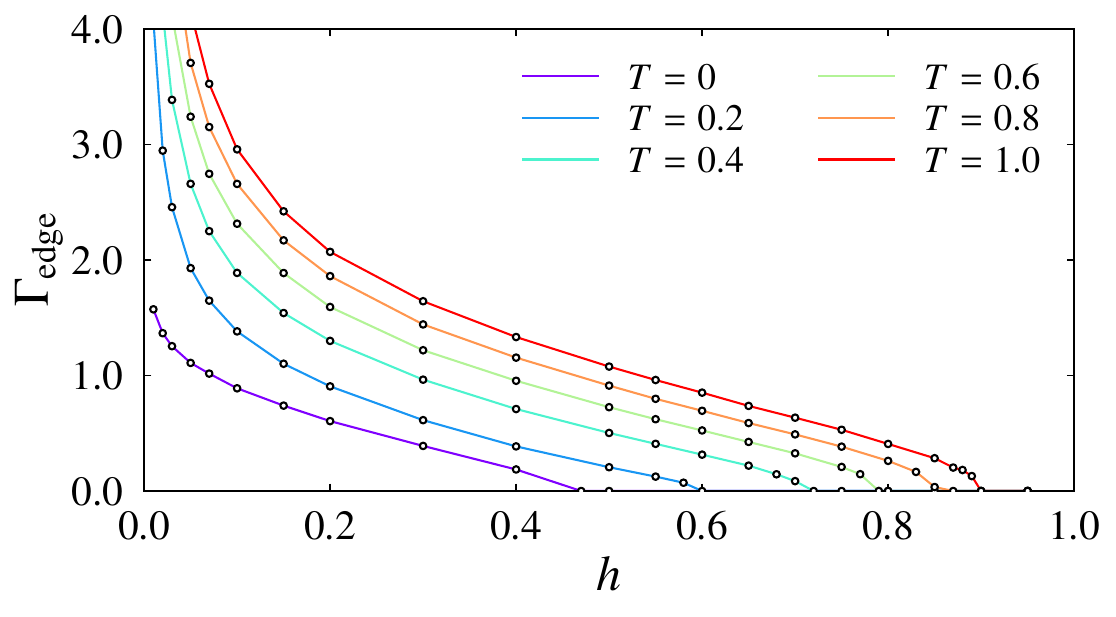}
    \caption{
      Magnetic-field dependence of the edge-mode damping rate $\Gamma_{\rm edge}$, which is defined in the main text, at several temperatures.
    }
    \label{fig:hdependence_edge}
  \end{center}
\end{figure}

\begin{figure}[t]
  \begin{center}
    \includegraphics[width=85mm,clip]{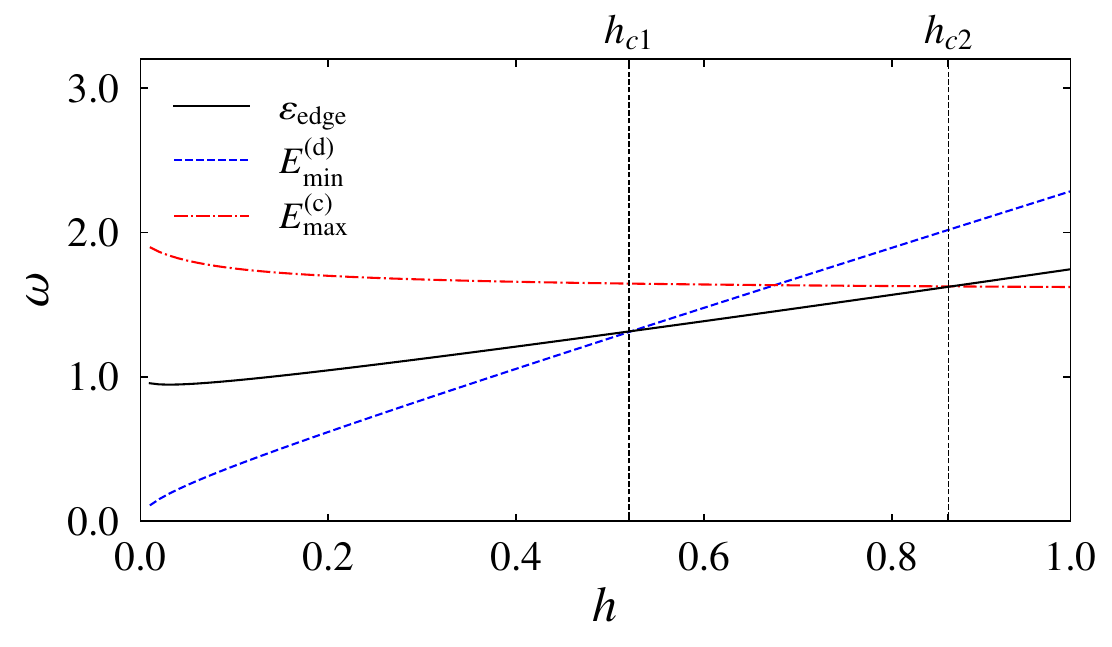}
    \caption{
     Magnetic field dependence of the energy of the edge-mode $\varepsilon_{\mathrm{edge}}$, the lowest energy of the two-magnon DOS $E_{\mathrm{min}}^{\text{(d)}}$,
     and the highest energy of the two-magnon collision DOS $E_{\mathrm{max}}^{\text{(c)}}$ 
     at $k_{a}=\pi$.
    }
    \label{fig:edge2dos}
  \end{center}
\end{figure}

In this section, we discuss the magnetic-field dependence of the magnon damping for the chiral edge mode.
As shown in Fig.~\ref{fig:h=0.1}(c), the two magnon branches along the zigzag edges are present between two bulk bands.
These chiral edge modes cross at $k_a=\pi$.
We focus on the magnon damping of the edge modes and introduce the damping rate for each edge mode at this point as $\Gamma_{\rm edge}$.
We have confirmed that the values of the damping rate for the two magnon edge modes are the same at $k_a=\pi$.
Figure~\ref{fig:hdependence_edge} shows the magnetic field dependence of $\Gamma_{\rm edge}$ at several temperatures.
At $T=0$, $\Gamma_{\rm edge}$ monotonically decreases with increasing $h$ and vanishes above $h\simeq 0.48$.
The condition for vanishing $\Gamma_{\rm edge}$ can be understood from the overlap between the chiral edge mode and two magnon DOS defined in Eq.~\eqref{eq:split_dos}.
As shown in Fig.~\ref{fig:2dos_h=0.1}(b), the chiral edge modes overlap with the two magnon continuum, indicating the finite lifetime of chiral edge magnons.
With increasing the magnetic field, the continuum is shifted to the higher-energy side and does not overlap with the edge modes, as shown in Figs.~\ref{fig:h=0.6and1.0}(c) and \ref{fig:h=0.6and1.0}(g).
Thus, we calculate the lowest energy of the continuum $D^{\text{(d)}}_{\bm{k}} (\omega)$ at $k_a=\pi$. Figure~\ref{fig:edge2dos} shows the magnetic field dependence of the lowest energy defined as $E_{\rm min}^{\text{(d)}}$.
We find that $E_{\rm min}^{\text{(d)}}$ is larger than the edge mode energy $\varepsilon_{\rm edge}=\varepsilon_{k_a=\pi}$ above $h_{c1}\simeq 0.52$.
This value is close to the critical field where $\Gamma_{\rm edge}$ vanishes at $T=0$, suggesting that the damping effect at zero temperature basically is understood from the decay to two magnons.
Meanwhile, the monotonic decrease of $\Gamma_{\rm edge}$ with increasing $h$ cannot be understood only from two magnon DOS.
This behavior originates from the matrix elements of the vertex $\bar{\mathcal{V}}_{\bm{k},\bm{q}\leftarrow \bm{p}}^{\eta,\eta'\leftarrow\eta''}$ and the smearing effects in $\Sigma_{\eta,\bm{k}}^{(\text{c})} (\omega,T)$ taken by the iDE calculations.

Next, we discuss the magnetic-field effect on the magnon damping of the edge modes at finite temperatures.
As shown in Fig.~\ref{fig:hdependence_edge}, $\Gamma_{\rm edge}$ increases with increasing temperature, and thermal fluctuations result in the shift to the high-field side without changing the overall $h$ dependence.
Accordingly, the critical field where $\Gamma_{\rm edge}$ vanishes shifts to the high-field side.
Since the magnon-damping effect originating from the self-energy $\Sigma_{\eta,\bm{k}}^{(\text{d})} (\omega,T)$ contributes only in $h\lesssim h_{c1}$, the shift is due to the collision process coming from $\Sigma_{\eta,\bm{k}}^{(\text{c})} (\omega,T)$.
This process is brought about when magnon energy overlaps with the two-magnon collision DOS presented in Eq.~\eqref{eq:collision_dos}.
As shown in Figs.~\ref{fig:h=0.6and1.0}(d) and \ref{fig:h=0.6and1.0}(h), the two-magnon collision DOS has a distribution centered at the zero energy.
To discuss the overlap with the chiral edge modes, we focus on the highest energy of two-magnon collision DOS at $k_a=\pi$, which we define as $E_{\rm max}^{\text{(c)}}$.
The magnetic field dependence of $E_{\rm max}^{\text{(c)}}$ is shown in Fig.~\ref{fig:edge2dos}.
We find that $E_{\rm max}^{\text{(c)}}$ and $\varepsilon_{\rm edge}$ cross at $h_{c2}\simeq 0.86$, which is larger than $h_{c1}$.
This result clearly indicates that thermal fluctuations give rise to the magnon damping of the chiral edge modes even when the noninteracting magnon picture for the edge modes is justified at zero temperature.

In the Kitaev candidate material $\alpha$-RuCl$_{3}$, the strength of the Kitaev interaction is considered to be $|2K|=100$~K~\cite{banerjee2016,laurell2020,sandilands2015,shdo2017,yadav2016}.
In this case, damping effects on the chiral edge magnons are inevitably present as long as we apply the magnetic fields with a Zeeman energy comparable to the exchange coupling, whose intensity is approximated as $\sim 100$~T.
Therefore, we conclude that, within the magnon picture, chiral edge modes possess a finite lifetime, and the damping effect is relevant in the pure Kitaev model under magnetic fields.
When we consider additional interactions such as Heisenberg and $\Gamma$ terms, the energy structures of two-magnon DOS and two-magnon collision DOS can be changed.
The additional interactions could stabilize the chiral edge modes if these two kinds of continuum do not overlap with chiral mode branches.
The effects of the additional terms remain as future work.

Finally, we note that topological thermal transport phenomena may occur even if the chiral edge mode possesses a finite lifetime.
The thermal Hall effect at low temperatures is mainly caused by the finite Berry curvatures defined at the low energy branches~\cite{matsumoto2011,matsumoto2014,murakami2017}.
At $h=0.1$, Figs.~\ref{fig:h=0.1}(b) and~\ref{fig:h=0.1}(d) show that the low energy magnons are stable in the presence of the magnon-magnon interactions despite the strong damping of the chiral edge magnon mode.
Therefore, if the low-energy magnons survive, the thermal Hall effect may be detectable regardless of the stability of chiral edge magnons.
To answer this question, one desires the formulation of the thermal Hall conductivity involved in the magnon damping in bulk systems.

Our calculation framework starts from a general Hamiltonian describing interactions between local degrees of freedom with any number of states and the MF ground state with any number of sublattices.
Thus, it is widely applicable to other systems with local degrees of freedom such as spin-orbital entangled systems with multipolar interactions~\cite{kusunose2001,nasu2021,nasu2022} and molecular-orbital crystals including spin-dimerized systems~\cite{attfield2015orbital,hiroi2015structural,Tsunetsugu2001,romhanyi2015}, and also to systems with the magnetic ordering characterized by long-period waves such as skyrmion crystals~\cite{Takeda2023_arxiv}.
Moreover, since our formalism is mapped onto bosonic Hamiltonian with no restriction of the number of local bosons, it can be applied to electronic models coupled with phonons, such as dynamical Jahn-Teller systems~\cite{Nasu2013}.
We also expect that our method is applicable to anticipating excitation spectra of collective modes in real materials.
It has been proposed that the general Hamiltonian in Eq.~\eqref{eq:H} based on the present study can be derived as a Kugel-Khomskii-type effective model for a Mott insulator from a realistic multiorbital Hubbard model~\cite{Iwazaki2023_arxiv}.
Combining this approach with our method could systematically demonstrate excitation spectra in Mott insulators based on first-principles calculations.

\section{Summary}
\label{summary}

In summary, we have proposed a framework that takes account of damping effects on collective excitations from a mean-field ground state in a general interacting model between local degrees of freedom.
We have introduced bosonic quasiparticles based on the flavor-wave theory.
We evaluate damping effects on the quasiparticles at finite temperatures by extending the imaginary Dyson equation method, which has been widely employed in analyzing various magnon systems.
We have found two contributions to quasiparticle damping processes in the self-energy: One occurs due to collisions with excited quasiparticles, and the other is a decay process into two quasiparticles.
The latter occurs even at zero temperature, but the former appears only at finite temperatures.
We have applied our method to the Kitaev model on a honeycomb lattice under a magnetic field, a well-known system with topologically nontrivial magnon bands.
We have validated the method by comparing the dynamical spin structure factor with that obtained by the continuous-time quantum Monte Carlo simulations.
We focus on chiral edge modes in a cluster with zigzag edges to examine damping effects on topological magnons.
We have demonstrated that the chiral edge modes are strongly damped in weak magnetic fields at zero temperature.
With increasing the intensity of the applied magnetic field, the lifetime of the chiral edge modes becomes longer and well-defined quasiparticle excitations.
We have also found that thermal fluctuations give rise to an increase in the damping rate of the chiral edge modes.
This effect originates from collisions with thermally excited magnons in the bulk.
Since the present approach starts from a general Hamiltonian, it can widely apply to other localized systems, such as spin-orbital and electron-phonon coupled systems.
We also expect an application to real materials by integrating with first-principles calculations giving a localized model.

\begin{acknowledgments}
The authors thank S.~Murakami, P.~McClarty, S.~Hoshino, A.~Ono, and R.~Iwazaki for fruitful discussions. 
Parts of the numerical calculations were performed in the supercomputing
systems in ISSP, the University of Tokyo.
This work was supported by Grant-in-Aid for Scientific Research from
JSPS, KAKENHI Grant No.~JP19K03742, JP20H00122, JP22H01175, JP23H01129, and JP23H04865,
and by JST, the establishment of university fellowships towards the creation of science technology innovation, Grant Number JPMJFS2102.
\end{acknowledgments}

\appendix

\section{Linear flavor-wave theory and Bogoliubov transformation}
\label{app:LSWT}

In this section, we show the details of the linear flavor-wave theory and Bogoliubov transformation given in Sec.~\ref{method:LSW}.
The $2N \times 2N$ Hermitian matrix ${\cal M}_{\bm{k}}^{}$ in the linear flavor-wave Hamiltonian $\mathcal{H}_{\text{LFW}}$ in Eq.~\eqref{eq:Hamil-A} is given by
\begin{align}
    {\cal M}_{\bm{k}}^{}=
    \begin{pmatrix}
        \bar{J}_{\bm{k}} + \bar{J}_{\bm{k}}^\dagger
        + \Delta E_{\rm diag}
        & J_{\bm{k}} +J_{-\bm{k}}^T\\
        J_{-\bm{k}}^* +J_{\bm{k}}^\dagger  &
        \bar{J}_{-\bm{k}}^* + \bar{J}_{-\bm{k}}^T
        + \Delta E_{\rm diag}
    \end{pmatrix}
\end{align}
where $\Delta E_{\rm diag}={\rm diag}\{\Delta E_1,\Delta E_2,\cdots,\Delta E_N\}$ with $\Delta E_\iota=\Delta E_m^l$ is a $N\times N$ diagonal matrix, and $J_{\bm{k}}$ and $\bar{J}_{\bm{k}}$ are defined as the following $N\times N$ diagonal matrices:
\begin{align}
    J_{\bm{k};\iota\iota'}=J_{\bm{k};(ml)(m'l')}
    = \frac{1}{2}\sum_{i\in l,j\in l^\prime}\sum_{\gamma\gamma^\prime} J_{ij}^{\gamma\gamma^\prime} e^{i\bm{k}\cdot(\bm{r}_{j}-\bm{r}_{i})} \bar{\mathcal{O}}^{l}_{\gamma m} \bar{\mathcal{O}}^{l^\prime}_{\gamma^\prime,m'},\\
    \bar{J}_{\bm{k};\iota\iota'}=\bar{J}_{\bm{k};(ml)(m'l')}
    = \frac{1}{2}\sum_{i\in l,j\in l^\prime}\sum_{\gamma\gamma^\prime} J_{ij}^{\gamma\gamma^\prime} e^{i\bm{k}\cdot(\bm{r}_{j}-\bm{r}_{i})} \bar{\mathcal{O}}^{l}_{\gamma m} \bar{\mathcal{O}}^{l^\prime *}_{\gamma^\prime,m'}.
\end{align}

To diagonalize this Hamiltonian into Eq.~\eqref{eq:Hamil-B}, we introduce the paraunitary matrix $T_{\bm{k}}^{}$ as,
\begin{align}
  \mathcal{A}_{\bm{k}}
  =
  T_{\bm{k}}^{} \mathcal{B}_{\bm{k}}
  =
  \left(\begin{array}{cc}
    U_{\bm{k}}^{} & V_{-\bm{k}}^{*}\\
    V_{\bm{k}}^{} & U_{-\bm{k}}^{*}
  \end{array}
  \right) \mathcal{B}_{\bm{k}},
\end{align}
or equivalently,
\begin{align}
  \label{eq:AtoB_1}
  a_{\iota,\bm{k}}  &=  \sum_{\eta}^{N}
  \Big(
    U_{\bm{k},\iota\eta}^{} b_{\eta,\bm{k}}^{}
    + V_{-\bm{k},\iota\eta}^{*} b_{\eta,-\bm{k}}^\dagger
  \Big),\\
  \label{eq:AtoB_2}
  a_{\iota,\bm{k}}^\dagger  &= \sum_{\eta}^{N} 
  \Big(
    U_{\bm{k},\iota\eta}^{*} b_{\eta, \bm{k}}^\dagger 
    + V_{-\bm{k},\iota\eta} b_{\eta, -\bm{k}}^{} 
  \Big).
\end{align}
Here, $U_{\bm{k}}$ and $V_{\bm{k}}$ are $N \times N$ unitary matrices.
We can find the paraunitary matrix $T_{\bm{k}}^{}$ by combining with the Cholesky decomposition and exact diagonalization~\cite{colpa}.

\begin{widetext}
\section{Nonlinear flavor-wave theory}
\label{app:NLSWT}

In this section, we present theoretical treatments for nonlinear terms in the flavor-wave theory.
Here, we focus on the cubic term, $\mathcal{H}_{3}'/\sqrt{\mathcal{S}}$ in Eq.~\eqref{eq:bosonic_H}, which is neglected in the linear flavor-wave theory.
By performing straightforward calculations, one can represent the cubic term as
\begin{align}
  \frac{1}{\sqrt{\mathcal{S}}}\mathcal{H}_{3}' &= \frac{1}{2\sqrt{\mathcal{S}}}\sum_{ll^\prime}^{M}\sum_{i\in l,j\in l^\prime}\sum_{\gamma\gamma^\prime}\sum_{mm'm''}^{\mathscr{N} - 1}
  J_{ij}^{\gamma\gamma^\prime} 
  \Big(
    \bar{\mathcal{O}}_{\gamma m}^{l} \delta \bar{\mathcal{O}}_{\gamma^\prime, m'm''}^{l^\prime} 
    a_{mi}^\dagger a_{m'j}^\dagger a_{m''j}^{}
    + 
    \delta \bar{\mathcal{O}}_{\gamma,m'm''}^{l} \bar{\mathcal{O}}_{\gamma^\prime m}^{l^\prime} 
    a_{m'i}^\dagger a_{m''i}^{} a_{mj}^\dagger 
    + \text{H.c.}
  \Big),
\end{align}
where
$\bar{\mathcal{O}}_{\gamma, m}^{l} = \braket{m;i|\mathcal{O}^\gamma|0;i}$ and 
$\delta\bar{\mathcal{O}}_{\gamma, m'm''}^{l} = \braket{m';i|\delta\mathcal{O}^\gamma|m'';i}$.
By using Eq.~\eqref{eq:fourier}, $\mathcal{H}_3'$ is expressed by
\begin{align}
  \frac{1}{\sqrt{\mathcal{S}}}\mathcal{H}_3' = \sqrt{\frac{M}{N_t\mathcal{S}}}\sum_{ll^\prime}^{M}
  \sum_{\bm{k}\bm{k}^\prime}\sum_{mm'm''}^{\mathscr{N} - 1}
  \Big(
    &J_{\bm{k};m,m'm''}^{l\delta l^\prime} a_{ml,\bm{k}}^{\dagger} a_{m'l^\prime,\bm{k}^\prime}^\dagger a_{m''l^\prime,\bm{k}+\bm{k}^\prime}^{}
    +J_{\bm{k};m,m'm''}^{\bar{l} \delta\bar{l}^\prime} a_{ml,-\bm{k}}^{} a_{m'l^\prime,\bm{k}^\prime}^\dagger a_{m''l^\prime,\bm{k}+\bm{k}^\prime}^{}\nonumber\\
    & \qquad\qquad
    +J_{\bm{k};m'm'',m}^{\delta l l^\prime} a_{m'l,\bm{k}^\prime}^\dagger a_{m''l,\bm{k}^\prime - \bm{k}}^{} a_{ml^\prime,-\bm{k}}^\dagger
    +J_{\bm{k};m'm'',m}^{\delta\bar{l} \bar{l}^\prime} a_{m''l, \bm{k}^\prime}^{\dagger} a_{m'l, \bm{k}^\prime - \bm{k}}^{} a_{ml^\prime,\bm{k}}^{}
  \Big), \label{eq:H3_2}
\end{align}
where the coefficients are given as follows:
\begin{align} 
  \begin{array}{ll}
  \displaystyle
  J_{\bm{k};m,m'm''}^{l \delta l^\prime} = \frac{1}{2}\sum_{i\in l,j\in l^\prime}\sum_{\gamma\gamma^\prime} J_{ij}^{\gamma\gamma^\prime} e^{i\bm{k}\cdot(\bm{r}_{j}-\bm{r}_{i})} \bar{\mathcal{O}}^{l}_{\gamma m} \delta \bar{\mathcal{O}}^{l^\prime}_{\gamma^\prime,m'm''},\qquad
  & \displaystyle
  J_{\bm{k};m,m'm''}^{l^\prime \delta l} = \frac{1}{2}\sum_{i\in l,j\in l^\prime}\sum_{\gamma\gamma^\prime} J_{ji}^{\gamma^\prime\gamma} e^{i\bm{k}\cdot(\bm{r}_{i}-\bm{r}_{j})} \bar{\mathcal{O}}^{l^\prime}_{\gamma^\prime m} \delta \bar{\mathcal{O}}^{l}_{\gamma,m'm''},\\[10pt]
  \displaystyle
  J_{\bm{k};m,m'm''}^{\bar{l} \delta l^\prime} = \frac{1}{2}\sum_{i\in l,j\in l^\prime}\sum_{\gamma\gamma^\prime} J_{ij}^{\gamma\gamma^\prime}e^{i\bm{k}\cdot(\bm{r}_{j}-\bm{r}_{i})} \bar{\mathcal{O}}_{\gamma m}^{l *}\, \delta \bar{\mathcal{O}}^{l^\prime}_{\gamma^\prime,m'm''},\qquad
  & \displaystyle
  J_{\bm{k};m,m'm''}^{\bar{l}^\prime \delta l} = \frac{1}{2}\sum_{i\in l,j\in l^\prime}\sum_{\gamma\gamma^\prime} J_{ji}^{\gamma^\prime\gamma}e^{i\bm{k}\cdot(\bm{r}_{i}-\bm{r}_{j})} \bar{\mathcal{O}}^{l^\prime *}_{\gamma^\prime m} \delta \bar{\mathcal{O}}_{\gamma,m'm''}^{l},\\[10pt]
  \displaystyle
  J_{\bm{k};m'm'',m}^{\delta l l^\prime} = \frac{1}{2}\sum_{i\in l,j\in l^\prime}\sum_{\gamma\gamma^\prime} J_{ij}^{\gamma\gamma^\prime}e^{i\bm{k}\cdot (\bm{r}_{j}-\bm{r}_{i})} \delta \bar{\mathcal{O}}^{l}_{\gamma,m'm''} \bar{\mathcal{O}}^{l^\prime}_{\gamma^\prime m},\qquad
  & \displaystyle
  J_{\bm{k};m'm'',m}^{\delta l^\prime l} = \frac{1}{2}\sum_{i\in l,j\in l^\prime}\sum_{\gamma\gamma^\prime} J_{ji}^{\gamma^\prime\gamma}e^{i\bm{k}\cdot (\bm{r}_{i}-\bm{r}_{j})} \delta \bar{\mathcal{O}}^{l^\prime}_{\gamma^\prime,m'm''} \bar{\mathcal{O}}^{l}_{\gamma m},\\[10pt]
  \displaystyle
  J_{\bm{k};m'm'',m}^{\delta l \bar{l}^\prime} = \frac{1}{2}\sum_{i\in l,j\in l^\prime}\sum_{\gamma\gamma^\prime} J_{ij}^{\gamma\gamma^\prime}e^{i\bm{k}\cdot (\bm{r}_{j}-\bm{r}_{i})}\, \delta \bar{\mathcal{O}}^{l}_{\gamma,m'm''} \bar{\mathcal{O}}^{l^\prime *}_{\gamma^\prime m},\qquad
  & \displaystyle
  J_{\bm{k};m'm'',m}^{\delta l^\prime \bar{l}} = \frac{1}{2}\sum_{i\in l,j\in l^\prime}\sum_{\gamma\gamma^\prime} J_{ji}^{\gamma^\prime\gamma}e^{i\bm{k}\cdot (\bm{r}_{i}-\bm{r}_{j})}\, \delta \bar{\mathcal{O}}^{l^\prime}_{\gamma^\prime,m'm''} \bar{\mathcal{O}}^{l *}_{\gamma m},\\[10pt]
  \displaystyle
  J_{\bm{k};m,m'm''}^{\bar{l} \delta\bar{l}^\prime} = \frac{1}{2}\sum_{i\in l,j\in l^\prime}\sum_{\gamma\gamma^\prime} J_{ij}^{\gamma\gamma^\prime}e^{i\bm{k}\cdot(\bm{r}_{j}-\bm{r}_{i})} \bar{\mathcal{O}}_{\gamma m}^{l *}\, \delta \bar{\mathcal{O}}^{l^\prime *}_{\gamma^\prime,m'm''},\qquad
  & \displaystyle
  J_{\bm{k};m,m'm''}^{\bar{l}^\prime \delta\bar{l}} = \frac{1}{2}\sum_{i\in l,j\in l^\prime}\sum_{\gamma\gamma^\prime} J_{ji}^{\gamma^\prime\gamma}e^{i\bm{k}\cdot(\bm{r}_{i}-\bm{r}_{j})} \bar{\mathcal{O}}^{l^\prime *}_{\gamma^\prime m} \delta \bar{\mathcal{O}}_{\gamma,m'm''}^{l *}.
  \end{array}
\end{align}
Using the relations between the above coefficients such as 
$J_{\bm{k};m,m'm''}^{ \bar{l} \delta\bar{l}'} = J_{-\bm{k};m,m'm''}^{l \delta l' *}=J_{\bm{k};m'm'',m}^{\delta l' l *}$, one can represent Eq.~\eqref{eq:H3_2} as
\begin{align}
  \frac{1}{\sqrt{\mathcal{S}}}\mathcal{H}_{3}' = \sqrt{\frac{M}{N_t\mathcal{S}}} \sum_{ll^\prime}^{M} \sum_{\bm{k}\bm{k}^\prime}\sum_{mm'm''}^{\mathscr{N} - 1}
  \Bigl(
    J_{\bm{k};m,m'm''}^{l \delta l^\prime} a_{ml,\bm{k}}^\dagger a_{m'l^\prime,\bm{k}^\prime}^\dagger a_{m''l^\prime, \bm{k}+\bm{k}^\prime}^{} 
    + J_{-\bm{k}^\prime;m'm'',m}^{\delta l l^\prime} a_{m'l,\bm{k}}^\dagger a_{m l^\prime,\bm{k}^\prime}^\dagger a_{m''l,\bm{k}+\bm{k}^\prime}^{} + \text{H.c.}
  \Bigr)
  \label{eq:H3_3}.
\end{align}
We introduce the following interaction vertex:
\begin{align}
    \mathcal{V}_{\bm{k},\bm{q}\leftarrow \bm{p}}^{\iota,\iota'\leftarrow \iota''} \equiv 
    J_{\bm{k};m,m'm''}^{l\delta l'} 
    \left[(\delta_{l', l'' } -\delta_{l,l''}) + 2\delta_{l,l''}\delta_{l',l''}\right]
    +
    J_{\bm{q};m',mm''}^{l' \delta l} 
    \left[(\delta_{l l''} - \delta_{l',l''}) + 2\delta_{l,l''}\delta_{l',l''}\right].
    \end{align}
The above cubic term is simplified as
\begin{align}
   \frac{1}{\sqrt{\mathcal{S}}}\mathcal{H}_{3}' = \frac{1}{2} \sqrt{\frac{M}{N_t\mathcal{S}}} \sum_{\iota\iota'\iota''}^{N}\sum_{\bm{k}\bm{q}}^{\bm{k}+\bm{q}=\bm{p}}
  \Bigl(
    \mathcal{V}_{\bm{k},\bm{q}\leftarrow \bm{p}}^{\iota,\iota' \leftarrow \iota''}
    a_{\iota, \bm{k}}^{\dagger} a_{\iota', \bm{q}}^\dagger a_{\iota'',\bm{p}}^{} + \text{H.c.}
  \Bigr) \label{eq:H3_HPboson},
\end{align}
where $\iota=(ml)$, $\iota'=(m'l')$, and $\iota''=(m''l'')$.
Similar expressions are derived for spin models in previous studies~\cite{mook2020,rau2018}.

By applying the Bogoliubov transformation, Eqs.~\eqref{eq:AtoB_1} and \eqref{eq:AtoB_2},
$\mathcal{H}_{3}'$ is represented by
\begin{align}
  \frac{1}{\sqrt{\mathcal{S}}}\mathcal{H}_3' = \frac{1}{\sqrt{\mathcal{S}}}\mathcal{H}_3^{(\text{d})} + \frac{1}{\sqrt{\mathcal{S}}}\mathcal{H}_3^{(\text{s})},
\end{align}
where
\begin{align}
  \label{appeq:decay}
  \frac{1}{\sqrt{\mathcal{S}}}\mathcal{H}_{3}^{(\text{d})} &= 
  \frac{1}{2!} \sqrt[]{\frac{M}{N_{t}\mathcal{S}}} \sum_{\eta\eta'\eta''}^{N}\sum_{\bm{k}\bm{q}\bm{p}}^{\bm{k}+\bm{q}=\bm{p}}
  \Biggl(
    \bar{\mathcal{V}}_{\bm{k},\bm{q}\leftarrow \bm{p}}^{\eta,\eta'\leftarrow\eta''} 
    b_{\eta,\bm{k}}^\dagger b_{\eta',\bm{q}}^\dagger b_{\eta'',\bm{p}}^{} + \text{H.c.}
  \Bigg),\\
  \label{appeq:source}
  \frac{1}{\sqrt{\mathcal{S}}}\mathcal{H}_{3}^{(\text{s})} &= 
  \frac{1}{3!} \sqrt[]{\frac{M}{N_{t}\mathcal{S}}} \sum_{\eta\eta'\eta''}^{N} \sum_{\bm{k}\bm{q}\bm{p}}^{\bm{k}+\bm{q}=-\bm{p}}
  \Biggl(
    \bar{\mathcal{W}}_{\bm{k},\bm{q},\bm{p}}^{\eta,\eta',\eta''} 
    b_{\eta,\bm{k}}^\dagger b_{\eta',\bm{q}}^\dagger b_{\eta'',\bm{p}}^{\dagger} + \text{H.c.}    
  \Bigg).
\end{align}
The interaction vertices in the above expressions are given by
\begin{align}
  \bar{\mathcal{V}}_{\bm{k},\bm{q}\leftarrow \bm{p}}^{\eta,\eta'\leftarrow\eta''} 
   =  \sum_{\iota\iota'\iota''}^{N}
  \Bigg\{
      &\mathcal{V}_{\bm{k},\bm{q}\leftarrow \bm{p}}^{\iota,\iota'\leftarrow \iota''} U_{\bm{k},\iota \eta}^{*} U_{\bm{q},\iota'\eta'}^{*} U_{\bm{p},\iota''\eta''}^{}
      + \Big(\mathcal{V}_{-\bm{k},-\bm{q}\leftarrow -\bm{p}}^{\iota,\iota'\leftarrow \iota''}\Big)^{*} V_{\bm{k},\iota\eta}^{*} V_{\bm{q},\iota'\eta'}^{*} V_{\bm{p},\iota''\eta''}^{}\nonumber\\
      + &\mathcal{V}_{\bm{q},-\bm{p}\leftarrow -\bm{k}}^{\iota,\iota'\leftarrow \iota''} U_{\bm{q},\iota\eta'}^{*} V_{\bm{p},\iota'\eta''} V_{\bm{k},\iota''\eta}^{*}
      + \Big(\mathcal{V}_{-\bm{q},\bm{p}\leftarrow \bm{k}}^{\iota,\iota'\leftarrow \iota''}\Big)^{*} V_{\bm{q},\iota\eta'}^{*} U_{\bm{p},\iota'\eta''}^{} U_{\bm{k},\iota''\eta}^{*}\nonumber\\
      + &\mathcal{V}_{-\bm{p},\bm{k}\leftarrow -\bm{q}}^{\iota,\iota'\leftarrow \iota''} V_{\bm{p},\iota,\eta''}^{} U_{\bm{k},\iota'\eta}^{*} V_{\bm{q},\iota''\eta'}^{*}
      + \Big(\mathcal{V}_{\bm{p},-\bm{k}\leftarrow \bm{q}}^{\iota,\iota'\leftarrow \iota''}\Big)^{*} U_{\bm{p},\iota\eta''}^{} V_{\bm{k},\iota'\eta}^{*} U_{\bm{q},\iota''\eta'}^{*}
  \Bigg\}, \label{appeq:Vbar}\\
  \bar{\mathcal{W}}_{\bm{k},\bm{q},\bm{p}}^{\eta,\eta',\eta''}
   =  \sum_{\iota\iota'\iota''}^{N}
  \Bigg\{
      &\mathcal{V}_{\bm{k},\bm{q}\leftarrow -\bm{p}}^{\iota,\iota'\leftarrow \iota''} U_{\bm{k},\iota\eta}^{*} U_{\bm{q},\iota'\eta'}^{*} V_{\bm{p},\iota''\eta''}^{*}
      + \Big(\mathcal{V}_{-\bm{k},-\bm{q}\leftarrow \bm{p}}^{\iota,\iota'\leftarrow \iota''}\Big)^{*} V_{\bm{k},\iota\eta}^{*} V_{\bm{q},\iota'\eta'}^{*} U_{\bm{p},\iota''\eta''}^{*}\nonumber\\
      + &\mathcal{V}_{\bm{q},\bm{p}\leftarrow -\bm{k}}^{\iota,\iota'\leftarrow \iota''} U_{\bm{q},\iota\eta'}^{*} U_{\bm{p},\iota'\eta''}^{*} V_{\bm{k},\iota''\eta}^{*}
      + \Big(\mathcal{V}_{-\bm{q},-\bm{p}\leftarrow \bm{k}}^{\iota,\iota'\leftarrow \iota''}\Big)^{*} V_{\bm{q},\iota\eta'}^{*} V_{\bm{p},\iota'\eta''}^{*} U_{\bm{k},\iota''\eta}^{*}\nonumber\\
      + &\mathcal{V}_{\bm{p},\bm{k}\leftarrow -\bm{q}}^{\iota,\iota'\leftarrow \iota''} U_{\bm{p},\iota\eta''}^{*} U_{\bm{k},\iota'\eta}^{*} V_{\bm{q},\iota''\eta'}^{*}
      + \Big(\mathcal{V}_{-\bm{p},-\bm{k}\leftarrow \bm{q}}^{\iota,\iota'\leftarrow \iota''}\Big)^{*} V_{\bm{p},\iota\eta''}^{*} V_{\bm{k},\iota'\eta}^{*} U_{\bm{q},\iota''\eta''}^{*}
  \Bigg\}\label{appeq:Wbar}.
\end{align}
These expressions are the same as those in previous studies~\cite{costa2000,mook2020}.
Figures~\ref{fig:vertices_diagram}(a) and~\ref{fig:vertices_diagram}(b) illustrate these vertices,
$\bar{\mathcal{V}}_{\bm{k},\bm{q}\leftarrow \bm{p}}^{\eta,\eta'\leftarrow\eta''}$
and
$\bar{\mathcal{W}}_{\bm{k},\bm{q},\bm{p}}^{\eta,\eta',\eta''}$, 
respectively.

To treat the anharmonic terms as perturbations up to order $1/\mathcal{S}$, we utilize the standard Green's function approach~\cite{mahan}.
In terms of the bosons $b_{\eta,\bm{k}}$, we define the temperature Green's function as follows:
\begin{align}
  \mathcal{G}_{\eta\eta',\bm{k}}(\tau) &\equiv -\big\langle T_{\tau} \mathcal{B}_{\eta,\bm{k}}(\tau)^{} \mathcal{B}_{\eta',\bm{k}}^\dagger \big\rangle
  \qquad\qquad \big( \eta,\eta' =1,2,\cdots,2N \big) \\
  \mathcal{G}_{\eta\eta',\bm{k}}(i\omega_n) &\equiv \int_{0}^\beta d\tau e^{i\omega_n\tau} \mathcal{G}_{\eta\eta'\bm{k}}(\tau),
\end{align}
where $T_{\tau}$ is the time-ordering operator in the imaginary time, $\omega_n=2n\pi/\beta$ is the Matsubara frequency with $n$ being integer, and $\langle \ \cdot\ \rangle$ stands for the thermal average.
$\beta=(k_BT)^{-1}$ is the inverse temperature, where $k_B$ is the Boltzmann constant.
The bare temperature Green's function is represented by

\begin{align}
  \mathcal{G}_{\bm{k}}^{(0)} (i\omega_n) \ = \ \frac{1}{i\omega_n\sigma_3 - \mathcal{E}_{\bm{k}}}.
\end{align}
The temperature Green's function up to the $1/\mathcal{S}$ correction with respect to the $\mathcal{H}_{\mathrm{LFW}}$~[see Eq.~\eqref{eq:bosonic_H}] can be written 
as~\cite{mahan}
\begin{align}
  \label{eq:Green perturbation}
  \mathcal{G}_{\eta\eta',\bm{k}} (\tau) \simeq 
  \mathcal{G}_{\eta\eta',\bm{k}}^{(0)} (\tau) + \frac{1}{\mathcal{S}}\int_{0}^\beta d\tau_1
  \Big\langle
  T_{\tau} \mathcal{H}'_4 (\tau_1) \mathcal{B}_{\eta,\bm{k}}^{}(\tau) \mathcal{B}_{\eta',\bm{k}}^\dagger
  \Big\rangle_{0}
  - \frac{1}{\mathcal{S}}\frac{1}{2!}
  \int_{0}^\beta d\tau_1 \int_{0}^\beta d\tau_2
  \Big\langle
  T_{\tau} \mathcal{H}'_{3}(\tau_1) \mathcal{H}'_{3}(\tau_2) \mathcal{B}_{\eta,\bm{k}}^{}(\tau) \mathcal{B}_{\eta',\bm{k}}^\dagger
  \Big\rangle_{0},
\end{align}
where $\mathcal{H}'_4$ is the four-quasiparticle interaction, and $\langle\ \cdot\ \rangle_{0}$ represents the thermal average in $\mathcal{H}_{\mathrm{LFW}}$.
We also introduce the self-energy, which is defined as
\begin{align}
  \label{appeq:self-energy}
  \Sigma_{\bm{k}}(\omega,T) &\equiv \Sigma_{\bm{k}} (i\omega_n\rightarrow \omega+i0^{+})\\
  \label{appeq:temperature self-energy}
  \Sigma_{\bm{k}} (i\omega_n) &\equiv  \Big[\mathcal{G}_{\bm{k}}^{(0)}(i\omega_n)\Big]^{-1} - \Big[\mathcal{G}_{\bm{k}} (i\omega_n)\Big]^{-1}.
\end{align}
By using the self-energy, the temperature and retarded Green's function can be written as~\cite{mahan}
\begin{align}
  \label{appeq:interaction temperature Green function}
  \mathcal{G}_{\bm{k}} (i\omega_n) &= \Big[ 1 - \mathcal{G}_{\bm{k}}^{(0)}(i\omega_n) \Sigma_{\bm{k}}(i\omega_n) \Big]^{-1} \mathcal{G}_{\bm{k}}^{(0)}(i\omega_n)
  = \Big[ i\omega_n\sigma_3 - \mathcal{E}_{\bm{k}} - \Sigma_{\bm{k}}(i\omega_n) \Big]^{-1},\\
  \label{appeq:interaction retarded Green function}
  G^{R}_{\bm{k}}(\omega,T) &\equiv \mathcal{G}_{\bm{k}}(i\omega_n\rightarrow \omega+i0^{+})
  = \Big[ (\omega+ i0^{+})\sigma_3 - \mathcal{E}_{\bm{k}} - \Sigma_{\bm{k}}(\omega,T) \Big]^{-1}.
\end{align}
Since we focus on the damping phenomena of the bosonic quasiparticles, we consider only the imaginary part of the self-energy within the $1/\mathcal{S}$ correction relative to $\mathcal{H}_{\mathrm{LFW}}$.
The lowest-order diagrams with the imaginary part are Figs.~\ref{fig:vertices_diagram}(c) and~\ref{fig:vertices_diagram}(d), which represent the spontaneous decays and collisions with thermally excited bosons, respectively.
Since multiple thermally excited bosons are required for the collision process,
only contributions from Fig.~\ref{fig:vertices_diagram}(d) are nonzero at zero temperature.
As in Refs.~\cite{winter2017_nc,mook2020}, we treat only diagonal terms of the self-energy, defined as $\tilde{\Sigma}_{\eta,\bm{k}}(i\omega_n)$.
In the following, we restrict the range of $\eta$ to $1,2,\cdots,N$.
Then, the temperature Grees's function are expressed by
\begin{align}
  \mathcal{G}_{\eta,\bm{k}}(i\omega_n) 
  =\frac{1}{i\omega_n - \varepsilon_{\eta,\bm{k}} - \tilde{\Sigma}_{\eta,\bm{k}}(i\omega_n)}.
\end{align}
From Figs.~\ref{fig:vertices_diagram}(c) and~\ref{fig:vertices_diagram}(d),
 $\tilde{\Sigma}_{\eta,\bm{k}} (\omega,T)$ can be written as
\begin{align}
  \tilde{\Sigma}_{\eta,\bm{k}} (\omega,T) &= \tilde{\Sigma}_{\eta,\bm{k}} (i\omega_n \rightarrow \omega + i0^{+})
  = \Sigma_{\eta,\bm{k}}^{(\text{c})} (\omega, T) +  \Sigma_{\eta,\bm{k}}^{(\text{d})} (\omega, T),
  \end{align}
where
\begin{align}
  \label{appeq:self_collision}
  \Sigma_{\eta,\bm{k}}^{(\text{c})} (\omega,T)
  &= \frac{M}{N_t \mathcal{S}} \sum_{\eta',\eta''}^{N} \sum_{\bm{q}}^{\mathrm{B.Z.}}
  \frac{\left| \bar{\mathcal{V}}_{\bm{k},\bm{q}\leftarrow \bm{k}+\bm{q}}^{\eta,\eta'\leftarrow\eta''} \right|^2}{\omega + \varepsilon_{\eta',\bm{q}} - \varepsilon_{\eta'',\bm{k}+\bm{q}} + i0^{+}}
  \left[
    g(\varepsilon_{\eta',\bm{q}}) - g(\varepsilon_{\eta'',\bm{k}+\bm{q}})
  \right]\qquad\qquad {\rm for}\ \ \omega \geq 0,\\
  \label{appeq:self_decay}
  \Sigma_{\eta,\bm{k}}^{(\text{d})} (\omega,T)
  &= \frac{1}{2} \frac{M}{N_t \mathcal{S}} \sum_{\eta',\eta''}^{N} \sum_{\bm{q}}^{\mathrm{B.Z.}}
  \frac{\left| \bar{\mathcal{V}}_{\bm{q},\bm{k}-\bm{q}\leftarrow \bm{k}}^{\eta',\eta''\leftarrow\eta} \right|^2}{\omega - \varepsilon_{\eta',\bm{q}} - \varepsilon_{\eta'',\bm{k}-\bm{q}} + i0^{+}}
  \left[
    g(\varepsilon_{\eta',\bm{q}}) + g(\varepsilon_{\eta'',\bm{k}-\bm{q}}) + 1
  \right].
\end{align}
The above quantities $\Sigma_{\eta,\bm{k}}^{(\text{c})}$ and $\Sigma_{\eta,\bm{k}}^{(\text{d})}$
correspond to the self-energies depicted by
Figs.~\ref{fig:vertices_diagram}(c) and~\ref{fig:vertices_diagram}(d), respectively.
The analytic continuation is performed under the condition 
that $\text{sgn}(\omega) \text{Im}\Sigma_{\bm{k}}(\omega) \leq 0$ is satisfied.
Note that we also consider the process $\Sigma_{\eta,\bm{k}}^{\text{(s)}} (\omega,T)$ shown in Figure~\ref{fig:source_diagram}, which is involved with the magnon interaction presented in Fig.~\ref{fig:vertices_diagram}(b).
This is given by
\begin{align}
  \label{eq:self_source}
  \Sigma_{\eta,\bm{k}}^{(\text{s})} (\omega,T)
  &= -\frac{1}{2} \frac{M}{N_t\mathcal{S}} \sum_{\eta',\eta''}^{N} \sum_{\bm{q}}^{\mathrm{B.Z.}}
  \frac{\left| \bar{\mathcal{W}}_{\bm{q},\bm{k}-\bm{q}, \bm{k}}^{\eta',\eta'', \eta} \right|^2}{\omega + \varepsilon_{\eta',\bm{q}} + \varepsilon_{\eta'',-\bm{k}-\bm{q}} - i0^{+}}
  \left[
    g(\varepsilon_{\eta',\bm{q}}) + g(\varepsilon_{\eta'',-\bm{k}-\bm{q}}) + 1
  \right].
\end{align}
Since $\varepsilon_{\eta',\bm{q}}$ and $\varepsilon_{\eta'',-\bm{k}-\bm{q}}$ are positive, the imaginary part of this quantity is zero.
We do not take it into account in the present scheme.

\begin{figure*}[t]
  \begin{center}
    \includegraphics[width=80mm,clip]{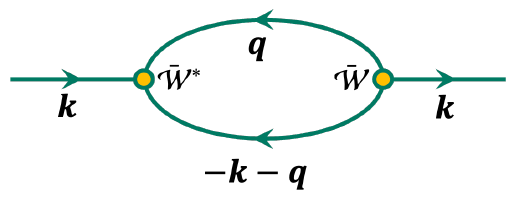}
    \caption{
    The lowest-order diagram involved with $\mathcal{H}_3^{\text{(s)}}$ defined in Eq.~\eqref{eq:source} [see also Fig.~\ref{fig:vertices_diagram}(b)].
    }
    \label{fig:source_diagram}
  \end{center}
\end{figure*}

\section{On-shell approximation}
\label{app:on-shell approximation}

In the on-shell approximation, $\omega$ in the self-energy $\tilde{\Sigma}_{\eta,\bm{k}} (\omega,T)$ is replaced to the one-particle energy $\varepsilon_{\eta,\bm{k}}$.
The on-shell approximation up to the $1/\mathcal{S}$ correction corresponds to the Born approximation. 
In this approximation, the damping rate $\Gamma_{\eta,\bm{k}}$ is evaluated as
\begin{align}
  \Gamma_{\eta,\bm{k}}(T)  &\simeq  - \text{Im}\tilde{\Sigma}_{\eta,\bm{k}} (\omega=\varepsilon_{\eta,\bm{k}},T)\nonumber\\
  \label{appeq:damping rate_onshell}
  &= \frac{M}{N_t \mathcal{S}} \sum_{\eta'\eta''}^{N} \sum_{\bm{q}}^{\mathrm{B.Z.}}
  \Bigg\{
    \frac{1}{2}\left| \bar{\mathcal{V}}_{\bm{q},\bm{k}-\bm{q}\leftarrow \bm{k}}^{\eta',\eta''\leftarrow\eta} \right|^2
    \left[
      g(\varepsilon_{\eta',\bm{q}}) + g(\varepsilon_{\eta'',\bm{k}-\bm{q}}) + 1
    \right] 
    \delta(\varepsilon_{\eta,\bm{k}} - \varepsilon_{\eta',\bm{q}} - \varepsilon_{\eta'',\bm{k}-\bm{q}})\nonumber\\
    &\qquad\qquad\qquad\qquad\qquad+
    \left| \bar{\mathcal{V}}_{\bm{k},\bm{q}\leftarrow \bm{k}+\bm{q}}^{\eta,\eta'\leftarrow\eta''} \right|^2
    \left[
      g(\varepsilon_{\eta',\bm{q}}) - g(\varepsilon_{\eta'',\bm{k}+\bm{q}})
    \right]
    \delta(\varepsilon_{\eta,\bm{k}} + \varepsilon_{\eta',\bm{q}} - \varepsilon_{\eta'',\bm{k}+\bm{q}})
  \Bigg\}.
\end{align}
We numerically confirmed that calculation results of the damping rate in the antiferromagnetic Heisenberg model on the square and tetragonal lattice are the same as those in the previous studies at $T=0$~\cite{mourigal2010,fuhrman2012}.
We note that since this approximation completely neglects the magnon-magnon interactions in the intermediate states of the self-energy, it shows non-analytic divergences~\cite{zhitomirsky2013}.

\section{Dynamical structure factor}
\label{app:Dsf}
In this appendix, we verify the validity of our method incorporated with the iDE approach at finite temperatures.
We calculate the dynamical structure factor in Eq.~\eqref{eq:kitaev}
and compare with the previous results computed by a continuous-time quantum Monte Carlo~(CTQMC) method in Ref.~\cite{yoshitake2020}.

The dynamical structure factor is defined as
\begin{align}
  S^{\gamma\gamma^\prime} (\bm{k},\omega) \ &\equiv \ \int_{-\infty}^{\infty} \frac{dt}{2\pi} e^{i\omega t} 
  \big\langle
    \delta\mathcal{O}^{\gamma}_{\bm{k}}(t) \delta\mathcal{O}^{\gamma^\prime}_{-\bm{k}}
  \big\rangle\\
  \label{eq:Dsf}
  &= \ \frac{1}{M}\sum_{ll'}^{M} \int_{-\infty}^{\infty} \frac{dt}{2\pi} e^{i\omega t} 
  \big\langle
    \delta\mathcal{O}^{\gamma}_{l,\bm{k}}(t) \delta\mathcal{O}^{\gamma^\prime}_{l',-\bm{k}}
  \big\rangle,
\end{align}
where $l$ and $l'$ are the sublattice indexes.
The Fourier transformation of the operator $\delta\mathcal{O}_{i}^{\gamma}$ introduced in Eq.~\eqref{eq:delO_1} is defined as
\begin{align}
  \delta\mathcal{O}_{\bm{k}}^{\gamma} \ &\equiv \sqrt{\frac{1}{N_t}}\sum_{i}  \delta\mathcal{O}_{i}^\gamma e^{-i\bm{k}\cdot\bm{r}_i} \ = \ \sqrt{\frac{1}{M}}\sum_{l}^{M} \delta\mathcal{O}_{l,\bm{k}}^\gamma,
\end{align}
with
\begin{align}
      \delta\mathcal{O}_{l,\bm{k}}^{\gamma} \ &\equiv \ \sqrt{\frac{M}{N_t}} \sum_{i\in l} \delta\mathcal{O}_{i}^{\gamma} e^{-i\bm{k}\cdot \bm{r}_{i}}.
\end{align}
To calculate Eq.~\eqref{eq:Dsf}, we introduce the following time-ordered correlation function:
\begin{align}
  \Phi^{\gamma\gamma^\prime} (\bm{k},\tau) \ &\equiv \  \frac{1}{M}\sum_{ll'}^{M}
  \big\langle
    T_{\tau}\delta\mathcal{O}_{l,\bm{k}}^{\gamma}(\tau) \delta\mathcal{O}_{l',-\bm{k}}^{\gamma^\prime}
  \big\rangle.
\end{align}
The Fourier transformation with respect to the imaginary time is given by
\begin{align}
  \Phi^{\gamma\gamma^\prime} (\bm{k},i\Omega) \ &= \ \int_{0}^\beta d\tau \Phi^{\gamma\gamma^\prime} (\bm{k},\tau) e^{i\Omega\tau}.
\end{align}
Here, we introduce the dynamical susceptibility as an analytic continuation from $\Phi^{\gamma\gamma^\prime} (\bm{k},i\Omega)$ as follows:
\begin{align}
    \chi^{\gamma\gamma'} (\bm{k},\omega) \equiv \Phi^{\gamma\gamma'} (\bm{k},i\Omega\to \omega + i0^{+}).
\end{align}
For $\omega > 0$, the diagonal components of Eq.~\eqref{eq:Dsf} is expressed as 
\begin{align}
  S^{\gamma\gamma} (\bm{k},\omega) &=  \frac{1}{\pi} {\rm Im}
    \chi^{\gamma\gamma} (\bm{k},\omega) 
\end{align}

In the present study, we only take into account the lowest-order contributions of $\delta\mathcal{O}_{i}^\gamma$ and approximate the retarded Green's function for bosonic quasiparticles as Eq.~\eqref{eq:retarded Green}.
By using this representation of the Green's function, $S^{\gamma\gamma} (\bm{k},\omega)$ is written as
\begin{align}
    \label{eq:dsf}
    S^{\gamma\gamma} (\bm{k},\omega) &= \frac{1}{M} \sum_{\eta}^{N}
    \left|\tilde{W}_{\gamma\bm{k}\eta}\right|^2 \frac{\Gamma_{\eta,\bn{k}}/\pi}{(\omega-\varepsilon_{\eta,\bn{k}})^2 + \Gamma_{\eta,\bm{k}}^2},
\end{align}
where
\begin{align}
    \tilde{W}_{\gamma\bm{k}\eta}
    \equiv \sum_{\iota}^{N}  
    \left( \bar{O}_{\gamma m}^{l} V_{\bm{k},\iota\eta} + \bar{O}_{\gamma m}^{l*} U_{\bm{k},\iota\eta}\right).
\end{align}

\begin{figure*}[t]
  \begin{center}
    \includegraphics[width=175mm,clip]{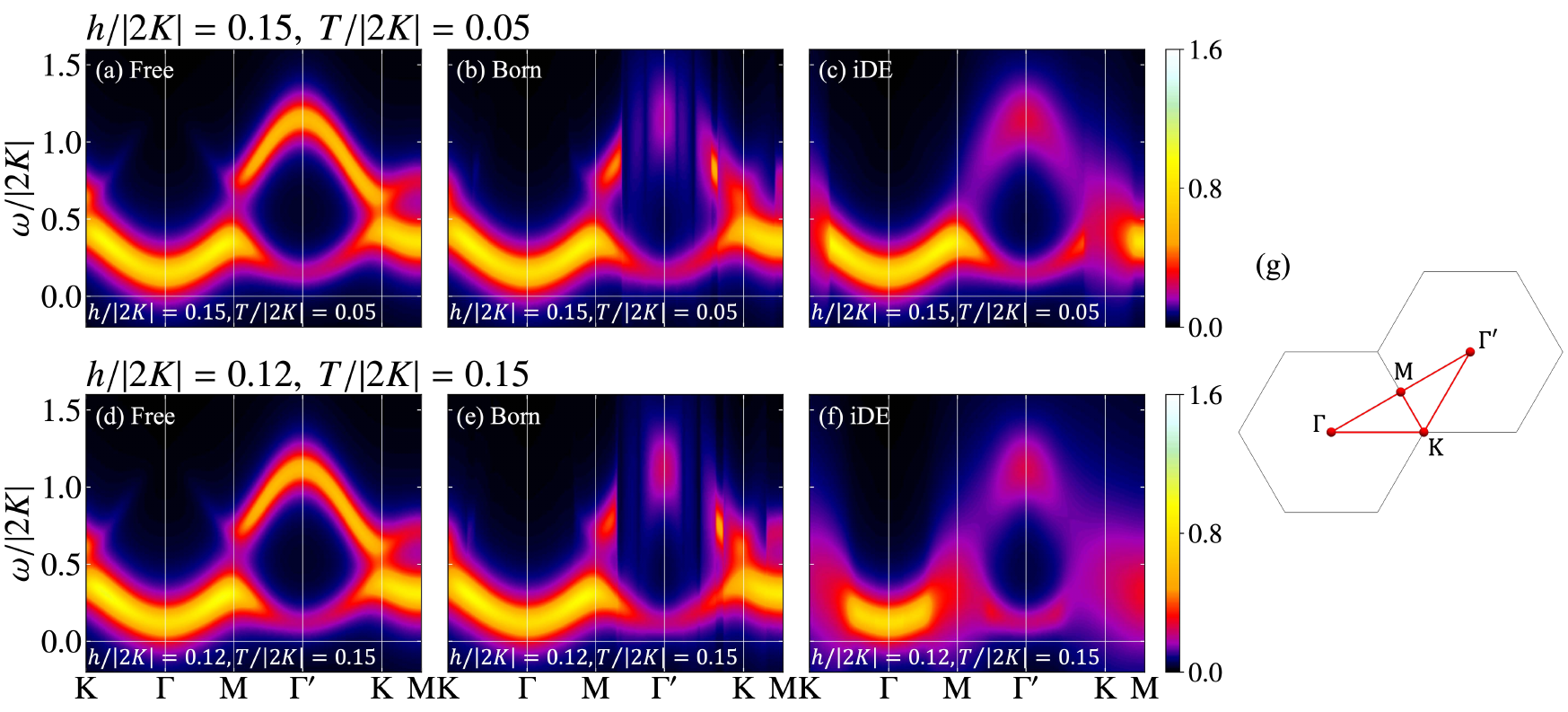}
    \caption{
      Dynamical structure factor $S(\bm{k},\omega)$ at $h/|2K|=0.15$ and $T/|2K|=0.05$ obtained by (a) the linear spin-wave theory and nonlinear theories with (b) the on-shell approximation and (c) the iDE approach.
      (d)--(f) Corresponding plots for $h/|2K|=0.12$ and $T/|2K|=0.15$.
      The smearing factor is introduced as 0.1.
      The parameters for upper and lower sides correspond to those for Figs.~3(h) and~3(e) in Ref.~\cite{yoshitake2020}, respectively.
      (g) First Brillouin zones centered at the $\Gamma$ point and adjoined one centered at the $\Gamma'$ point.
      The red lines represent the lines connected with the high-symmetry points used in (a)--(f).
    }
    \label{fig:Dsf}
  \end{center}
\end{figure*}

We calculate the trace of the dynamical structure factor, $ S(\bm{k},\omega) = \sum_{\gamma} 
S^{\gamma\gamma}(\bm{k},\omega)$.
Figure~\ref{fig:Dsf} shows $S(\bm{k},\omega)$ at $(h/|2K|, T/|2K|)=(0.15, 0.05)$ and $(0.12, 0.15)$.
We choose these parameters to compare the results with the previous ones in Ref.~\cite{yoshitake2020}.
Figures~\ref{fig:Dsf}(b), and \ref{fig:Dsf}(e) show $S(\bm{k},\omega)$ obtained by the on-shell Born approximation.
We find that the spectral weight around the $\Gamma'$ point is smeared due to the magnon damping compared to the results calculated by the linear spin-wave theory [Figs~\ref{fig:Dsf}(a), and \ref{fig:Dsf}(d)].
In addition, unnatural suppression in the high-energy region is observed.
This behavior is due to the artifact intrinsic to the on-shell Born approximation as mentioned in Sec.~\ref{app:on-shell approximation}.
On the other hand, in the results by the iDE approach, such unnatural behavior is not observed, as shown in Figs.~\ref{fig:Dsf}(c), and \ref{fig:Dsf}(f).
Furthermore, these spectral structures are in good agreement with those obtained by CTQMC simulations shown in Fig.~3(h) and 3(e) in Ref.~\cite{yoshitake2020}.
In particular, the present approach can reproduce the broadening of the excitation spectra at the $\Gamma$ and $\Gamma'$ points by thermal fluctuations.
This is in contrast to the results by the on-shell Born approximation, which are largely intact against thermal fluctuations [Figs~\ref{fig:Dsf}(b), and \ref{fig:Dsf}(e)].
These results indicate that the damping effect is essential in evaluating the dynamic structure factors in the Kitaev model under magnetic fields, and our present approach can treat this effect appropriately at finite temperatures.

\end{widetext}
\bibliography{refs}

\begin{thebibliography}{93}%
\makeatletter
\providecommand \@ifxundefined [1]{%
 \@ifx{#1\undefined}
}%
\providecommand \@ifnum [1]{%
 \ifnum #1\expandafter \@firstoftwo
 \else \expandafter \@secondoftwo
 \fi
}%
\providecommand \@ifx [1]{%
 \ifx #1\expandafter \@firstoftwo
 \else \expandafter \@secondoftwo
 \fi
}%
\providecommand \natexlab [1]{#1}%
\providecommand \enquote  [1]{``#1''}%
\providecommand \bibnamefont  [1]{#1}%
\providecommand \bibfnamefont [1]{#1}%
\providecommand \citenamefont [1]{#1}%
\providecommand \href@noop [0]{\@secondoftwo}%
\providecommand \href [0]{\begingroup \@sanitize@url \@href}%
\providecommand \@href[1]{\@@startlink{#1}\@@href}%
\providecommand \@@href[1]{\endgroup#1\@@endlink}%
\providecommand \@sanitize@url [0]{\catcode `\\12\catcode `\$12\catcode `\&12\catcode `\#12\catcode `\^12\catcode `\_12\catcode `\%12\relax}%
\providecommand \@@startlink[1]{}%
\providecommand \@@endlink[0]{}%
\providecommand \url  [0]{\begingroup\@sanitize@url \@url }%
\providecommand \@url [1]{\endgroup\@href {#1}{\urlprefix }}%
\providecommand \urlprefix  [0]{URL }%
\providecommand \Eprint [0]{\href }%
\providecommand \doibase [0]{https://doi.org/}%
\providecommand \selectlanguage [0]{\@gobble}%
\providecommand \bibinfo  [0]{\@secondoftwo}%
\providecommand \bibfield  [0]{\@secondoftwo}%
\providecommand \translation [1]{[#1]}%
\providecommand \BibitemOpen [0]{}%
\providecommand \bibitemStop [0]{}%
\providecommand \bibitemNoStop [0]{.\EOS\space}%
\providecommand \EOS [0]{\spacefactor3000\relax}%
\providecommand \BibitemShut  [1]{\csname bibitem#1\endcsname}%
\let\auto@bib@innerbib\@empty
\bibitem [{\citenamefont {Thouless}\ \emph {et~al.}(1982)\citenamefont {Thouless}, \citenamefont {Kohmoto}, \citenamefont {Nightingale},\ and\ \citenamefont {den Nijs}}]{thouless1982}%
  \BibitemOpen
  \bibfield  {author} {\bibinfo {author} {\bibfnamefont {D.~J.}\ \bibnamefont {Thouless}}, \bibinfo {author} {\bibfnamefont {M.}~\bibnamefont {Kohmoto}}, \bibinfo {author} {\bibfnamefont {M.~P.}\ \bibnamefont {Nightingale}},\ and\ \bibinfo {author} {\bibfnamefont {M.}~\bibnamefont {den Nijs}},\ }\bibfield  {title} {\bibinfo {title} {Quantized {H}all {C}onductance in a {T}wo-{D}imensional {P}eriodic {P}otential},\ }\href {https://doi.org/10.1103/PhysRevLett.49.405} {\bibfield  {journal} {\bibinfo  {journal} {Phys. Rev. Lett.}\ }\textbf {\bibinfo {volume} {49}},\ \bibinfo {pages} {405} (\bibinfo {year} {1982})}\BibitemShut {NoStop}%
\bibitem [{\citenamefont {Kohmoto}(1985)}]{kohmoto1985}%
  \BibitemOpen
  \bibfield  {author} {\bibinfo {author} {\bibfnamefont {M.}~\bibnamefont {Kohmoto}},\ }\bibfield  {title} {\bibinfo {title} {Topological invariant and the quantization of the {H}all conductance},\ }\href {https://doi.org/https://doi.org/10.1016/0003-4916(85)90148-4} {\bibfield  {journal} {\bibinfo  {journal} {Ann. Phys.}\ }\textbf {\bibinfo {volume} {160}},\ \bibinfo {pages} {343} (\bibinfo {year} {1985})}\BibitemShut {NoStop}%
\bibitem [{\citenamefont {Katsura}\ \emph {et~al.}(2010)\citenamefont {Katsura}, \citenamefont {Nagaosa},\ and\ \citenamefont {Lee}}]{katsura2010}%
  \BibitemOpen
  \bibfield  {author} {\bibinfo {author} {\bibfnamefont {H.}~\bibnamefont {Katsura}}, \bibinfo {author} {\bibfnamefont {N.}~\bibnamefont {Nagaosa}},\ and\ \bibinfo {author} {\bibfnamefont {P.~A.}\ \bibnamefont {Lee}},\ }\bibfield  {title} {\bibinfo {title} {Theory of the {T}hermal {H}all {E}ffect in {Q}uantum {M}agnets},\ }\href {https://doi.org/10.1103/PhysRevLett.104.066403} {\bibfield  {journal} {\bibinfo  {journal} {Phys. Rev. Lett.}\ }\textbf {\bibinfo {volume} {104}},\ \bibinfo {pages} {066403} (\bibinfo {year} {2010})}\BibitemShut {NoStop}%
\bibitem [{\citenamefont {Onose}\ \emph {et~al.}(2010)\citenamefont {Onose}, \citenamefont {Ideue}, \citenamefont {Katsura}, \citenamefont {Shiomi}, \citenamefont {Nagaosa},\ and\ \citenamefont {Tokura}}]{onose2010}%
  \BibitemOpen
  \bibfield  {author} {\bibinfo {author} {\bibfnamefont {Y.}~\bibnamefont {Onose}}, \bibinfo {author} {\bibfnamefont {T.}~\bibnamefont {Ideue}}, \bibinfo {author} {\bibfnamefont {H.}~\bibnamefont {Katsura}}, \bibinfo {author} {\bibfnamefont {Y.}~\bibnamefont {Shiomi}}, \bibinfo {author} {\bibfnamefont {N.}~\bibnamefont {Nagaosa}},\ and\ \bibinfo {author} {\bibfnamefont {Y.}~\bibnamefont {Tokura}},\ }\bibfield  {title} {\bibinfo {title} {Observation of the {M}agnon {H}all {E}ffect},\ }\href {https://doi.org/10.1126/science.1188260} {\bibfield  {journal} {\bibinfo  {journal} {Science}\ }\textbf {\bibinfo {volume} {329}},\ \bibinfo {pages} {297} (\bibinfo {year} {2010})}\BibitemShut {NoStop}%
\bibitem [{\citenamefont {McClarty}(2022)}]{mcclarty2022}%
  \BibitemOpen
  \bibfield  {author} {\bibinfo {author} {\bibfnamefont {P.~A.}\ \bibnamefont {McClarty}},\ }\bibfield  {title} {\bibinfo {title} {Topological {M}agnons: {A} {R}eview},\ }\href {https://doi.org/10.1146/annurev-conmatphys-031620-104715} {\bibfield  {journal} {\bibinfo  {journal} {Annu. Rev. Condens. Matter Phys.}\ }\textbf {\bibinfo {volume} {13}},\ \bibinfo {pages} {171} (\bibinfo {year} {2022})}\BibitemShut {NoStop}%
\bibitem [{\citenamefont {Berry}(1984)}]{berry1984}%
  \BibitemOpen
  \bibfield  {author} {\bibinfo {author} {\bibfnamefont {M.~V.}\ \bibnamefont {Berry}},\ }\bibfield  {title} {\bibinfo {title} {Quantal phase factors accompanying adiabatic changes},\ }\href {https://doi.org/10.1098/rspa.1984.0023} {\bibfield  {journal} {\bibinfo  {journal} {Proc. R. Soc. London. A. Math. Phys. Sci.}\ }\textbf {\bibinfo {volume} {392}},\ \bibinfo {pages} {45} (\bibinfo {year} {1984})}\BibitemShut {NoStop}%
\bibitem [{\citenamefont {Owerre}(2016)}]{owerre2016}%
  \BibitemOpen
  \bibfield  {author} {\bibinfo {author} {\bibfnamefont {S.~A.}\ \bibnamefont {Owerre}},\ }\bibfield  {title} {\bibinfo {title} {A first theoretical realization of honeycomb topological magnon insulator},\ }\href {https://doi.org/10.1088/0953-8984/28/38/386001} {\bibfield  {journal} {\bibinfo  {journal} {Journal of Physics: Condensed Matter}\ }\textbf {\bibinfo {volume} {28}},\ \bibinfo {pages} {386001} (\bibinfo {year} {2016})}\BibitemShut {NoStop}%
\bibitem [{\citenamefont {Owerre}(2017)}]{owerre2017}%
  \BibitemOpen
  \bibfield  {author} {\bibinfo {author} {\bibfnamefont {S.~A.}\ \bibnamefont {Owerre}},\ }\bibfield  {title} {\bibinfo {title} {Topological thermal {H}all effect in frustrated kagome antiferromagnets},\ }\href {https://doi.org/10.1103/PhysRevB.95.014422} {\bibfield  {journal} {\bibinfo  {journal} {Phys. Rev. B}\ }\textbf {\bibinfo {volume} {95}},\ \bibinfo {pages} {014422} (\bibinfo {year} {2017})}\BibitemShut {NoStop}%
\bibitem [{\citenamefont {Laurell}\ and\ \citenamefont {Fiete}(2018)}]{laurell2018}%
  \BibitemOpen
  \bibfield  {author} {\bibinfo {author} {\bibfnamefont {P.}~\bibnamefont {Laurell}}\ and\ \bibinfo {author} {\bibfnamefont {G.~A.}\ \bibnamefont {Fiete}},\ }\bibfield  {title} {\bibinfo {title} {Magnon thermal {H}all effect in kagome antiferromagnets with {D}zyaloshinskii-{M}oriya interactions},\ }\href {https://doi.org/10.1103/PhysRevB.98.094419} {\bibfield  {journal} {\bibinfo  {journal} {Phys. Rev. B}\ }\textbf {\bibinfo {volume} {98}},\ \bibinfo {pages} {094419} (\bibinfo {year} {2018})}\BibitemShut {NoStop}%
\bibitem [{\citenamefont {McClarty}\ \emph {et~al.}(2018)\citenamefont {McClarty}, \citenamefont {Dong}, \citenamefont {Gohlke}, \citenamefont {Rau}, \citenamefont {Pollmann}, \citenamefont {Moessner},\ and\ \citenamefont {Penc}}]{mcclarty2018}%
  \BibitemOpen
  \bibfield  {author} {\bibinfo {author} {\bibfnamefont {P.~A.}\ \bibnamefont {McClarty}}, \bibinfo {author} {\bibfnamefont {X.-Y.}\ \bibnamefont {Dong}}, \bibinfo {author} {\bibfnamefont {M.}~\bibnamefont {Gohlke}}, \bibinfo {author} {\bibfnamefont {J.~G.}\ \bibnamefont {Rau}}, \bibinfo {author} {\bibfnamefont {F.}~\bibnamefont {Pollmann}}, \bibinfo {author} {\bibfnamefont {R.}~\bibnamefont {Moessner}},\ and\ \bibinfo {author} {\bibfnamefont {K.}~\bibnamefont {Penc}},\ }\bibfield  {title} {\bibinfo {title} {Topological magnons in {K}itaev magnets at high fields},\ }\href {https://doi.org/10.1103/PhysRevB.98.060404} {\bibfield  {journal} {\bibinfo  {journal} {Phys. Rev. B}\ }\textbf {\bibinfo {volume} {98}},\ \bibinfo {pages} {060404} (\bibinfo {year} {2018})}\BibitemShut {NoStop}%
\bibitem [{\citenamefont {Zhang}\ \emph {et~al.}(shed)\citenamefont {Zhang}, \citenamefont {Gao},\ and\ \citenamefont {Chen}}]{zhang2023_arxiv}%
  \BibitemOpen
  \bibfield  {author} {\bibinfo {author} {\bibfnamefont {X.-T.}\ \bibnamefont {Zhang}}, \bibinfo {author} {\bibfnamefont {Y.~H.}\ \bibnamefont {Gao}},\ and\ \bibinfo {author} {\bibfnamefont {G.}~\bibnamefont {Chen}},\ }\bibfield  {title} {\bibinfo {title} {Thermal {H}all effects in quantum magnets},\ }\href {https://arxiv.org/abs/2305.04830} {\bibfield  {journal} {\bibinfo  {journal} {arXiv:2007.06757}\ } (\bibinfo {year} {unpublished})}\BibitemShut {NoStop}%
\bibitem [{\citenamefont {Zhang}\ \emph {et~al.}(2021{\natexlab{a}})\citenamefont {Zhang}, \citenamefont {Zhu}, \citenamefont {Go}, \citenamefont {Lux}, \citenamefont {dos Santos}, \citenamefont {Lounis}, \citenamefont {Su}, \citenamefont {Bl\"ugel},\ and\ \citenamefont {Mokrousov}}]{chuan_zhang2021_prb}%
  \BibitemOpen
  \bibfield  {author} {\bibinfo {author} {\bibfnamefont {L.-C.}\ \bibnamefont {Zhang}}, \bibinfo {author} {\bibfnamefont {F.}~\bibnamefont {Zhu}}, \bibinfo {author} {\bibfnamefont {D.}~\bibnamefont {Go}}, \bibinfo {author} {\bibfnamefont {F.~R.}\ \bibnamefont {Lux}}, \bibinfo {author} {\bibfnamefont {F.~J.}\ \bibnamefont {dos Santos}}, \bibinfo {author} {\bibfnamefont {S.}~\bibnamefont {Lounis}}, \bibinfo {author} {\bibfnamefont {Y.}~\bibnamefont {Su}}, \bibinfo {author} {\bibfnamefont {S.}~\bibnamefont {Bl\"ugel}},\ and\ \bibinfo {author} {\bibfnamefont {Y.}~\bibnamefont {Mokrousov}},\ }\bibfield  {title} {\bibinfo {title} {Interplay of {D}zyaloshinskii-{M}oriya and {K}itaev interactions for magnonic properties of {H}eisenberg-{K}itaev honeycomb ferromagnets},\ }\href {https://doi.org/10.1103/PhysRevB.103.134414} {\bibfield  {journal} {\bibinfo  {journal} {Phys. Rev. B}\ }\textbf {\bibinfo {volume} {103}},\ \bibinfo {pages} {134414} (\bibinfo {year} {2021}{\natexlab{a}})}\BibitemShut {NoStop}%
\bibitem [{\citenamefont {Ideue}\ \emph {et~al.}(2012)\citenamefont {Ideue}, \citenamefont {Onose}, \citenamefont {Katsura}, \citenamefont {Shiomi}, \citenamefont {Ishiwata}, \citenamefont {Nagaosa},\ and\ \citenamefont {Tokura}}]{ideue2012}%
  \BibitemOpen
  \bibfield  {author} {\bibinfo {author} {\bibfnamefont {T.}~\bibnamefont {Ideue}}, \bibinfo {author} {\bibfnamefont {Y.}~\bibnamefont {Onose}}, \bibinfo {author} {\bibfnamefont {H.}~\bibnamefont {Katsura}}, \bibinfo {author} {\bibfnamefont {Y.}~\bibnamefont {Shiomi}}, \bibinfo {author} {\bibfnamefont {S.}~\bibnamefont {Ishiwata}}, \bibinfo {author} {\bibfnamefont {N.}~\bibnamefont {Nagaosa}},\ and\ \bibinfo {author} {\bibfnamefont {Y.}~\bibnamefont {Tokura}},\ }\bibfield  {title} {\bibinfo {title} {Effect of lattice geometry on magnon {H}all effect in ferromagnetic insulators},\ }\href {https://doi.org/10.1103/PhysRevB.85.134411} {\bibfield  {journal} {\bibinfo  {journal} {Phys. Rev. B}\ }\textbf {\bibinfo {volume} {85}},\ \bibinfo {pages} {134411} (\bibinfo {year} {2012})}\BibitemShut {NoStop}%
\bibitem [{\citenamefont {Hirschberger}\ \emph {et~al.}(2015{\natexlab{a}})\citenamefont {Hirschberger}, \citenamefont {Krizan}, \citenamefont {Cava},\ and\ \citenamefont {Ong}}]{hirschberger2015_science}%
  \BibitemOpen
  \bibfield  {author} {\bibinfo {author} {\bibfnamefont {M.}~\bibnamefont {Hirschberger}}, \bibinfo {author} {\bibfnamefont {J.~W.}\ \bibnamefont {Krizan}}, \bibinfo {author} {\bibfnamefont {R.~J.}\ \bibnamefont {Cava}},\ and\ \bibinfo {author} {\bibfnamefont {N.~P.}\ \bibnamefont {Ong}},\ }\bibfield  {title} {\bibinfo {title} {Large thermal {H}all conductivity of neutral spin excitations in a frustrated quantum magnet},\ }\href {https://doi.org/10.1126/science.1257340} {\bibfield  {journal} {\bibinfo  {journal} {Science}\ }\textbf {\bibinfo {volume} {348}},\ \bibinfo {pages} {106} (\bibinfo {year} {2015}{\natexlab{a}})}\BibitemShut {NoStop}%
\bibitem [{\citenamefont {Czajka}\ \emph {et~al.}(2023)\citenamefont {Czajka}, \citenamefont {Gao}, \citenamefont {Hirschberger}, \citenamefont {Lampen-Kelley}, \citenamefont {Banerjee}, \citenamefont {Quirk}, \citenamefont {Mandrus}, \citenamefont {Nagler},\ and\ \citenamefont {Ong}}]{czajka2023}%
  \BibitemOpen
  \bibfield  {author} {\bibinfo {author} {\bibfnamefont {P.}~\bibnamefont {Czajka}}, \bibinfo {author} {\bibfnamefont {T.}~\bibnamefont {Gao}}, \bibinfo {author} {\bibfnamefont {M.}~\bibnamefont {Hirschberger}}, \bibinfo {author} {\bibfnamefont {P.}~\bibnamefont {Lampen-Kelley}}, \bibinfo {author} {\bibfnamefont {A.}~\bibnamefont {Banerjee}}, \bibinfo {author} {\bibfnamefont {N.}~\bibnamefont {Quirk}}, \bibinfo {author} {\bibfnamefont {D.~G.}\ \bibnamefont {Mandrus}}, \bibinfo {author} {\bibfnamefont {S.~E.}\ \bibnamefont {Nagler}},\ and\ \bibinfo {author} {\bibfnamefont {N.~P.}\ \bibnamefont {Ong}},\ }\bibfield  {title} {\bibinfo {title} {Planar thermal {H}all effect of topological bosons in the {K}itaev magnet $\alpha$-{RuCl}$_3$},\ }\href {https://doi.org/10.1038/s41563-022-01397-w} {\bibfield  {journal} {\bibinfo  {journal} {Nat. Mater.}\ }\textbf {\bibinfo {volume} {22}},\ \bibinfo {pages} {36} (\bibinfo {year} {2023})}\BibitemShut {NoStop}%
\bibitem [{\citenamefont {Zhang}\ \emph {et~al.}(2021{\natexlab{b}})\citenamefont {Zhang}, \citenamefont {Xu}, \citenamefont {Carnahan}, \citenamefont {Sretenovic}, \citenamefont {Suri}, \citenamefont {Xiao},\ and\ \citenamefont {Ke}}]{zhang2021_prl}%
  \BibitemOpen
  \bibfield  {author} {\bibinfo {author} {\bibfnamefont {H.}~\bibnamefont {Zhang}}, \bibinfo {author} {\bibfnamefont {C.}~\bibnamefont {Xu}}, \bibinfo {author} {\bibfnamefont {C.}~\bibnamefont {Carnahan}}, \bibinfo {author} {\bibfnamefont {M.}~\bibnamefont {Sretenovic}}, \bibinfo {author} {\bibfnamefont {N.}~\bibnamefont {Suri}}, \bibinfo {author} {\bibfnamefont {D.}~\bibnamefont {Xiao}},\ and\ \bibinfo {author} {\bibfnamefont {X.}~\bibnamefont {Ke}},\ }\bibfield  {title} {\bibinfo {title} {Anomalous {T}hermal {H}all {E}ffect in an {I}nsulating van der {W}aals {M}agnet},\ }\href {https://doi.org/10.1103/PhysRevLett.127.247202} {\bibfield  {journal} {\bibinfo  {journal} {Phys. Rev. Lett.}\ }\textbf {\bibinfo {volume} {127}},\ \bibinfo {pages} {247202} (\bibinfo {year} {2021}{\natexlab{b}})}\BibitemShut {NoStop}%
\bibitem [{\citenamefont {Hirschberger}\ \emph {et~al.}(2015{\natexlab{b}})\citenamefont {Hirschberger}, \citenamefont {Chisnell}, \citenamefont {Lee},\ and\ \citenamefont {Ong}}]{hirschberger2015_prl}%
  \BibitemOpen
  \bibfield  {author} {\bibinfo {author} {\bibfnamefont {M.}~\bibnamefont {Hirschberger}}, \bibinfo {author} {\bibfnamefont {R.}~\bibnamefont {Chisnell}}, \bibinfo {author} {\bibfnamefont {Y.~S.}\ \bibnamefont {Lee}},\ and\ \bibinfo {author} {\bibfnamefont {N.~P.}\ \bibnamefont {Ong}},\ }\bibfield  {title} {\bibinfo {title} {Thermal {H}all {E}ffect of {S}pin {E}xcitations in a {K}agome {M}agnet},\ }\href {https://doi.org/10.1103/PhysRevLett.115.106603} {\bibfield  {journal} {\bibinfo  {journal} {Phys. Rev. Lett.}\ }\textbf {\bibinfo {volume} {115}},\ \bibinfo {pages} {106603} (\bibinfo {year} {2015}{\natexlab{b}})}\BibitemShut {NoStop}%
\bibitem [{\citenamefont {Akazawa}\ \emph {et~al.}(2020)\citenamefont {Akazawa}, \citenamefont {Shimozawa}, \citenamefont {Kittaka}, \citenamefont {Sakakibara}, \citenamefont {Okuma}, \citenamefont {Hiroi}, \citenamefont {Lee}, \citenamefont {Kawashima}, \citenamefont {Han},\ and\ \citenamefont {Yamashita}}]{Akazawa2020}%
  \BibitemOpen
  \bibfield  {author} {\bibinfo {author} {\bibfnamefont {M.}~\bibnamefont {Akazawa}}, \bibinfo {author} {\bibfnamefont {M.}~\bibnamefont {Shimozawa}}, \bibinfo {author} {\bibfnamefont {S.}~\bibnamefont {Kittaka}}, \bibinfo {author} {\bibfnamefont {T.}~\bibnamefont {Sakakibara}}, \bibinfo {author} {\bibfnamefont {R.}~\bibnamefont {Okuma}}, \bibinfo {author} {\bibfnamefont {Z.}~\bibnamefont {Hiroi}}, \bibinfo {author} {\bibfnamefont {H.-Y.}\ \bibnamefont {Lee}}, \bibinfo {author} {\bibfnamefont {N.}~\bibnamefont {Kawashima}}, \bibinfo {author} {\bibfnamefont {J.~H.}\ \bibnamefont {Han}},\ and\ \bibinfo {author} {\bibfnamefont {M.}~\bibnamefont {Yamashita}},\ }\bibfield  {title} {\bibinfo {title} {{Thermal Hall Effects of Spins and Phonons in Kagome Antiferromagnet Cd-Kapellasite}},\ }\href {https://doi.org/10.1103/PhysRevX.10.041059} {\bibfield  {journal} {\bibinfo  {journal} {Phys. Rev. X}\ }\textbf {\bibinfo {volume} {10}},\ \bibinfo {pages} {041059} (\bibinfo {year} {2020})}\BibitemShut {NoStop}%
\bibitem [{\citenamefont {Shindou}\ \emph {et~al.}(2013)\citenamefont {Shindou}, \citenamefont {Matsumoto}, \citenamefont {Murakami},\ and\ \citenamefont {Ohe}}]{shindou2013}%
  \BibitemOpen
  \bibfield  {author} {\bibinfo {author} {\bibfnamefont {R.}~\bibnamefont {Shindou}}, \bibinfo {author} {\bibfnamefont {R.}~\bibnamefont {Matsumoto}}, \bibinfo {author} {\bibfnamefont {S.}~\bibnamefont {Murakami}},\ and\ \bibinfo {author} {\bibfnamefont {J.-i.}\ \bibnamefont {Ohe}},\ }\bibfield  {title} {\bibinfo {title} {Topological chiral magnonic edge mode in a magnonic crystal},\ }\href {https://doi.org/10.1103/PhysRevB.87.174427} {\bibfield  {journal} {\bibinfo  {journal} {Phys. Rev. B}\ }\textbf {\bibinfo {volume} {87}},\ \bibinfo {pages} {174427} (\bibinfo {year} {2013})}\BibitemShut {NoStop}%
\bibitem [{\citenamefont {Matsumoto}\ \emph {et~al.}(2014)\citenamefont {Matsumoto}, \citenamefont {Sindou},\ and\ \citenamefont {Murakami}}]{matsumoto2014}%
  \BibitemOpen
  \bibfield  {author} {\bibinfo {author} {\bibfnamefont {R.}~\bibnamefont {Matsumoto}}, \bibinfo {author} {\bibfnamefont {R.}~\bibnamefont {Sindou}},\ and\ \bibinfo {author} {\bibfnamefont {S.}~\bibnamefont {Murakami}},\ }\bibfield  {title} {\bibinfo {title} {Thermal {H}all effect magnons in magnets with dipolar interaction},\ }\href {https://journals.aps.org/prb/abstract/10.1103/PhysRevB.89.054420} {\bibfield  {journal} {\bibinfo  {journal} {Phys. Rev. B}\ }\textbf {\bibinfo {volume} {89}},\ \bibinfo {pages} {054420} (\bibinfo {year} {2014})}\BibitemShut {NoStop}%
\bibitem [{\citenamefont {Romh{\'a}nyi}\ \emph {et~al.}(2015)\citenamefont {Romh{\'a}nyi}, \citenamefont {Penc},\ and\ \citenamefont {Ganesh}}]{romhanyi2015}%
  \BibitemOpen
  \bibfield  {author} {\bibinfo {author} {\bibfnamefont {J.}~\bibnamefont {Romh{\'a}nyi}}, \bibinfo {author} {\bibfnamefont {K.}~\bibnamefont {Penc}},\ and\ \bibinfo {author} {\bibfnamefont {R.}~\bibnamefont {Ganesh}},\ }\bibfield  {title} {\bibinfo {title} {Hall effect of triplons in a dimerized quantum magnet},\ }\href {https://doi.org/10.1038/ncomms7805} {\bibfield  {journal} {\bibinfo  {journal} {Nat. Commun.}\ }\textbf {\bibinfo {volume} {6}},\ \bibinfo {pages} {6805} (\bibinfo {year} {2015})}\BibitemShut {NoStop}%
\bibitem [{\citenamefont {McClarty}\ \emph {et~al.}(2017)\citenamefont {McClarty}, \citenamefont {Kr{\"u}ger}, \citenamefont {Guidi}, \citenamefont {Parker}, \citenamefont {Refson}, \citenamefont {Parker}, \citenamefont {Prabhakaran},\ and\ \citenamefont {Coldea}}]{mcclarty2017topological}%
  \BibitemOpen
  \bibfield  {author} {\bibinfo {author} {\bibfnamefont {P.~A.}\ \bibnamefont {McClarty}}, \bibinfo {author} {\bibfnamefont {F.}~\bibnamefont {Kr{\"u}ger}}, \bibinfo {author} {\bibfnamefont {T.}~\bibnamefont {Guidi}}, \bibinfo {author} {\bibfnamefont {S.}~\bibnamefont {Parker}}, \bibinfo {author} {\bibfnamefont {K.}~\bibnamefont {Refson}}, \bibinfo {author} {\bibfnamefont {A.}~\bibnamefont {Parker}}, \bibinfo {author} {\bibfnamefont {D.}~\bibnamefont {Prabhakaran}},\ and\ \bibinfo {author} {\bibfnamefont {R.}~\bibnamefont {Coldea}},\ }\bibfield  {title} {\bibinfo {title} {{Topological triplon modes and bound states in a Shastry--Sutherland magnet}},\ }\href {https://doi.org/10.1038/nphys4117} {\bibfield  {journal} {\bibinfo  {journal} {Nat. Phys.}\ }\textbf {\bibinfo {volume} {13}},\ \bibinfo {pages} {736} (\bibinfo {year} {2017})}\BibitemShut {NoStop}%
\bibitem [{\citenamefont {Zayed}\ \emph {et~al.}(2014)\citenamefont {Zayed}, \citenamefont {R\"uegg}, \citenamefont {Str\"assle}, \citenamefont {Stuhr}, \citenamefont {Roessli}, \citenamefont {Ay}, \citenamefont {Mesot}, \citenamefont {Link}, \citenamefont {Pomjakushina}, \citenamefont {Stingaciu}, \citenamefont {Conder},\ and\ \citenamefont {R\o{}nnow}}]{zayed2014}%
  \BibitemOpen
  \bibfield  {author} {\bibinfo {author} {\bibfnamefont {M.~E.}\ \bibnamefont {Zayed}}, \bibinfo {author} {\bibfnamefont {C.}~\bibnamefont {R\"uegg}}, \bibinfo {author} {\bibfnamefont {T.}~\bibnamefont {Str\"assle}}, \bibinfo {author} {\bibfnamefont {U.}~\bibnamefont {Stuhr}}, \bibinfo {author} {\bibfnamefont {B.}~\bibnamefont {Roessli}}, \bibinfo {author} {\bibfnamefont {M.}~\bibnamefont {Ay}}, \bibinfo {author} {\bibfnamefont {J.}~\bibnamefont {Mesot}}, \bibinfo {author} {\bibfnamefont {P.}~\bibnamefont {Link}}, \bibinfo {author} {\bibfnamefont {E.}~\bibnamefont {Pomjakushina}}, \bibinfo {author} {\bibfnamefont {M.}~\bibnamefont {Stingaciu}}, \bibinfo {author} {\bibfnamefont {K.}~\bibnamefont {Conder}},\ and\ \bibinfo {author} {\bibfnamefont {H.~M.}\ \bibnamefont {R\o{}nnow}},\ }\bibfield  {title} {\bibinfo {title} {Correlated {D}ecay of {T}riplet {E}xcitations in the {S}hastry-{S}utherland {C}ompound {${\mathbf{\text{SrCu}}}_{2}({\mathrm{BO}}_{3}{)}_{2}$}},\ }\href {https://doi.org/10.1103/PhysRevLett.113.067201} {\bibfield  {journal} {\bibinfo  {journal} {Phys. Rev. Lett.}\ }\textbf {\bibinfo {volume} {113}},\ \bibinfo {pages} {067201} (\bibinfo {year} {2014})}\BibitemShut {NoStop}%
\bibitem [{\citenamefont {Joshi}\ \emph {et~al.}(1999)\citenamefont {Joshi}, \citenamefont {Ma}, \citenamefont {Mila}, \citenamefont {Shi},\ and\ \citenamefont {Zhang}}]{joshi1999}%
  \BibitemOpen
  \bibfield  {author} {\bibinfo {author} {\bibfnamefont {A.}~\bibnamefont {Joshi}}, \bibinfo {author} {\bibfnamefont {M.}~\bibnamefont {Ma}}, \bibinfo {author} {\bibfnamefont {F.}~\bibnamefont {Mila}}, \bibinfo {author} {\bibfnamefont {D.~N.}\ \bibnamefont {Shi}},\ and\ \bibinfo {author} {\bibfnamefont {F.~C.}\ \bibnamefont {Zhang}},\ }\bibfield  {title} {\bibinfo {title} {Elementary excitations in magnetically ordered systems with orbital degeneracy},\ }\href {https://journals.aps.org/prb/abstract/10.1103/PhysRevB.60.6584} {\bibfield  {journal} {\bibinfo  {journal} {Phys. Rev. B}\ }\textbf {\bibinfo {volume} {60}},\ \bibinfo {pages} {6584} (\bibinfo {year} {1999})}\BibitemShut {NoStop}%
\bibitem [{\citenamefont {L\"auchli}\ \emph {et~al.}(2006)\citenamefont {L\"auchli}, \citenamefont {Mila},\ and\ \citenamefont {Penc}}]{Lauchli2006}%
  \BibitemOpen
  \bibfield  {author} {\bibinfo {author} {\bibfnamefont {A.}~\bibnamefont {L\"auchli}}, \bibinfo {author} {\bibfnamefont {F.}~\bibnamefont {Mila}},\ and\ \bibinfo {author} {\bibfnamefont {K.}~\bibnamefont {Penc}},\ }\bibfield  {title} {\bibinfo {title} {Quadrupolar {P}hases of the ${S=1}$ {B}ilinear-{B}iquadratic {H}eisenberg {M}odel on the {T}riangular {L}attice},\ }\href {https://doi.org/10.1103/PhysRevLett.97.087205} {\bibfield  {journal} {\bibinfo  {journal} {Phys. Rev. Lett.}\ }\textbf {\bibinfo {volume} {97}},\ \bibinfo {pages} {087205} (\bibinfo {year} {2006})}\BibitemShut {NoStop}%
\bibitem [{\citenamefont {Tsunetsugu}\ and\ \citenamefont {Arikawa}(2006)}]{Tsunetsugu2006}%
  \BibitemOpen
  \bibfield  {author} {\bibinfo {author} {\bibfnamefont {H.}~\bibnamefont {Tsunetsugu}}\ and\ \bibinfo {author} {\bibfnamefont {M.}~\bibnamefont {Arikawa}},\ }\bibfield  {title} {\bibinfo {title} {Spin {N}ematic {P}hase in ${S=1}$ {T}riangular {A}ntiferromagnets},\ }\href {https://doi.org/10.1143/JPSJ.75.083701} {\bibfield  {journal} {\bibinfo  {journal} {J. Phys. Soc. Jpn.}\ }\textbf {\bibinfo {volume} {75}},\ \bibinfo {pages} {083701} (\bibinfo {year} {2006})}\BibitemShut {NoStop}%
\bibitem [{\citenamefont {Kim}\ \emph {et~al.}(2017)\citenamefont {Kim}, \citenamefont {Penc}, \citenamefont {Nataf},\ and\ \citenamefont {Mila}}]{Kim_flavor-wave2017}%
  \BibitemOpen
  \bibfield  {author} {\bibinfo {author} {\bibfnamefont {F.~H.}\ \bibnamefont {Kim}}, \bibinfo {author} {\bibfnamefont {K.}~\bibnamefont {Penc}}, \bibinfo {author} {\bibfnamefont {P.}~\bibnamefont {Nataf}},\ and\ \bibinfo {author} {\bibfnamefont {F.}~\bibnamefont {Mila}},\ }\bibfield  {title} {\bibinfo {title} {Linear flavor-wave theory for fully antisymmetric {SU}(${N}$) irreducible representations},\ }\href {https://doi.org/10.1103/PhysRevB.96.205142} {\bibfield  {journal} {\bibinfo  {journal} {Phys. Rev. B}\ }\textbf {\bibinfo {volume} {96}},\ \bibinfo {pages} {205142} (\bibinfo {year} {2017})}\BibitemShut {NoStop}%
\bibitem [{\citenamefont {Zhitomirsky}\ and\ \citenamefont {Chernyshev}(1999)}]{zhitomirsky1999}%
  \BibitemOpen
  \bibfield  {author} {\bibinfo {author} {\bibfnamefont {M.~E.}\ \bibnamefont {Zhitomirsky}}\ and\ \bibinfo {author} {\bibfnamefont {A.~L.}\ \bibnamefont {Chernyshev}},\ }\bibfield  {title} {\bibinfo {title} {Instability of {A}ntiferromagnetic {M}agnons in {S}trong {F}ields},\ }\href {https://doi.org/10.1103/PhysRevLett.82.4536} {\bibfield  {journal} {\bibinfo  {journal} {Phys. Rev. Lett.}\ }\textbf {\bibinfo {volume} {82}},\ \bibinfo {pages} {4536} (\bibinfo {year} {1999})}\BibitemShut {NoStop}%
\bibitem [{\citenamefont {Zhitomirsky}\ and\ \citenamefont {Chernyshev}(2013)}]{zhitomirsky2013}%
  \BibitemOpen
  \bibfield  {author} {\bibinfo {author} {\bibfnamefont {M.~E.}\ \bibnamefont {Zhitomirsky}}\ and\ \bibinfo {author} {\bibfnamefont {A.~L.}\ \bibnamefont {Chernyshev}},\ }\bibfield  {title} {\bibinfo {title} {Colloquium: {S}pontaneous magnon decays},\ }\href {https://doi.org/10.1103/RevModPhys.85.219} {\bibfield  {journal} {\bibinfo  {journal} {Rev. Mod. Phys.}\ }\textbf {\bibinfo {volume} {85}},\ \bibinfo {pages} {219} (\bibinfo {year} {2013})}\BibitemShut {NoStop}%
\bibitem [{\citenamefont {Mourigal}\ \emph {et~al.}(2010{\natexlab{a}})\citenamefont {Mourigal}, \citenamefont {Zhitomirsky},\ and\ \citenamefont {Chernyshev}}]{mourigal2010}%
  \BibitemOpen
  \bibfield  {author} {\bibinfo {author} {\bibfnamefont {M.}~\bibnamefont {Mourigal}}, \bibinfo {author} {\bibfnamefont {M.~E.}\ \bibnamefont {Zhitomirsky}},\ and\ \bibinfo {author} {\bibfnamefont {A.~L.}\ \bibnamefont {Chernyshev}},\ }\bibfield  {title} {\bibinfo {title} {Field-induced decay dynamics in square-lattice antiferromagnets},\ }\href {https://doi.org/10.1103/PhysRevB.82.144402} {\bibfield  {journal} {\bibinfo  {journal} {Phys. Rev. B}\ }\textbf {\bibinfo {volume} {82}},\ \bibinfo {pages} {144402} (\bibinfo {year} {2010}{\natexlab{a}})}\BibitemShut {NoStop}%
\bibitem [{\citenamefont {Fuhrman}\ \emph {et~al.}(2012)\citenamefont {Fuhrman}, \citenamefont {Mourigal}, \citenamefont {Zhitomirsky},\ and\ \citenamefont {Chernyshev}}]{fuhrman2012}%
  \BibitemOpen
  \bibfield  {author} {\bibinfo {author} {\bibfnamefont {W.~T.}\ \bibnamefont {Fuhrman}}, \bibinfo {author} {\bibfnamefont {M.}~\bibnamefont {Mourigal}}, \bibinfo {author} {\bibfnamefont {M.~E.}\ \bibnamefont {Zhitomirsky}},\ and\ \bibinfo {author} {\bibfnamefont {A.~L.}\ \bibnamefont {Chernyshev}},\ }\bibfield  {title} {\bibinfo {title} {Dynamical structure factor of quasi-two-dimensional antiferromagnet in high fields},\ }\href {https://doi.org/10.1103/PhysRevB.85.184405} {\bibfield  {journal} {\bibinfo  {journal} {Phys. Rev. B}\ }\textbf {\bibinfo {volume} {85}},\ \bibinfo {pages} {184405} (\bibinfo {year} {2012})}\BibitemShut {NoStop}%
\bibitem [{\citenamefont {Mook}\ \emph {et~al.}(2021)\citenamefont {Mook}, \citenamefont {Plekhanov}, \citenamefont {Klinovaja},\ and\ \citenamefont {Loss}}]{mook2021}%
  \BibitemOpen
  \bibfield  {author} {\bibinfo {author} {\bibfnamefont {A.}~\bibnamefont {Mook}}, \bibinfo {author} {\bibfnamefont {K.}~\bibnamefont {Plekhanov}}, \bibinfo {author} {\bibfnamefont {J.}~\bibnamefont {Klinovaja}},\ and\ \bibinfo {author} {\bibfnamefont {D.}~\bibnamefont {Loss}},\ }\bibfield  {title} {\bibinfo {title} {Interaction-{S}tabilized {T}opological {M}agnon {I}nsulator in {F}erromagnets},\ }\href {https://doi.org/10.1103/PhysRevX.11.021061} {\bibfield  {journal} {\bibinfo  {journal} {Phys. Rev. X}\ }\textbf {\bibinfo {volume} {11}},\ \bibinfo {pages} {021061} (\bibinfo {year} {2021})}\BibitemShut {NoStop}%
\bibitem [{\citenamefont {Chernyshev}\ and\ \citenamefont {Maksimov}(2016)}]{chernyshev2016}%
  \BibitemOpen
  \bibfield  {author} {\bibinfo {author} {\bibfnamefont {A.~L.}\ \bibnamefont {Chernyshev}}\ and\ \bibinfo {author} {\bibfnamefont {P.~A.}\ \bibnamefont {Maksimov}},\ }\bibfield  {title} {\bibinfo {title} {Damped {T}opological {M}agnons in the {K}agome-{L}attice {F}erromagnets},\ }\href {https://doi.org/10.1103/PhysRevLett.117.187203} {\bibfield  {journal} {\bibinfo  {journal} {Phys. Rev. Lett.}\ }\textbf {\bibinfo {volume} {117}},\ \bibinfo {pages} {187203} (\bibinfo {year} {2016})}\BibitemShut {NoStop}%
\bibitem [{\citenamefont {Suetsugu}\ \emph {et~al.}(2022)\citenamefont {Suetsugu}, \citenamefont {Yokoi}, \citenamefont {Totsuka}, \citenamefont {Ono}, \citenamefont {Tanaka}, \citenamefont {Kasahara}, \citenamefont {Kasahara}, \citenamefont {Chengchao}, \citenamefont {Kageyama},\ and\ \citenamefont {Matsuda}}]{suetsugu2022}%
  \BibitemOpen
  \bibfield  {author} {\bibinfo {author} {\bibfnamefont {S.}~\bibnamefont {Suetsugu}}, \bibinfo {author} {\bibfnamefont {T.}~\bibnamefont {Yokoi}}, \bibinfo {author} {\bibfnamefont {K.}~\bibnamefont {Totsuka}}, \bibinfo {author} {\bibfnamefont {T.}~\bibnamefont {Ono}}, \bibinfo {author} {\bibfnamefont {I.}~\bibnamefont {Tanaka}}, \bibinfo {author} {\bibfnamefont {S.}~\bibnamefont {Kasahara}}, \bibinfo {author} {\bibfnamefont {Y.}~\bibnamefont {Kasahara}}, \bibinfo {author} {\bibfnamefont {Z.}~\bibnamefont {Chengchao}}, \bibinfo {author} {\bibfnamefont {H.}~\bibnamefont {Kageyama}},\ and\ \bibinfo {author} {\bibfnamefont {Y.}~\bibnamefont {Matsuda}},\ }\bibfield  {title} {\bibinfo {title} {Intrinsic suppression of the topological thermal {H}all effect in an exactly solvable quantum magnet},\ }\href {https://doi.org/10.1103/PhysRevB.105.024415} {\bibfield  {journal} {\bibinfo  {journal} {Phys. Rev. B}\ }\textbf {\bibinfo {volume} {105}},\ \bibinfo {pages} {024415} (\bibinfo {year} {2022})}\BibitemShut {NoStop}%
\bibitem [{\citenamefont {Choi}\ \emph {et~al.}(2023)\citenamefont {Choi}, \citenamefont {Yang}, \citenamefont {Park},\ and\ \citenamefont {Park}}]{choi2023}%
  \BibitemOpen
  \bibfield  {author} {\bibinfo {author} {\bibfnamefont {Y.}~\bibnamefont {Choi}}, \bibinfo {author} {\bibfnamefont {H.}~\bibnamefont {Yang}}, \bibinfo {author} {\bibfnamefont {J.}~\bibnamefont {Park}},\ and\ \bibinfo {author} {\bibfnamefont {J.-G.}\ \bibnamefont {Park}},\ }\bibfield  {title} {\bibinfo {title} {Sizable suppression of magnon {H}all effect by magnon damping in {Cr}$_{2}${Ge}$_{2}${Te}$_{6}$},\ }\href {https://doi.org/10.1103/PhysRevB.107.184434} {\bibfield  {journal} {\bibinfo  {journal} {Phys. Rev. B}\ }\textbf {\bibinfo {volume} {107}},\ \bibinfo {pages} {184434} (\bibinfo {year} {2023})}\BibitemShut {NoStop}%
\bibitem [{\citenamefont {Kitaev}(2006)}]{kitaev}%
  \BibitemOpen
  \bibfield  {author} {\bibinfo {author} {\bibfnamefont {A.}~\bibnamefont {Kitaev}},\ }\bibfield  {title} {\bibinfo {title} {Anyons in an exactly solved model and beyond},\ }\href {https://doi.org/https://doi.org/10.1016/j.aop.2005.10.005} {\bibfield  {journal} {\bibinfo  {journal} {Ann. Phys.}\ }\textbf {\bibinfo {volume} {321}},\ \bibinfo {pages} {2} (\bibinfo {year} {2006})}\BibitemShut {NoStop}%
\bibitem [{\citenamefont {Joshi}(2018)}]{joshi2018}%
  \BibitemOpen
  \bibfield  {author} {\bibinfo {author} {\bibfnamefont {D.~G.}\ \bibnamefont {Joshi}},\ }\bibfield  {title} {\bibinfo {title} {Topological excitations in the ferromagnetic {K}itaev-{H}eisenberg model},\ }\href {https://doi.org/10.1103/PhysRevB.98.060405} {\bibfield  {journal} {\bibinfo  {journal} {Phys. Rev. B}\ }\textbf {\bibinfo {volume} {98}},\ \bibinfo {pages} {060405} (\bibinfo {year} {2018})}\BibitemShut {NoStop}%
\bibitem [{\citenamefont {Li}\ and\ \citenamefont {Okamoto}(2022)}]{li2022_prb}%
  \BibitemOpen
  \bibfield  {author} {\bibinfo {author} {\bibfnamefont {S.}~\bibnamefont {Li}}\ and\ \bibinfo {author} {\bibfnamefont {S.}~\bibnamefont {Okamoto}},\ }\bibfield  {title} {\bibinfo {title} {Thermal {H}all effect in the {K}itaev-{H}eisenberg system with spin-phonon coupling},\ }\href {https://doi.org/10.1103/PhysRevB.106.024413} {\bibfield  {journal} {\bibinfo  {journal} {Phys. Rev. B}\ }\textbf {\bibinfo {volume} {106}},\ \bibinfo {pages} {024413} (\bibinfo {year} {2022})}\BibitemShut {NoStop}%
\bibitem [{\citenamefont {Koyama}\ and\ \citenamefont {Nasu}(2021)}]{koyama2021}%
  \BibitemOpen
  \bibfield  {author} {\bibinfo {author} {\bibfnamefont {S.}~\bibnamefont {Koyama}}\ and\ \bibinfo {author} {\bibfnamefont {J.}~\bibnamefont {Nasu}},\ }\bibfield  {title} {\bibinfo {title} {Field-angle dependence of thermal {H}all conductivity in a magnetically ordered {K}itaev-{H}eisenberg system},\ }\href {https://doi.org/10.1103/PhysRevB.104.075121} {\bibfield  {journal} {\bibinfo  {journal} {Phys. Rev. B}\ }\textbf {\bibinfo {volume} {104}},\ \bibinfo {pages} {075121} (\bibinfo {year} {2021})}\BibitemShut {NoStop}%
\bibitem [{\citenamefont {Zhang}\ \emph {et~al.}(2021{\natexlab{c}})\citenamefont {Zhang}, \citenamefont {Chern},\ and\ \citenamefont {Kim}}]{zhang2021_prb}%
  \BibitemOpen
  \bibfield  {author} {\bibinfo {author} {\bibfnamefont {E.~Z.}\ \bibnamefont {Zhang}}, \bibinfo {author} {\bibfnamefont {L.~E.}\ \bibnamefont {Chern}},\ and\ \bibinfo {author} {\bibfnamefont {Y.~B.}\ \bibnamefont {Kim}},\ }\bibfield  {title} {\bibinfo {title} {Topological magnons for thermal {H}all transport in frustrated magnets with bond-dependent interactions},\ }\href {https://doi.org/10.1103/PhysRevB.103.174402} {\bibfield  {journal} {\bibinfo  {journal} {Phys. Rev. B}\ }\textbf {\bibinfo {volume} {103}},\ \bibinfo {pages} {174402} (\bibinfo {year} {2021}{\natexlab{c}})}\BibitemShut {NoStop}%
\bibitem [{\citenamefont {Chern}\ \emph {et~al.}(2021)\citenamefont {Chern}, \citenamefont {Zhang},\ and\ \citenamefont {Kim}}]{chern2021_prl}%
  \BibitemOpen
  \bibfield  {author} {\bibinfo {author} {\bibfnamefont {L.~E.}\ \bibnamefont {Chern}}, \bibinfo {author} {\bibfnamefont {E.~Z.}\ \bibnamefont {Zhang}},\ and\ \bibinfo {author} {\bibfnamefont {Y.~B.}\ \bibnamefont {Kim}},\ }\bibfield  {title} {\bibinfo {title} {Sign {S}tructure of {T}hermal {H}all {C}onductivity and {T}opological {M}agnons for {I}n-{P}lane {F}ield {P}olarized {K}itaev {M}agnets},\ }\href {https://doi.org/10.1103/PhysRevLett.126.147201} {\bibfield  {journal} {\bibinfo  {journal} {Phys. Rev. Lett.}\ }\textbf {\bibinfo {volume} {126}},\ \bibinfo {pages} {147201} (\bibinfo {year} {2021})}\BibitemShut {NoStop}%
\bibitem [{\citenamefont {Chern}\ \emph {et~al.}(2020)\citenamefont {Chern}, \citenamefont {Kaneko}, \citenamefont {Lee},\ and\ \citenamefont {Kim}}]{chern2020_prr}%
  \BibitemOpen
  \bibfield  {author} {\bibinfo {author} {\bibfnamefont {L.~E.}\ \bibnamefont {Chern}}, \bibinfo {author} {\bibfnamefont {R.}~\bibnamefont {Kaneko}}, \bibinfo {author} {\bibfnamefont {H.-Y.}\ \bibnamefont {Lee}},\ and\ \bibinfo {author} {\bibfnamefont {Y.~B.}\ \bibnamefont {Kim}},\ }\bibfield  {title} {\bibinfo {title} {Magnetic field induced competing phases in spin-orbital entangled {K}itaev magnets},\ }\href {https://doi.org/10.1103/PhysRevResearch.2.013014} {\bibfield  {journal} {\bibinfo  {journal} {Phys. Rev. Res.}\ }\textbf {\bibinfo {volume} {2}},\ \bibinfo {pages} {013014} (\bibinfo {year} {2020})}\BibitemShut {NoStop}%
\bibitem [{\citenamefont {Janssen}\ and\ \citenamefont {Vojta}(2019)}]{janssen2019_cm}%
  \BibitemOpen
  \bibfield  {author} {\bibinfo {author} {\bibfnamefont {L.}~\bibnamefont {Janssen}}\ and\ \bibinfo {author} {\bibfnamefont {M.}~\bibnamefont {Vojta}},\ }\bibfield  {title} {\bibinfo {title} {Heisenberg{\textendash}{K}itaev physics in magnetic fields},\ }\href {https://doi.org/10.1088/1361-648x/ab283e} {\bibfield  {journal} {\bibinfo  {journal} {J. Phys.: Condens. Matter}\ }\textbf {\bibinfo {volume} {31}},\ \bibinfo {pages} {423002} (\bibinfo {year} {2019})}\BibitemShut {NoStop}%
\bibitem [{\citenamefont {Cookmeyer}\ and\ \citenamefont {Moore}(2018)}]{cookmeyer2018}%
  \BibitemOpen
  \bibfield  {author} {\bibinfo {author} {\bibfnamefont {T.}~\bibnamefont {Cookmeyer}}\ and\ \bibinfo {author} {\bibfnamefont {J.~E.}\ \bibnamefont {Moore}},\ }\bibfield  {title} {\bibinfo {title} {Spin-wave analysis of the low-temperature thermal {H}all effect in the candidate {K}itaev spin liquid $\alpha$-{RuCl}$_{3}$},\ }\href {https://doi.org/10.1103/PhysRevB.98.060412} {\bibfield  {journal} {\bibinfo  {journal} {Phys. Rev. B}\ }\textbf {\bibinfo {volume} {98}},\ \bibinfo {pages} {060412} (\bibinfo {year} {2018})}\BibitemShut {NoStop}%
\bibitem [{\citenamefont {Zhang}\ \emph {et~al.}(2023)\citenamefont {Zhang}, \citenamefont {Wilke},\ and\ \citenamefont {Kim}}]{zhang2023}%
  \BibitemOpen
  \bibfield  {author} {\bibinfo {author} {\bibfnamefont {E.~Z.}\ \bibnamefont {Zhang}}, \bibinfo {author} {\bibfnamefont {R.~H.}\ \bibnamefont {Wilke}},\ and\ \bibinfo {author} {\bibfnamefont {Y.~B.}\ \bibnamefont {Kim}},\ }\bibfield  {title} {\bibinfo {title} {Spin excitation continuum to topological magnon crossover and thermal {H}all conductivity in {K}itaev magnets},\ }\href {https://doi.org/10.1103/PhysRevB.107.184418} {\bibfield  {journal} {\bibinfo  {journal} {Phys. Rev. B}\ }\textbf {\bibinfo {volume} {107}},\ \bibinfo {pages} {184418} (\bibinfo {year} {2023})}\BibitemShut {NoStop}%
\bibitem [{\citenamefont {Kee}(2023)}]{young2023}%
  \BibitemOpen
  \bibfield  {author} {\bibinfo {author} {\bibfnamefont {H.-Y.}\ \bibnamefont {Kee}},\ }\bibfield  {title} {\bibinfo {title} {Thermal {H}all conductivity of $\alpha$-{RuCl}$_{3}$},\ }\href {https://doi.org/10.1038/s41563-022-01444-6} {\bibfield  {journal} {\bibinfo  {journal} {Nat. Mater.}\ }\textbf {\bibinfo {volume} {22}},\ \bibinfo {pages} {6} (\bibinfo {year} {2023})}\BibitemShut {NoStop}%
\bibitem [{\citenamefont {Hentrich}\ \emph {et~al.}(2019)\citenamefont {Hentrich}, \citenamefont {Roslova}, \citenamefont {Isaeva}, \citenamefont {Doert}, \citenamefont {Brenig}, \citenamefont {B\"uchner},\ and\ \citenamefont {Hess}}]{hentrich2019}%
  \BibitemOpen
  \bibfield  {author} {\bibinfo {author} {\bibfnamefont {R.}~\bibnamefont {Hentrich}}, \bibinfo {author} {\bibfnamefont {M.}~\bibnamefont {Roslova}}, \bibinfo {author} {\bibfnamefont {A.}~\bibnamefont {Isaeva}}, \bibinfo {author} {\bibfnamefont {T.}~\bibnamefont {Doert}}, \bibinfo {author} {\bibfnamefont {W.}~\bibnamefont {Brenig}}, \bibinfo {author} {\bibfnamefont {B.}~\bibnamefont {B\"uchner}},\ and\ \bibinfo {author} {\bibfnamefont {C.}~\bibnamefont {Hess}},\ }\bibfield  {title} {\bibinfo {title} {Large thermal {H}all effect in ${\alpha}$-{RuCl}$_{3}$: Evidence for heat transport by {K}itaev-{H}eisenberg paramagnons},\ }\href {https://doi.org/10.1103/PhysRevB.99.085136} {\bibfield  {journal} {\bibinfo  {journal} {Phys. Rev. B}\ }\textbf {\bibinfo {volume} {99}},\ \bibinfo {pages} {085136} (\bibinfo {year} {2019})}\BibitemShut {NoStop}%
\bibitem [{\citenamefont {Kasahara}\ \emph {et~al.}(2018{\natexlab{a}})\citenamefont {Kasahara}, \citenamefont {Sugii}, \citenamefont {Ohnishi}, \citenamefont {Shimozawa}, \citenamefont {Yamashita}, \citenamefont {Kurita}, \citenamefont {Tanaka}, \citenamefont {Nasu}, \citenamefont {Motome}, \citenamefont {Shibauchi},\ and\ \citenamefont {Matsuda}}]{kasahara2018_prl}%
  \BibitemOpen
  \bibfield  {author} {\bibinfo {author} {\bibfnamefont {Y.}~\bibnamefont {Kasahara}}, \bibinfo {author} {\bibfnamefont {K.}~\bibnamefont {Sugii}}, \bibinfo {author} {\bibfnamefont {T.}~\bibnamefont {Ohnishi}}, \bibinfo {author} {\bibfnamefont {M.}~\bibnamefont {Shimozawa}}, \bibinfo {author} {\bibfnamefont {M.}~\bibnamefont {Yamashita}}, \bibinfo {author} {\bibfnamefont {N.}~\bibnamefont {Kurita}}, \bibinfo {author} {\bibfnamefont {H.}~\bibnamefont {Tanaka}}, \bibinfo {author} {\bibfnamefont {J.}~\bibnamefont {Nasu}}, \bibinfo {author} {\bibfnamefont {Y.}~\bibnamefont {Motome}}, \bibinfo {author} {\bibfnamefont {T.}~\bibnamefont {Shibauchi}},\ and\ \bibinfo {author} {\bibfnamefont {Y.}~\bibnamefont {Matsuda}},\ }\bibfield  {title} {\bibinfo {title} {Unusual {T}hermal {H}all {E}ffect in a {K}itaev {S}pin {L}iquid {C}andidate $\ensuremath{\alpha}$-{RuCl}$_{3}$},\ }\href {https://doi.org/10.1103/PhysRevLett.120.217205} {\bibfield  {journal} {\bibinfo  {journal} {Phys. Rev. Lett.}\ }\textbf {\bibinfo {volume} {120}},\ \bibinfo {pages} {217205} (\bibinfo {year} {2018}{\natexlab{a}})}\BibitemShut {NoStop}%
\bibitem [{\citenamefont {Kasahara}\ \emph {et~al.}(2018{\natexlab{b}})\citenamefont {Kasahara}, \citenamefont {Ohnishi}, \citenamefont {Mizukami}, \citenamefont {Tanaka}, \citenamefont {Ma}, \citenamefont {Sugii}, \citenamefont {Kurita}, \citenamefont {Tanaka}, \citenamefont {Nasu}, \citenamefont {Motome}, \citenamefont {Shibauchi},\ and\ \citenamefont {Matsuda}}]{kasahara2018_nature}%
  \BibitemOpen
  \bibfield  {author} {\bibinfo {author} {\bibfnamefont {Y.}~\bibnamefont {Kasahara}}, \bibinfo {author} {\bibfnamefont {T.}~\bibnamefont {Ohnishi}}, \bibinfo {author} {\bibfnamefont {Y.}~\bibnamefont {Mizukami}}, \bibinfo {author} {\bibfnamefont {O.}~\bibnamefont {Tanaka}}, \bibinfo {author} {\bibfnamefont {S.}~\bibnamefont {Ma}}, \bibinfo {author} {\bibfnamefont {K.}~\bibnamefont {Sugii}}, \bibinfo {author} {\bibfnamefont {N.}~\bibnamefont {Kurita}}, \bibinfo {author} {\bibfnamefont {H.}~\bibnamefont {Tanaka}}, \bibinfo {author} {\bibfnamefont {J.}~\bibnamefont {Nasu}}, \bibinfo {author} {\bibfnamefont {Y.}~\bibnamefont {Motome}}, \bibinfo {author} {\bibfnamefont {T.}~\bibnamefont {Shibauchi}},\ and\ \bibinfo {author} {\bibfnamefont {Y.}~\bibnamefont {Matsuda}},\ }\bibfield  {title} {\bibinfo {title} {Majorana quantization and half-integer thermal quantum {H}all effect in a {K}itaev spin liquid},\ }\href {https://doi.org/10.1038/s41586-018-0274-0} {\bibfield  {journal} {\bibinfo  {journal} {Nature}\ }\textbf {\bibinfo {volume} {559}},\ \bibinfo {pages} {227} (\bibinfo {year} {2018}{\natexlab{b}})}\BibitemShut {NoStop}%
\bibitem [{\citenamefont {Yokoi}\ \emph {et~al.}(2021)\citenamefont {Yokoi}, \citenamefont {Ma}, \citenamefont {Kasahara}, \citenamefont {Kasahara}, \citenamefont {Shibauchi}, \citenamefont {Kurita}, \citenamefont {Tanaka}, \citenamefont {Nasu}, \citenamefont {Motome}, \citenamefont {Hickey}, \citenamefont {Trebst},\ and\ \citenamefont {Matsuda}}]{yokoi2021}%
  \BibitemOpen
  \bibfield  {author} {\bibinfo {author} {\bibfnamefont {T.}~\bibnamefont {Yokoi}}, \bibinfo {author} {\bibfnamefont {S.}~\bibnamefont {Ma}}, \bibinfo {author} {\bibfnamefont {Y.}~\bibnamefont {Kasahara}}, \bibinfo {author} {\bibfnamefont {S.}~\bibnamefont {Kasahara}}, \bibinfo {author} {\bibfnamefont {T.}~\bibnamefont {Shibauchi}}, \bibinfo {author} {\bibfnamefont {N.}~\bibnamefont {Kurita}}, \bibinfo {author} {\bibfnamefont {H.}~\bibnamefont {Tanaka}}, \bibinfo {author} {\bibfnamefont {J.}~\bibnamefont {Nasu}}, \bibinfo {author} {\bibfnamefont {Y.}~\bibnamefont {Motome}}, \bibinfo {author} {\bibfnamefont {C.}~\bibnamefont {Hickey}}, \bibinfo {author} {\bibfnamefont {S.}~\bibnamefont {Trebst}},\ and\ \bibinfo {author} {\bibfnamefont {Y.}~\bibnamefont {Matsuda}},\ }\bibfield  {title} {\bibinfo {title} {Half-integer quantized anomalous thermal {H}all effect in the {K}itaev material candidate $\alpha$-{RuCl}$_3$},\ }\href {https://doi.org/10.1126/science.aay5551} {\bibfield  {journal} {\bibinfo  {journal} {Science}\ }\textbf {\bibinfo {volume} {373}},\ \bibinfo {pages} {568} (\bibinfo {year} {2021})}\BibitemShut {NoStop}%
\bibitem [{\citenamefont {Yamashita}\ \emph {et~al.}(2020)\citenamefont {Yamashita}, \citenamefont {Gouchi}, \citenamefont {Uwatoko}, \citenamefont {Kurita},\ and\ \citenamefont {Tanaka}}]{yamashita2020}%
  \BibitemOpen
  \bibfield  {author} {\bibinfo {author} {\bibfnamefont {M.}~\bibnamefont {Yamashita}}, \bibinfo {author} {\bibfnamefont {J.}~\bibnamefont {Gouchi}}, \bibinfo {author} {\bibfnamefont {Y.}~\bibnamefont {Uwatoko}}, \bibinfo {author} {\bibfnamefont {N.}~\bibnamefont {Kurita}},\ and\ \bibinfo {author} {\bibfnamefont {H.}~\bibnamefont {Tanaka}},\ }\bibfield  {title} {\bibinfo {title} {Sample dependence of half-integer quantized thermal {H}all effect in the {K}itaev spin-liquid candidate ${\alpha}$-{RuCl}$_{3}$},\ }\href {https://doi.org/10.1103/PhysRevB.102.220404} {\bibfield  {journal} {\bibinfo  {journal} {Phys. Rev. B}\ }\textbf {\bibinfo {volume} {102}},\ \bibinfo {pages} {220404} (\bibinfo {year} {2020})}\BibitemShut {NoStop}%
\bibitem [{\citenamefont {Czajka}\ \emph {et~al.}(2021)\citenamefont {Czajka}, \citenamefont {Gao}, \citenamefont {Hirschberger}, \citenamefont {Lampen-Kelley}, \citenamefont {Banerjee}, \citenamefont {Yan}, \citenamefont {Mandrus}, \citenamefont {Nagler},\ and\ \citenamefont {Ong}}]{czajka2021_nature}%
  \BibitemOpen
  \bibfield  {author} {\bibinfo {author} {\bibfnamefont {P.}~\bibnamefont {Czajka}}, \bibinfo {author} {\bibfnamefont {T.}~\bibnamefont {Gao}}, \bibinfo {author} {\bibfnamefont {M.}~\bibnamefont {Hirschberger}}, \bibinfo {author} {\bibfnamefont {P.}~\bibnamefont {Lampen-Kelley}}, \bibinfo {author} {\bibfnamefont {A.}~\bibnamefont {Banerjee}}, \bibinfo {author} {\bibfnamefont {J.}~\bibnamefont {Yan}}, \bibinfo {author} {\bibfnamefont {D.~G.}\ \bibnamefont {Mandrus}}, \bibinfo {author} {\bibfnamefont {S.~E.}\ \bibnamefont {Nagler}},\ and\ \bibinfo {author} {\bibfnamefont {N.~P.}\ \bibnamefont {Ong}},\ }\bibfield  {title} {\bibinfo {title} {{Oscillations of the thermal conductivity in the spin-liquid state of $\alpha$-{RuCl}$_3$}},\ }\href {https://doi.org/10.1038/s41567-021-01243-x} {\bibfield  {journal} {\bibinfo  {journal} {Nature Physics}\ }\textbf {\bibinfo {volume} {17}},\ \bibinfo {pages} {915} (\bibinfo {year} {2021})}\BibitemShut {NoStop}%
\bibitem [{\citenamefont {Bruin}\ \emph {et~al.}(2022)\citenamefont {Bruin}, \citenamefont {Claus}, \citenamefont {Matsumoto}, \citenamefont {Kurita}, \citenamefont {Tanaka},\ and\ \citenamefont {Takagi}}]{bruin2022}%
  \BibitemOpen
  \bibfield  {author} {\bibinfo {author} {\bibfnamefont {J.~A.~N.}\ \bibnamefont {Bruin}}, \bibinfo {author} {\bibfnamefont {R.~R.}\ \bibnamefont {Claus}}, \bibinfo {author} {\bibfnamefont {Y.}~\bibnamefont {Matsumoto}}, \bibinfo {author} {\bibfnamefont {N.}~\bibnamefont {Kurita}}, \bibinfo {author} {\bibfnamefont {H.}~\bibnamefont {Tanaka}},\ and\ \bibinfo {author} {\bibfnamefont {H.}~\bibnamefont {Takagi}},\ }\bibfield  {title} {\bibinfo {title} {Robustness of the thermal {H}all effect close to half-quantization in $\alpha$-{RuCl}$_3$},\ }\href {https://doi.org/10.1038/s41567-021-01501-y} {\bibfield  {journal} {\bibinfo  {journal} {Nature Physics}\ }\textbf {\bibinfo {volume} {18}},\ \bibinfo {pages} {401} (\bibinfo {year} {2022})}\BibitemShut {NoStop}%
\bibitem [{\citenamefont {Kasahara}\ \emph {et~al.}(2022)\citenamefont {Kasahara}, \citenamefont {Suetsugu}, \citenamefont {Asaba}, \citenamefont {Kasahara}, \citenamefont {Shibauchi}, \citenamefont {Kurita}, \citenamefont {Tanaka},\ and\ \citenamefont {Matsuda}}]{kasahara2022}%
  \BibitemOpen
  \bibfield  {author} {\bibinfo {author} {\bibfnamefont {Y.}~\bibnamefont {Kasahara}}, \bibinfo {author} {\bibfnamefont {S.}~\bibnamefont {Suetsugu}}, \bibinfo {author} {\bibfnamefont {T.}~\bibnamefont {Asaba}}, \bibinfo {author} {\bibfnamefont {S.}~\bibnamefont {Kasahara}}, \bibinfo {author} {\bibfnamefont {T.}~\bibnamefont {Shibauchi}}, \bibinfo {author} {\bibfnamefont {N.}~\bibnamefont {Kurita}}, \bibinfo {author} {\bibfnamefont {H.}~\bibnamefont {Tanaka}},\ and\ \bibinfo {author} {\bibfnamefont {Y.}~\bibnamefont {Matsuda}},\ }\bibfield  {title} {\bibinfo {title} {Quantized and unquantized thermal {H}all conductance of the {K}itaev spin liquid candidate ${\alpha}$-{RuCl}$_{3}$},\ }\href {https://doi.org/10.1103/PhysRevB.106.L060410} {\bibfield  {journal} {\bibinfo  {journal} {Phys. Rev. B}\ }\textbf {\bibinfo {volume} {106}},\ \bibinfo {pages} {L060410} (\bibinfo {year} {2022})}\BibitemShut {NoStop}%
\bibitem [{\citenamefont {Lefran\ifmmode~\mbox{\c{c}}\else \c{c}\fi{}ois}\ \emph {et~al.}(2022)\citenamefont {Lefran\ifmmode~\mbox{\c{c}}\else \c{c}\fi{}ois}, \citenamefont {Grissonnanche}, \citenamefont {Baglo}, \citenamefont {Lampen-Kelley}, \citenamefont {Yan}, \citenamefont {Balz}, \citenamefont {Mandrus}, \citenamefont {Nagler}, \citenamefont {Kim}, \citenamefont {Kim}, \citenamefont {Doiron-Leyraud},\ and\ \citenamefont {Taillefer}}]{lefrancois2022}%
  \BibitemOpen
  \bibfield  {author} {\bibinfo {author} {\bibfnamefont {E.}~\bibnamefont {Lefran\ifmmode~\mbox{\c{c}}\else \c{c}\fi{}ois}}, \bibinfo {author} {\bibfnamefont {G.}~\bibnamefont {Grissonnanche}}, \bibinfo {author} {\bibfnamefont {J.}~\bibnamefont {Baglo}}, \bibinfo {author} {\bibfnamefont {P.}~\bibnamefont {Lampen-Kelley}}, \bibinfo {author} {\bibfnamefont {J.-Q.}\ \bibnamefont {Yan}}, \bibinfo {author} {\bibfnamefont {C.}~\bibnamefont {Balz}}, \bibinfo {author} {\bibfnamefont {D.}~\bibnamefont {Mandrus}}, \bibinfo {author} {\bibfnamefont {S.~E.}\ \bibnamefont {Nagler}}, \bibinfo {author} {\bibfnamefont {S.}~\bibnamefont {Kim}}, \bibinfo {author} {\bibfnamefont {Y.-J.}\ \bibnamefont {Kim}}, \bibinfo {author} {\bibfnamefont {N.}~\bibnamefont {Doiron-Leyraud}},\ and\ \bibinfo {author} {\bibfnamefont {L.}~\bibnamefont {Taillefer}},\ }\bibfield  {title} {\bibinfo {title} {Evidence of a {P}honon {H}all {E}ffect in the {K}itaev {S}pin {L}iquid {C}andidate $\ensuremath{\alpha}\text{\ensuremath{-}}${RuCl}$_{3}$},\ }\href {https://doi.org/10.1103/PhysRevX.12.021025} {\bibfield  {journal} {\bibinfo  {journal} {Phys. Rev. X}\ }\textbf {\bibinfo {volume} {12}},\ \bibinfo {pages} {021025} (\bibinfo {year} {2022})}\BibitemShut {NoStop}%
\bibitem [{\citenamefont {Imamura}\ \emph {et~al.}(shed)\citenamefont {Imamura}, \citenamefont {Suetsugu}, \citenamefont {Mizukami}, \citenamefont {Yoshida}, \citenamefont {Hashimoto}, \citenamefont {Ohtsuka}, \citenamefont {Kasahara}, \citenamefont {Kurita}, \citenamefont {Tanaka}, \citenamefont {Noh}, \citenamefont {Nasu}, \citenamefont {Moon}, \citenamefont {Matsuda},\ and\ \citenamefont {Shibauchi}}]{imamura2023_arxiv}%
  \BibitemOpen
  \bibfield  {author} {\bibinfo {author} {\bibfnamefont {K.}~\bibnamefont {Imamura}}, \bibinfo {author} {\bibfnamefont {S.}~\bibnamefont {Suetsugu}}, \bibinfo {author} {\bibfnamefont {Y.}~\bibnamefont {Mizukami}}, \bibinfo {author} {\bibfnamefont {Y.}~\bibnamefont {Yoshida}}, \bibinfo {author} {\bibfnamefont {K.}~\bibnamefont {Hashimoto}}, \bibinfo {author} {\bibfnamefont {K.}~\bibnamefont {Ohtsuka}}, \bibinfo {author} {\bibfnamefont {Y.}~\bibnamefont {Kasahara}}, \bibinfo {author} {\bibfnamefont {N.}~\bibnamefont {Kurita}}, \bibinfo {author} {\bibfnamefont {H.}~\bibnamefont {Tanaka}}, \bibinfo {author} {\bibfnamefont {P.}~\bibnamefont {Noh}}, \bibinfo {author} {\bibfnamefont {J.}~\bibnamefont {Nasu}}, \bibinfo {author} {\bibfnamefont {E.~G.}\ \bibnamefont {Moon}}, \bibinfo {author} {\bibfnamefont {Y.}~\bibnamefont {Matsuda}},\ and\ \bibinfo {author} {\bibfnamefont {T.}~\bibnamefont {Shibauchi}},\ }\bibfield  {title} {\bibinfo {title} {{Majorana-fermion origin of the planar thermal Hall effect in the Kitaev magnet $\alpha$-RuCl$_3$}},\ }\href {https://arxiv.org/abs/2305.10619} {\bibfield  {journal} {\bibinfo  {journal} {arxiv:2305.10619}\ } (\bibinfo {year} {unpublished})}\BibitemShut {NoStop}%
\bibitem [{\citenamefont {Winter}\ \emph {et~al.}(2017)\citenamefont {Winter}, \citenamefont {Riedl}, \citenamefont {Maksimov}, \citenamefont {Chernyshev}, \citenamefont {Honecker},\ and\ \citenamefont {Valent{\'\i}}}]{winter2017_nc}%
  \BibitemOpen
  \bibfield  {author} {\bibinfo {author} {\bibfnamefont {S.~M.}\ \bibnamefont {Winter}}, \bibinfo {author} {\bibfnamefont {K.}~\bibnamefont {Riedl}}, \bibinfo {author} {\bibfnamefont {P.~A.}\ \bibnamefont {Maksimov}}, \bibinfo {author} {\bibfnamefont {A.~L.}\ \bibnamefont {Chernyshev}}, \bibinfo {author} {\bibfnamefont {A.}~\bibnamefont {Honecker}},\ and\ \bibinfo {author} {\bibfnamefont {R.}~\bibnamefont {Valent{\'\i}}},\ }\bibfield  {title} {\bibinfo {title} {Breakdown of magnons in a strongly spin-orbital coupled magnet},\ }\href {https://doi.org/10.1038/s41467-017-01177-0} {\bibfield  {journal} {\bibinfo  {journal} {Nat. Commun.}\ }\textbf {\bibinfo {volume} {8}},\ \bibinfo {pages} {1152} (\bibinfo {year} {2017})}\BibitemShut {NoStop}%
\bibitem [{\citenamefont {Maksimov}\ and\ \citenamefont {Chernyshev}(2020)}]{maksimov2020}%
  \BibitemOpen
  \bibfield  {author} {\bibinfo {author} {\bibfnamefont {P.~A.}\ \bibnamefont {Maksimov}}\ and\ \bibinfo {author} {\bibfnamefont {A.~L.}\ \bibnamefont {Chernyshev}},\ }\bibfield  {title} {\bibinfo {title} {Rethinking $\ensuremath{\alpha}\text{\ensuremath{-}}${RuCl}$_{3}$},\ }\href {https://doi.org/10.1103/PhysRevResearch.2.033011} {\bibfield  {journal} {\bibinfo  {journal} {Phys. Rev. Res.}\ }\textbf {\bibinfo {volume} {2}},\ \bibinfo {pages} {033011} (\bibinfo {year} {2020})}\BibitemShut {NoStop}%
\bibitem [{\citenamefont {Kusunose}\ and\ \citenamefont {Kuramoto}(2001)}]{kusunose2001}%
  \BibitemOpen
  \bibfield  {author} {\bibinfo {author} {\bibfnamefont {H.}~\bibnamefont {Kusunose}}\ and\ \bibinfo {author} {\bibfnamefont {Y.}~\bibnamefont {Kuramoto}},\ }\bibfield  {title} {\bibinfo {title} {Spin-{O}rbital {W}ave {E}xcitations in {O}rbitally {D}egenerate {E}xchange {M}odel with {M}ultipolar {I}nteractions},\ }\href {https://journals.jps.jp/doi/10.1143/JPSJ.70.3076} {\bibfield  {journal} {\bibinfo  {journal} {J. Phys. Soc. Jpn.}\ }\textbf {\bibinfo {volume} {70}},\ \bibinfo {pages} {3076} (\bibinfo {year} {2001})}\BibitemShut {NoStop}%
\bibitem [{\citenamefont {Nasu}\ and\ \citenamefont {Naka}(2021)}]{nasu2021}%
  \BibitemOpen
  \bibfield  {author} {\bibinfo {author} {\bibfnamefont {J.}~\bibnamefont {Nasu}}\ and\ \bibinfo {author} {\bibfnamefont {M.}~\bibnamefont {Naka}},\ }\bibfield  {title} {\bibinfo {title} {Spin {S}eebeck effect in nonmagnetic excitonic insulators},\ }\href {https://doi.org/10.1103/PhysRevB.103.L121104} {\bibfield  {journal} {\bibinfo  {journal} {Phys. Rev. B}\ }\textbf {\bibinfo {volume} {103}},\ \bibinfo {pages} {L121104} (\bibinfo {year} {2021})}\BibitemShut {NoStop}%
\bibitem [{\citenamefont {Nasu}\ and\ \citenamefont {Hayami}(2022)}]{nasu2022}%
  \BibitemOpen
  \bibfield  {author} {\bibinfo {author} {\bibfnamefont {J.}~\bibnamefont {Nasu}}\ and\ \bibinfo {author} {\bibfnamefont {S.}~\bibnamefont {Hayami}},\ }\bibfield  {title} {\bibinfo {title} {Antisymmetric thermopolarization by electric toroidicity},\ }\href {https://doi.org/10.1103/PhysRevB.105.245125} {\bibfield  {journal} {\bibinfo  {journal} {Phys. Rev. B}\ }\textbf {\bibinfo {volume} {105}},\ \bibinfo {pages} {245125} (\bibinfo {year} {2022})}\BibitemShut {NoStop}%
\bibitem [{\citenamefont {Colpa}(1978)}]{colpa}%
  \BibitemOpen
  \bibfield  {author} {\bibinfo {author} {\bibfnamefont {J.}~\bibnamefont {Colpa}},\ }\bibfield  {title} {\bibinfo {title} {Diagonalization of the quadratic boson hamiltonian},\ }\href {https://doi.org/https://doi.org/10.1016/0378-4371(78)90160-7} {\bibfield  {journal} {\bibinfo  {journal} {Physica}\ }\textbf {\bibinfo {volume} {93A}},\ \bibinfo {pages} {327} (\bibinfo {year} {1978})}\BibitemShut {NoStop}%
\bibitem [{\citenamefont {Mourigal}\ \emph {et~al.}(2010{\natexlab{b}})\citenamefont {Mourigal}, \citenamefont {Zhitomirsky},\ and\ \citenamefont {Chernyshev}}]{zhitomirsky2010}%
  \BibitemOpen
  \bibfield  {author} {\bibinfo {author} {\bibfnamefont {M.}~\bibnamefont {Mourigal}}, \bibinfo {author} {\bibfnamefont {M.~E.}\ \bibnamefont {Zhitomirsky}},\ and\ \bibinfo {author} {\bibfnamefont {A.~L.}\ \bibnamefont {Chernyshev}},\ }\bibfield  {title} {\bibinfo {title} {Field-induced decay dynamics in square-lattice antiferromagnets},\ }\href {https://doi.org/10.1103/PhysRevB.82.144402} {\bibfield  {journal} {\bibinfo  {journal} {Phys. Rev. B}\ }\textbf {\bibinfo {volume} {82}},\ \bibinfo {pages} {144402} (\bibinfo {year} {2010}{\natexlab{b}})}\BibitemShut {NoStop}%
\bibitem [{\citenamefont {Chernyshev}\ and\ \citenamefont {Zhitomirsky}(2009)}]{chernyshev2009}%
  \BibitemOpen
  \bibfield  {author} {\bibinfo {author} {\bibfnamefont {A.~L.}\ \bibnamefont {Chernyshev}}\ and\ \bibinfo {author} {\bibfnamefont {M.~E.}\ \bibnamefont {Zhitomirsky}},\ }\bibfield  {title} {\bibinfo {title} {Spin waves in a triangular lattice antiferromagnet: {D}ecays, spectrum renormalization, and singularities},\ }\href {https://doi.org/10.1103/PhysRevB.79.144416} {\bibfield  {journal} {\bibinfo  {journal} {Phys. Rev. B}\ }\textbf {\bibinfo {volume} {79}},\ \bibinfo {pages} {144416} (\bibinfo {year} {2009})}\BibitemShut {NoStop}%
\bibitem [{\citenamefont {Mahan}(2000)}]{mahan}%
  \BibitemOpen
  \bibfield  {author} {\bibinfo {author} {\bibfnamefont {G.~D.}\ \bibnamefont {Mahan}},\ }\href@noop {} {\bibfield  {journal} {\bibinfo  {journal} {{\it Many-Particle Physics}, 3rd ed.}\ } (\bibinfo {year} {2000})}\BibitemShut {NoStop}%
\bibitem [{\citenamefont {Maksimov}\ \emph {et~al.}(2016)\citenamefont {Maksimov}, \citenamefont {Zhitomirsky},\ and\ \citenamefont {Chernyshev}}]{maksimov2016_prl}%
  \BibitemOpen
  \bibfield  {author} {\bibinfo {author} {\bibfnamefont {P.~A.}\ \bibnamefont {Maksimov}}, \bibinfo {author} {\bibfnamefont {M.~E.}\ \bibnamefont {Zhitomirsky}},\ and\ \bibinfo {author} {\bibfnamefont {A.~L.}\ \bibnamefont {Chernyshev}},\ }\bibfield  {title} {\bibinfo {title} {{Field-induced decays in XXZ triangular-lattice antiferromagnets}},\ }\href {https://doi.org/10.1103/PhysRevB.94.140407} {\bibfield  {journal} {\bibinfo  {journal} {Phys. Rev. B}\ }\textbf {\bibinfo {volume} {94}},\ \bibinfo {pages} {140407} (\bibinfo {year} {2016})}\BibitemShut {NoStop}%
\bibitem [{\citenamefont {Pershoguba}\ \emph {et~al.}(2018)\citenamefont {Pershoguba}, \citenamefont {Banerjee}, \citenamefont {Lashley}, \citenamefont {Park}, \citenamefont {\AA{}gren}, \citenamefont {Aeppli},\ and\ \citenamefont {Balatsky}}]{pershoguba2018}%
  \BibitemOpen
  \bibfield  {author} {\bibinfo {author} {\bibfnamefont {S.~S.}\ \bibnamefont {Pershoguba}}, \bibinfo {author} {\bibfnamefont {S.}~\bibnamefont {Banerjee}}, \bibinfo {author} {\bibfnamefont {J.~C.}\ \bibnamefont {Lashley}}, \bibinfo {author} {\bibfnamefont {J.}~\bibnamefont {Park}}, \bibinfo {author} {\bibfnamefont {H.}~\bibnamefont {\AA{}gren}}, \bibinfo {author} {\bibfnamefont {G.}~\bibnamefont {Aeppli}},\ and\ \bibinfo {author} {\bibfnamefont {A.~V.}\ \bibnamefont {Balatsky}},\ }\bibfield  {title} {\bibinfo {title} {Dirac {M}agnons in {H}oneycomb {F}erromagnets},\ }\href {https://doi.org/10.1103/PhysRevX.8.011010} {\bibfield  {journal} {\bibinfo  {journal} {Phys. Rev. X}\ }\textbf {\bibinfo {volume} {8}},\ \bibinfo {pages} {011010} (\bibinfo {year} {2018})}\BibitemShut {NoStop}%
\bibitem [{\citenamefont {Maksimov}\ and\ \citenamefont {Chernyshev}(2016)}]{maksimov2016_prb}%
  \BibitemOpen
  \bibfield  {author} {\bibinfo {author} {\bibfnamefont {P.~A.}\ \bibnamefont {Maksimov}}\ and\ \bibinfo {author} {\bibfnamefont {A.~L.}\ \bibnamefont {Chernyshev}},\ }\bibfield  {title} {\bibinfo {title} {Field-induced dynamical properties of the $\mathit{XXZ}$ model on a honeycomb lattice},\ }\href {https://doi.org/10.1103/PhysRevB.93.014418} {\bibfield  {journal} {\bibinfo  {journal} {Phys. Rev. B}\ }\textbf {\bibinfo {volume} {93}},\ \bibinfo {pages} {014418} (\bibinfo {year} {2016})}\BibitemShut {NoStop}%
\bibitem [{\citenamefont {Smit}\ \emph {et~al.}(2020)\citenamefont {Smit}, \citenamefont {Keupert}, \citenamefont {Tsyplyatyev}, \citenamefont {Maksimov}, \citenamefont {Chernyshev},\ and\ \citenamefont {Kopietz}}]{smit2020}%
  \BibitemOpen
  \bibfield  {author} {\bibinfo {author} {\bibfnamefont {R.~L.}\ \bibnamefont {Smit}}, \bibinfo {author} {\bibfnamefont {S.}~\bibnamefont {Keupert}}, \bibinfo {author} {\bibfnamefont {O.}~\bibnamefont {Tsyplyatyev}}, \bibinfo {author} {\bibfnamefont {P.~A.}\ \bibnamefont {Maksimov}}, \bibinfo {author} {\bibfnamefont {A.~L.}\ \bibnamefont {Chernyshev}},\ and\ \bibinfo {author} {\bibfnamefont {P.}~\bibnamefont {Kopietz}},\ }\bibfield  {title} {\bibinfo {title} {Magnon damping in the zigzag phase of the {K}itaev-{H}eisenberg-$\mathrm{\ensuremath{\Gamma}}$ model on a honeycomb lattice},\ }\href {https://doi.org/10.1103/PhysRevB.101.054424} {\bibfield  {journal} {\bibinfo  {journal} {Phys. Rev. B}\ }\textbf {\bibinfo {volume} {101}},\ \bibinfo {pages} {054424} (\bibinfo {year} {2020})}\BibitemShut {NoStop}%
\bibitem [{\citenamefont {Maksimov}\ and\ \citenamefont {Chernyshev}(2022)}]{maksimov2022}%
  \BibitemOpen
  \bibfield  {author} {\bibinfo {author} {\bibfnamefont {P.~A.}\ \bibnamefont {Maksimov}}\ and\ \bibinfo {author} {\bibfnamefont {A.~L.}\ \bibnamefont {Chernyshev}},\ }\bibfield  {title} {\bibinfo {title} {{Easy-plane anisotropic-exchange magnets on a honeycomb lattice: Quantum effects and dealing with them}},\ }\href {https://doi.org/10.1103/PhysRevB.106.214411} {\bibfield  {journal} {\bibinfo  {journal} {Phys. Rev. B}\ }\textbf {\bibinfo {volume} {106}},\ \bibinfo {pages} {214411} (\bibinfo {year} {2022})}\BibitemShut {NoStop}%
\bibitem [{\citenamefont {Rau}\ \emph {et~al.}(2019)\citenamefont {Rau}, \citenamefont {Moessner},\ and\ \citenamefont {McClarty}}]{rau2019}%
  \BibitemOpen
  \bibfield  {author} {\bibinfo {author} {\bibfnamefont {J.~G.}\ \bibnamefont {Rau}}, \bibinfo {author} {\bibfnamefont {R.}~\bibnamefont {Moessner}},\ and\ \bibinfo {author} {\bibfnamefont {P.~A.}\ \bibnamefont {McClarty}},\ }\bibfield  {title} {\bibinfo {title} {Magnon interactions in the frustrated pyrochlore ferromagnet {${\mathrm{Yb}}_{2}{\mathrm{Ti}}_{2}{\mathrm{O}}_{7}$}},\ }\href {https://doi.org/10.1103/PhysRevB.100.104423} {\bibfield  {journal} {\bibinfo  {journal} {Phys. Rev. B}\ }\textbf {\bibinfo {volume} {100}},\ \bibinfo {pages} {104423} (\bibinfo {year} {2019})}\BibitemShut {NoStop}%
\bibitem [{\citenamefont {Yoshitake}\ \emph {et~al.}(2020)\citenamefont {Yoshitake}, \citenamefont {Nasu}, \citenamefont {Kato},\ and\ \citenamefont {Motome}}]{yoshitake2020}%
  \BibitemOpen
  \bibfield  {author} {\bibinfo {author} {\bibfnamefont {J.}~\bibnamefont {Yoshitake}}, \bibinfo {author} {\bibfnamefont {J.}~\bibnamefont {Nasu}}, \bibinfo {author} {\bibfnamefont {Y.}~\bibnamefont {Kato}},\ and\ \bibinfo {author} {\bibfnamefont {Y.}~\bibnamefont {Motome}},\ }\bibfield  {title} {\bibinfo {title} {Majorana-magnon crossover by a magnetic field in the {K}itaev model: {C}ontinuous-time quantum {M}onte {C}arlo study},\ }\href {https://doi.org/10.1103/PhysRevB.101.100408} {\bibfield  {journal} {\bibinfo  {journal} {Phys. Rev. B}\ }\textbf {\bibinfo {volume} {101}},\ \bibinfo {pages} {100408} (\bibinfo {year} {2020})}\BibitemShut {NoStop}%
\bibitem [{\citenamefont {Motome}\ and\ \citenamefont {Nasu}(2020)}]{Motome2020rev}%
  \BibitemOpen
  \bibfield  {author} {\bibinfo {author} {\bibfnamefont {Y.}~\bibnamefont {Motome}}\ and\ \bibinfo {author} {\bibfnamefont {J.}~\bibnamefont {Nasu}},\ }\bibfield  {title} {\bibinfo {title} {Hunting {M}ajorana {F}ermions in {K}itaev {M}agnets},\ }\href {https://doi.org/10.7566/JPSJ.89.012002} {\bibfield  {journal} {\bibinfo  {journal} {J. Phys. Soc. Jpn.}\ }\textbf {\bibinfo {volume} {89}},\ \bibinfo {pages} {012002} (\bibinfo {year} {2020})}\BibitemShut {NoStop}%
\bibitem [{\citenamefont {Zhang}\ \emph {et~al.}(2022)\citenamefont {Zhang}, \citenamefont {Hal{\'a}sz},\ and\ \citenamefont {Batista}}]{zhang2022}%
  \BibitemOpen
  \bibfield  {author} {\bibinfo {author} {\bibfnamefont {S.-S.}\ \bibnamefont {Zhang}}, \bibinfo {author} {\bibfnamefont {G.~B.}\ \bibnamefont {Hal{\'a}sz}},\ and\ \bibinfo {author} {\bibfnamefont {C.~D.}\ \bibnamefont {Batista}},\ }\bibfield  {title} {\bibinfo {title} {Theory of the {K}itaev model in a {$[$}111{$]$} magnetic field},\ }\href {https://doi.org/10.1038/s41467-022-28014-3} {\bibfield  {journal} {\bibinfo  {journal} {Nat. Commun,}\ }\textbf {\bibinfo {volume} {13}},\ \bibinfo {pages} {399} (\bibinfo {year} {2022})}\BibitemShut {NoStop}%
\bibitem [{\citenamefont {Hickey}\ and\ \citenamefont {Trebst}(2019)}]{hickey2019}%
  \BibitemOpen
  \bibfield  {author} {\bibinfo {author} {\bibfnamefont {C.}~\bibnamefont {Hickey}}\ and\ \bibinfo {author} {\bibfnamefont {S.}~\bibnamefont {Trebst}},\ }\bibfield  {title} {\bibinfo {title} {Emergence of a field-driven {U}(1) spin liquid in the {K}itaev honeycomb model},\ }\href {https://doi.org/10.1038/s41467-019-08459-9} {\bibfield  {journal} {\bibinfo  {journal} {Nat. Commun.}\ }\textbf {\bibinfo {volume} {10}},\ \bibinfo {pages} {530} (\bibinfo {year} {2019})}\BibitemShut {NoStop}%
\bibitem [{\citenamefont {Jackeli}\ and\ \citenamefont {Khaliullin}(2009)}]{jackeli2009}%
  \BibitemOpen
  \bibfield  {author} {\bibinfo {author} {\bibfnamefont {G.}~\bibnamefont {Jackeli}}\ and\ \bibinfo {author} {\bibfnamefont {G.}~\bibnamefont {Khaliullin}},\ }\bibfield  {title} {\bibinfo {title} {Mott {I}nsulators in the {S}trong {S}pin-{O}rbit {C}oupling {L}imit: {F}rom {H}eisenberg to a {Q}uantum {C}ompass and {K}itaev {M}odels},\ }\href {https://doi.org/10.1103/PhysRevLett.102.017205} {\bibfield  {journal} {\bibinfo  {journal} {Phys. Rev. Lett.}\ }\textbf {\bibinfo {volume} {102}},\ \bibinfo {pages} {017205} (\bibinfo {year} {2009})}\BibitemShut {NoStop}%
\bibitem [{\citenamefont {Motome}\ \emph {et~al.}(2020)\citenamefont {Motome}, \citenamefont {Sano}, \citenamefont {Jang}, \citenamefont {Sugita},\ and\ \citenamefont {Kato}}]{motome2020_iop}%
  \BibitemOpen
  \bibfield  {author} {\bibinfo {author} {\bibfnamefont {Y.}~\bibnamefont {Motome}}, \bibinfo {author} {\bibfnamefont {R.}~\bibnamefont {Sano}}, \bibinfo {author} {\bibfnamefont {S.}~\bibnamefont {Jang}}, \bibinfo {author} {\bibfnamefont {Y.}~\bibnamefont {Sugita}},\ and\ \bibinfo {author} {\bibfnamefont {Y.}~\bibnamefont {Kato}},\ }\bibfield  {title} {\bibinfo {title} {Materials design of {K}itaev spin liquids beyond the {J}ackeli--{K}haliullin mechanism},\ }\href {https://doi.org/10.1088/1361-648X/ab8525} {\bibfield  {journal} {\bibinfo  {journal} {J. Phys.: Condens. Matter}\ }\textbf {\bibinfo {volume} {32}},\ \bibinfo {pages} {404001} (\bibinfo {year} {2020})}\BibitemShut {NoStop}%
\bibitem [{\citenamefont {Banerjee}\ \emph {et~al.}(2016)\citenamefont {Banerjee}, \citenamefont {Bridges}, \citenamefont {Yan}, \citenamefont {Aczel}, \citenamefont {Li}, \citenamefont {Stone}, \citenamefont {Granroth}, \citenamefont {Lumsden}, \citenamefont {Yiu}, \citenamefont {Knolle}, \citenamefont {Bhattacharjee}, \citenamefont {Kovrizhin}, \citenamefont {Moessner}, \citenamefont {Tennant}, \citenamefont {Mandrus},\ and\ \citenamefont {Nagler}}]{banerjee2016}%
  \BibitemOpen
  \bibfield  {author} {\bibinfo {author} {\bibfnamefont {A.}~\bibnamefont {Banerjee}}, \bibinfo {author} {\bibfnamefont {C.~A.}\ \bibnamefont {Bridges}}, \bibinfo {author} {\bibfnamefont {J.~Q.}\ \bibnamefont {Yan}}, \bibinfo {author} {\bibfnamefont {A.~A.}\ \bibnamefont {Aczel}}, \bibinfo {author} {\bibfnamefont {L.}~\bibnamefont {Li}}, \bibinfo {author} {\bibfnamefont {M.~B.}\ \bibnamefont {Stone}}, \bibinfo {author} {\bibfnamefont {G.~E.}\ \bibnamefont {Granroth}}, \bibinfo {author} {\bibfnamefont {M.~D.}\ \bibnamefont {Lumsden}}, \bibinfo {author} {\bibfnamefont {Y.}~\bibnamefont {Yiu}}, \bibinfo {author} {\bibfnamefont {J.}~\bibnamefont {Knolle}}, \bibinfo {author} {\bibfnamefont {S.}~\bibnamefont {Bhattacharjee}}, \bibinfo {author} {\bibfnamefont {D.~L.}\ \bibnamefont {Kovrizhin}}, \bibinfo {author} {\bibfnamefont {R.}~\bibnamefont {Moessner}}, \bibinfo {author} {\bibfnamefont {D.~A.}\ \bibnamefont {Tennant}}, \bibinfo {author} {\bibfnamefont {D.~G.}\ \bibnamefont {Mandrus}},\ and\ \bibinfo {author} {\bibfnamefont {S.~E.}\ \bibnamefont {Nagler}},\ }\bibfield  {title} {\bibinfo {title} {Proximate {K}itaev quantum spin liquid behaviour in a honeycomb magnet},\ }\href {https://doi.org/10.1038/nmat4604} {\bibfield  {journal} {\bibinfo  {journal} {Nat. Mater.}\ }\textbf {\bibinfo {volume} {15}},\ \bibinfo {pages} {733} (\bibinfo {year} {2016})}\BibitemShut {NoStop}%
\bibitem [{\citenamefont {Laurell}\ and\ \citenamefont {Okamoto}(2020)}]{laurell2020}%
  \BibitemOpen
  \bibfield  {author} {\bibinfo {author} {\bibfnamefont {P.}~\bibnamefont {Laurell}}\ and\ \bibinfo {author} {\bibfnamefont {S.}~\bibnamefont {Okamoto}},\ }\bibfield  {title} {\bibinfo {title} {{Dynamical and thermal magnetic properties of the Kitaev spin liquid candidate $\alpha$-{RuCl}$_3$}},\ }\href {https://doi.org/10.1038/s41535-019-0203-y} {\bibfield  {journal} {\bibinfo  {journal} {npj Quantum Mater.}\ }\textbf {\bibinfo {volume} {5}},\ \bibinfo {pages} {2} (\bibinfo {year} {2020})}\BibitemShut {NoStop}%
\bibitem [{\citenamefont {Sandilands}\ \emph {et~al.}(2015)\citenamefont {Sandilands}, \citenamefont {Tian}, \citenamefont {Plumb}, \citenamefont {Kim},\ and\ \citenamefont {Burch}}]{sandilands2015}%
  \BibitemOpen
  \bibfield  {author} {\bibinfo {author} {\bibfnamefont {L.~J.}\ \bibnamefont {Sandilands}}, \bibinfo {author} {\bibfnamefont {Y.}~\bibnamefont {Tian}}, \bibinfo {author} {\bibfnamefont {K.~W.}\ \bibnamefont {Plumb}}, \bibinfo {author} {\bibfnamefont {Y.-J.}\ \bibnamefont {Kim}},\ and\ \bibinfo {author} {\bibfnamefont {K.~S.}\ \bibnamefont {Burch}},\ }\bibfield  {title} {\bibinfo {title} {Scattering {C}ontinuum and {P}ossible {F}ractionalized {E}xcitations in $\ensuremath{\alpha}\text{\ensuremath{-}}${RuCl}$_{3}$},\ }\href {https://doi.org/10.1103/PhysRevLett.114.147201} {\bibfield  {journal} {\bibinfo  {journal} {Phys. Rev. Lett.}\ }\textbf {\bibinfo {volume} {114}},\ \bibinfo {pages} {147201} (\bibinfo {year} {2015})}\BibitemShut {NoStop}%
\bibitem [{\citenamefont {Do}\ \emph {et~al.}(2017)\citenamefont {Do}, \citenamefont {Park}, \citenamefont {Yoshitake}, \citenamefont {Nasu}, \citenamefont {Motome}, \citenamefont {Kwon}, \citenamefont {Adroja}, \citenamefont {Voneshen}, \citenamefont {Kim}, \citenamefont {Jang}, \citenamefont {Park}, \citenamefont {Choi},\ and\ \citenamefont {Ji}}]{shdo2017}%
  \BibitemOpen
  \bibfield  {author} {\bibinfo {author} {\bibfnamefont {S.-H.}\ \bibnamefont {Do}}, \bibinfo {author} {\bibfnamefont {S.-Y.}\ \bibnamefont {Park}}, \bibinfo {author} {\bibfnamefont {J.}~\bibnamefont {Yoshitake}}, \bibinfo {author} {\bibfnamefont {J.}~\bibnamefont {Nasu}}, \bibinfo {author} {\bibfnamefont {Y.}~\bibnamefont {Motome}}, \bibinfo {author} {\bibfnamefont {Y.~S.}\ \bibnamefont {Kwon}}, \bibinfo {author} {\bibfnamefont {D.~T.}\ \bibnamefont {Adroja}}, \bibinfo {author} {\bibfnamefont {D.~J.}\ \bibnamefont {Voneshen}}, \bibinfo {author} {\bibfnamefont {K.}~\bibnamefont {Kim}}, \bibinfo {author} {\bibfnamefont {T.~H.}\ \bibnamefont {Jang}}, \bibinfo {author} {\bibfnamefont {J.~H.}\ \bibnamefont {Park}}, \bibinfo {author} {\bibfnamefont {K.-Y.}\ \bibnamefont {Choi}},\ and\ \bibinfo {author} {\bibfnamefont {S.}~\bibnamefont {Ji}},\ }\bibfield  {title} {\bibinfo {title} {Majorana fermions in the {K}itaev quantum spin system $\alpha$-{RuCl}$_3$},\ }\href {https://doi.org/10.1038/nphys4264} {\bibfield  {journal} {\bibinfo  {journal} {Nature Physics}\ }\textbf {\bibinfo {volume} {13}},\ \bibinfo {pages} {1079} (\bibinfo {year} {2017})}\BibitemShut {NoStop}%
\bibitem [{\citenamefont {Yadav}\ \emph {et~al.}(2016)\citenamefont {Yadav}, \citenamefont {Bogdanov}, \citenamefont {Katukuri}, \citenamefont {Nishimoto}, \citenamefont {van~den Brink},\ and\ \citenamefont {Hozoi}}]{yadav2016}%
  \BibitemOpen
  \bibfield  {author} {\bibinfo {author} {\bibfnamefont {R.}~\bibnamefont {Yadav}}, \bibinfo {author} {\bibfnamefont {N.~A.}\ \bibnamefont {Bogdanov}}, \bibinfo {author} {\bibfnamefont {V.~M.}\ \bibnamefont {Katukuri}}, \bibinfo {author} {\bibfnamefont {S.}~\bibnamefont {Nishimoto}}, \bibinfo {author} {\bibfnamefont {J.}~\bibnamefont {van~den Brink}},\ and\ \bibinfo {author} {\bibfnamefont {L.}~\bibnamefont {Hozoi}},\ }\bibfield  {title} {\bibinfo {title} {Kitaev exchange and field-induced quantum spin-liquid states in honeycomb $\alpha$-{RuCl}$_3$},\ }\href {https://doi.org/10.1038/srep37925} {\bibfield  {journal} {\bibinfo  {journal} {Sci. Rep.}\ }\textbf {\bibinfo {volume} {6}},\ \bibinfo {pages} {37925} (\bibinfo {year} {2016})}\BibitemShut {NoStop}%
\bibitem [{\citenamefont {Matsumoto}\ and\ \citenamefont {Murakami}(2011)}]{matsumoto2011}%
  \BibitemOpen
  \bibfield  {author} {\bibinfo {author} {\bibfnamefont {R.}~\bibnamefont {Matsumoto}}\ and\ \bibinfo {author} {\bibfnamefont {S.}~\bibnamefont {Murakami}},\ }\bibfield  {title} {\bibinfo {title} {Theoretical {P}rediction of a {R}otating {M}agnon {W}ave {P}acket in {F}erromagnets},\ }\href {https://doi.org/10.1103/PhysRevLett.106.197202} {\bibfield  {journal} {\bibinfo  {journal} {Phys. Rev. Lett.}\ }\textbf {\bibinfo {volume} {106}},\ \bibinfo {pages} {197202} (\bibinfo {year} {2011})}\BibitemShut {NoStop}%
\bibitem [{\citenamefont {Murakami}\ and\ \citenamefont {Okamoto}(2017)}]{murakami2017}%
  \BibitemOpen
  \bibfield  {author} {\bibinfo {author} {\bibfnamefont {S.}~\bibnamefont {Murakami}}\ and\ \bibinfo {author} {\bibfnamefont {A.}~\bibnamefont {Okamoto}},\ }\bibfield  {title} {\bibinfo {title} {Thermal {H}all {E}ffect of {M}agnons},\ }\href {https://journals.jps.jp/doi/10.7566/JPSJ.86.011010} {\bibfield  {journal} {\bibinfo  {journal} {J. Phys. Soc. Jpn.}\ }\textbf {\bibinfo {volume} {86}},\ \bibinfo {pages} {011010} (\bibinfo {year} {2017})}\BibitemShut {NoStop}%
\bibitem [{\citenamefont {Attfield}(2015)}]{attfield2015orbital}%
  \BibitemOpen
  \bibfield  {author} {\bibinfo {author} {\bibfnamefont {J.~P.}\ \bibnamefont {Attfield}},\ }\bibfield  {title} {\bibinfo {title} {Orbital molecules in electronic materials},\ }\href {https://dx.doi.org/10.1063/1.4913736} {\bibfield  {journal} {\bibinfo  {journal} {APL Mater.}\ }\textbf {\bibinfo {volume} {3}} (\bibinfo {year} {2015})}\BibitemShut {NoStop}%
\bibitem [{\citenamefont {Hiroi}(2015)}]{hiroi2015structural}%
  \BibitemOpen
  \bibfield  {author} {\bibinfo {author} {\bibfnamefont {Z.}~\bibnamefont {Hiroi}},\ }\bibfield  {title} {\bibinfo {title} {Structural instability of the rutile compounds and its relevance to the metal--insulator transition of {VO$_2$}},\ }\href {https://doi.org/10.1016/j.progsolidstchem.2015.02.001} {\bibfield  {journal} {\bibinfo  {journal} {Prog. Solid. State Ch.}\ }\textbf {\bibinfo {volume} {43}},\ \bibinfo {pages} {47} (\bibinfo {year} {2015})}\BibitemShut {NoStop}%
\bibitem [{\citenamefont {Tsunetsugu}(2001)}]{Tsunetsugu2001}%
  \BibitemOpen
  \bibfield  {author} {\bibinfo {author} {\bibfnamefont {H.}~\bibnamefont {Tsunetsugu}},\ }\bibfield  {title} {\bibinfo {title} {Spin-singlet order in a pyrochlore antiferromagnet},\ }\href {https://doi.org/10.1103/PhysRevB.65.024415} {\bibfield  {journal} {\bibinfo  {journal} {Phys. Rev. B}\ }\textbf {\bibinfo {volume} {65}},\ \bibinfo {pages} {024415} (\bibinfo {year} {2001})}\BibitemShut {NoStop}%
\bibitem [{\citenamefont {Takeda}\ \emph {et~al.}(shed)\citenamefont {Takeda}, \citenamefont {Kawano}, \citenamefont {Tamura}, \citenamefont {Akazawa}, \citenamefont {Yan}, \citenamefont {Waki}, \citenamefont {Nakamura}, \citenamefont {Sato}, \citenamefont {Narumi}, \citenamefont {Hagiwara}, \citenamefont {Yamashita},\ and\ \citenamefont {Hotta}}]{Takeda2023_arxiv}%
  \BibitemOpen
  \bibfield  {author} {\bibinfo {author} {\bibfnamefont {H.}~\bibnamefont {Takeda}}, \bibinfo {author} {\bibfnamefont {M.}~\bibnamefont {Kawano}}, \bibinfo {author} {\bibfnamefont {K.}~\bibnamefont {Tamura}}, \bibinfo {author} {\bibfnamefont {M.}~\bibnamefont {Akazawa}}, \bibinfo {author} {\bibfnamefont {J.}~\bibnamefont {Yan}}, \bibinfo {author} {\bibfnamefont {T.}~\bibnamefont {Waki}}, \bibinfo {author} {\bibfnamefont {H.}~\bibnamefont {Nakamura}}, \bibinfo {author} {\bibfnamefont {K.}~\bibnamefont {Sato}}, \bibinfo {author} {\bibfnamefont {Y.}~\bibnamefont {Narumi}}, \bibinfo {author} {\bibfnamefont {M.}~\bibnamefont {Hagiwara}}, \bibinfo {author} {\bibfnamefont {M.}~\bibnamefont {Yamashita}},\ and\ \bibinfo {author} {\bibfnamefont {C.}~\bibnamefont {Hotta}},\ }\bibfield  {title} {\bibinfo {title} {Emergent {SU}(3) magnons and thermal {Hall} effect in the antiferromagnetic skyrmion lattice},\ }\href {https://arxiv.org/abs/2304.08029} {\bibfield  {journal} {\bibinfo  {journal} {arXiv:2304.08029}\ } (\bibinfo {year} {unpublished})}\BibitemShut {NoStop}%
\bibitem [{\citenamefont {Nasu}\ and\ \citenamefont {Ishihara}(2013)}]{Nasu2013}%
  \BibitemOpen
  \bibfield  {author} {\bibinfo {author} {\bibfnamefont {J.}~\bibnamefont {Nasu}}\ and\ \bibinfo {author} {\bibfnamefont {S.}~\bibnamefont {Ishihara}},\ }\bibfield  {title} {\bibinfo {title} {Vibronic excitation dynamics in orbitally degenerate correlated electron system},\ }\href {https://doi.org/10.1103/PhysRevB.88.205110} {\bibfield  {journal} {\bibinfo  {journal} {Phys. Rev. B}\ }\textbf {\bibinfo {volume} {88}},\ \bibinfo {pages} {205110} (\bibinfo {year} {2013})}\BibitemShut {NoStop}%
\bibitem [{\citenamefont {Iwazaki}\ \emph {et~al.}(shed)\citenamefont {Iwazaki}, \citenamefont {Shinaoka},\ and\ \citenamefont {Hoshino}}]{Iwazaki2023_arxiv}%
  \BibitemOpen
  \bibfield  {author} {\bibinfo {author} {\bibfnamefont {R.}~\bibnamefont {Iwazaki}}, \bibinfo {author} {\bibfnamefont {H.}~\bibnamefont {Shinaoka}},\ and\ \bibinfo {author} {\bibfnamefont {S.}~\bibnamefont {Hoshino}},\ }\bibfield  {title} {\bibinfo {title} {{Material-based analysis of spin-orbital Mott insulators}},\ }\href {https://arxiv.org/abs/2301.09824} {\bibfield  {journal} {\bibinfo  {journal} {arXiv:2301.09824}\ } (\bibinfo {year} {unpublished})}\BibitemShut {NoStop}%
\bibitem [{\citenamefont {Mook}\ \emph {et~al.}(2020)\citenamefont {Mook}, \citenamefont {Klinovaja},\ and\ \citenamefont {Loss}}]{mook2020}%
  \BibitemOpen
  \bibfield  {author} {\bibinfo {author} {\bibfnamefont {A.}~\bibnamefont {Mook}}, \bibinfo {author} {\bibfnamefont {J.}~\bibnamefont {Klinovaja}},\ and\ \bibinfo {author} {\bibfnamefont {D.}~\bibnamefont {Loss}},\ }\bibfield  {title} {\bibinfo {title} {Quantum damping of skyrmion crystal eigenmodes due to spontaneous quasiparticle decay},\ }\href {https://doi.org/10.1103/PhysRevResearch.2.033491} {\bibfield  {journal} {\bibinfo  {journal} {Phys. Rev. Res.}\ }\textbf {\bibinfo {volume} {2}},\ \bibinfo {pages} {033491} (\bibinfo {year} {2020})}\BibitemShut {NoStop}%
\bibitem [{\citenamefont {Rau}\ \emph {et~al.}(2018)\citenamefont {Rau}, \citenamefont {McClarty},\ and\ \citenamefont {Moessner}}]{rau2018}%
  \BibitemOpen
  \bibfield  {author} {\bibinfo {author} {\bibfnamefont {J.~G.}\ \bibnamefont {Rau}}, \bibinfo {author} {\bibfnamefont {P.~A.}\ \bibnamefont {McClarty}},\ and\ \bibinfo {author} {\bibfnamefont {R.}~\bibnamefont {Moessner}},\ }\bibfield  {title} {\bibinfo {title} {Pseudo-{G}oldstone {G}aps and {O}rder-by-{Q}uantum {D}isorder in {F}rustrated {M}agnets},\ }\href {https://doi.org/10.1103/PhysRevLett.121.237201} {\bibfield  {journal} {\bibinfo  {journal} {Phys. Rev. Lett.}\ }\textbf {\bibinfo {volume} {121}},\ \bibinfo {pages} {237201} (\bibinfo {year} {2018})}\BibitemShut {NoStop}%
\bibitem [{\citenamefont {Costa~Filho}\ \emph {et~al.}(2000)\citenamefont {Costa~Filho}, \citenamefont {Cottam},\ and\ \citenamefont {Farias}}]{costa2000}%
  \BibitemOpen
  \bibfield  {author} {\bibinfo {author} {\bibfnamefont {R.~N.}\ \bibnamefont {Costa~Filho}}, \bibinfo {author} {\bibfnamefont {M.~G.}\ \bibnamefont {Cottam}},\ and\ \bibinfo {author} {\bibfnamefont {G.~A.}\ \bibnamefont {Farias}},\ }\bibfield  {title} {\bibinfo {title} {Microscopic theory of dipole-exchange spin waves in ferromagnetic films: Linear and nonlinear processes},\ }\href {https://doi.org/10.1103/PhysRevB.62.6545} {\bibfield  {journal} {\bibinfo  {journal} {Phys. Rev. B}\ }\textbf {\bibinfo {volume} {62}},\ \bibinfo {pages} {6545} (\bibinfo {year} {2000})}\BibitemShut {NoStop}%
\end{thebibliography}%
\end{document}